\documentclass[10pt,a4paper]{article}  
\usepackage[T1]{fontenc}
\usepackage[utf8]{inputenc}
\usepackage{amsfonts}
\usepackage{amssymb}
\usepackage[italian, english ]{babel}
\usepackage{graphicx}
\usepackage{colortbl}                
\usepackage{latexsym}
\usepackage{mathtools}
\usepackage{amsmath}
\usepackage{amsthm}
\usepackage{empheq}
\usepackage{slashed}
\usepackage{cases}
\usepackage{subfigure}
\usepackage{titlesec}
\usepackage{bbold}
\usepackage{float}
\usepackage[affil-it]{authblk}

\usepackage{circuitikz}
\usetikzlibrary{circuits.ee.IEC, circuits.logic.US, circuits.logic.IEC}
\usetikzlibrary{patterns}
\usepackage{graphicx}
\usepackage{enumerate}
\usepackage{enumitem}
\usepackage[toc,page]{appendix}
\usepackage{caption}
\usepackage{empheq}
\usepackage{multirow}
\usepackage[version=3]{mhchem}
\usepackage{braket}
\usepackage{centernot}
\usepackage{relsize}
\usepackage{amsmath}
\usepackage{geometry}
\usepackage{lmodern}
\usepackage[tone]{tipa}
\usepackage{array}
\usepackage{bm}
\usepackage{slashed}
\usepackage{multicol}
\usepackage{amscd}
\usepackage{xcolor}

\usepackage{tcolorbox}
\usepackage{lipsum}
\tcbuselibrary{skins,breakable}
\usetikzlibrary{shadings,shadows}
\usepackage{tikz-cd}
\usepackage{tipa}
\usepackage{upgreek}

    {\endtcolorbox}

    {\endtcolorbox}

    {\endtcolorbox}

	\titleformat{\part}
	[display]
	{\centering\normalfont\Huge\bfseries}
	{\vspace{3pt}\titlerule[2pt]\vspace{3pt}\MakeUppercase{\partname} \thepart}
	{0pt}
	{\titlerule[2pt]\vspace{50pt}\huge\MakeUppercase}

	\titlespacing*{\part}{0pt}{50pt}{20pt}

\catcode`,\active

\catcode`\,12

\newcommand{\nn}{\noindent}
\newcommand{\de}{\partial}

\newcommand{\La}{\mathcal{L}}

\newcommand{\R}{\mathbb{R}}

\newcommand{\Ol}{\mathcal{O}}

\newcommand{\Q}{\mathbb{Q}}
\newcommand{\Z}{\mathbb{Z}}

\newcommand{\C}{\mathbb{C}}

\newcommand{\Pro}{\mathbb{P}}
\usepackage{jheppub}

\title{Wall crossing structure from quantum phenomena to Feynman Integrals }
\author[1]{R.Angius}
\author[2,3,4]{S.L.Cacciatori}
\author[2,3]{A.Massidda}

\affiliation[1]{Insituto de Fis\'ica Te\'orica IFT-UAM/CSIC,\protect\\ C/Nicol\'as Cabrera 13-15,Campus de Cantoblanco, 28049 Madrid, Spain}
\affiliation[2]{DiSAT, Universit\`a dell'Insubria, via Valleggio 11, 22100, Como, Italy}
\affiliation[3]{INFN, Sezione di Milano, via Celoria 16, 20133, Milano, Italy}
\affiliation[4]{Como Lake centre for AstroPhysics (CLAP), DiSAT, Universit\`a dell’Insubria, via Valleggio 11, 22100 Como, Italy}

\emailAdd{roberta.angius@csic.es}
\emailAdd{sergio.cacciatori@uninsubria.it}
\emailAdd{amassidda@uninsubria.it}

\abstract{
A growing body of evidence suggests that the complexity of Feynman integrals is best understood through geometry. 
Recent mathematical developments \cite{Kontsevich:2024mks} have illuminated the role of exponential integrals as periods of twisted de Rham cocycles over Betti cycles, providing a structured approach to tackle this problem in many situations. 
In this paper, we apply these concepts to show how families of physically relevant integrals, ranging from exponentials to logarithmic multivalued functions, can be recast as twisted periods of differential forms over homology cycles. In the case of holomorphic exponents, we provide explicit decompositions as thimble expansions and reveal a geometric wall-crossing structure behind the analytic continuation in parameters.
We then show that the generalization to multivalued functions provides the right framework to describe Feynman integrals in the Baikov representation, where the multivaluedness is governed by the logarithm of the Baikov polynomial. In this context, the thimble decomposition aligns with the decomposition into Master Integrals. We highlight how the wall-crossing structure allows for a sharp count of independent Master Integrals (or periods), circumventing complications arising from Stokes phenomena. Additionally, we study the large-parameter expansions of these integrals, whose coefficients correspond to periods of standard (co-)homology associated with families of algebraic varieties, and which reveal the dominant basis elements in different sectors of the wall crossing structure. This unifies perturbative expansions and geometric representation theory under a single cohomological framework.}

\begin{document}
\maketitle

\section{Introduction}
Computing integrals is one of the main activities of physicists: evaluate fluxes of electric or magnetic fields, determine expectation values for processes in quantum mechanics, calculate averages in statistical physics, or infinite dimensional integrals in defining quantum field theories, and so on and so forth. More broadly, many physical problems reduce to solving systems of nonlinear partial differential equations \cite{Kotikov1991123,KOTIKOV1991158,Lunev_1994,Remiddi:1997ny,Argeri:2007up} (see \cite{Grimm:2024tbg,Grimm:2025zhv} for recent developments in reduction methods). In very few cases, this can be done exactly, but generically it must be done perturbatively with various techniques. This is how Feynman integrals, for example, find their origin. The ever-increasing precision with which modern technologies allow us to test theory against experiments requires ever-increasing precision in solving the equations that characterize the theory. In most cases, this means being able to compute more and more difficult integrals.
Nevertheless, certain kinds of integrals are easier than others. This is what we learn from Stokes' theorem: when we consider fluxes of closed forms over closed cycles, we can deform the cycle and add exact forms to the integrands without changing the final result. Here is where homology and cohomology start playing a central role.\\
In the simplest cases, mathematically, these integrals correspond to periods of elements of a de Rham cohomology over closed cycles of a dual homology (closed manifolds up to boundaries). However, more general cases can be considered by suitably identifying the correct cohomology theory entering the game. An important example is the case of hypergeometric type integrals and their generalizations, like GKZ equation systems and Euler integrals, where it has been realized that they can be again understood in terms of co-homology after replacing the de Rham cohomology with its twisted version, and the dual cohomology with one realized in terms of open twisted cycles \cite{aomoto2011theory,yoshida2013hypergeometric,MANA:MANA19941660122,MANA:MANA19941680111,matsumoto1994,cho1995,AOMOTO1997119,matsumoto1998,adolphson_sperber_1997,Mimachi2003,Ohara98intersectionnumbers,Mimachi2004,OST2003,Goto:2013Laur,Yoshiaki-GOTO2015203,gotomatsumoto2015,matsubaraheo2019algorithm,Goto2022homology,Matsubara-Heo:2020lzo,Matsubara-Heo:2021dtm}.\\
In \cite{Mastrolia:2018uzb}, observing a similar underlying structure in Feynman integrals in QFT on Minkowski spacetime, it was proposed that the same strategies be applied to the computation of scattering amplitudes.  The key advantage lies in the fact that both homology and cohomology rings are finitely generated and endowed with a non-degenerate internal product, the topological intersection product for homology and its dual in cohomology, defined between closed forms. In this way, a given Feynman integral can be identified as an element of a finite dimensional vector space endowed with a double structure suitable for projecting any vector to a given basis, given precisely by the intersection products of (co)homology. Moreover, when a Hodge structure is available, important quadratic relations, generalizing the Riemann quadratic relations, can be determined, see \cite{Cacciatori:2021nli,Cattani:2008ec}. The identification of Feynman integrals with periods of a suitable cohomology thus allows to replace the integration by part identities (IBP) (\cite{Chetyrkin:1981qh,Laporta:2001dd}) with intersection projection, that provide a more systematic procedure, both for determining a basis of master integrals (MI) and to decompose the vector space accordingly, as well as for finding differential equations and quadratic identities satisfied by the MI, see \cite{Frellesvig:2019kgj,Mizera:2019vvs,Frellesvig:2019uqt,Frellesvig:2020qot,Mizera:2016jhj,Mizera:2017rqa,Kaderli:2019dny,Kalyanapuram:2020vil,Ma:2021cxg,Weinzierl:2020nhw,Gasparotto:2022mmp,Chen:2022lzr,Giroux:2022wav,Ahmed:2023htp,Duhr:2024bzt,Crisanti:2024onv} for several successful applications to physics. The efficiency of this strategy depends on identifying the appropriate cohomology and efficient ways to calculate the intersection product \cite{Brunello:2023rpq,KeijiMatsumoto2024,Caron-Huot:2021xqj,Caron-Huot:2021iev,Chestnov:2022alh,Chestnov:2022xsy,Chestnov:2024mnw,brunello2024}. Recently, \cite{Cacciatori:2022mbi}, it has been shown that intersection theory plays a role in more general situations, beyond the realm of Feynman integrals, for integrals involving generalized orthogonal polynomials and computations of matrix elements in quantum mechanics, Green's functions in field theories, higher-order moments of probability distributions, suggesting a deep intertwinement between physics, geometry, and statistics.\footnote{It is not trivially expected an arbitrary generalization of the applicability of these methods to any situation, given that ``being a period'' is a special situation, see \cite{Kontsevich:2001kza} for an introduction to this idea.}\\
Driven by emerging perspectives that reveal integration theory as a unifying language in mathematical physics, in the present paper we started investigating a quite general approach that can be applied to several questions of physical interest, like Feynman integrals, Fourier integrals \cite{Brunello2}, and the examples investigated in \cite{Cacciatori:2022mbi}: the exponential integrals.  Our analysis will be based on \cite{Kontsevich:2024mks}, where exponential integrals, and the related wall-crossing structures, are analyzed in light of a series of isomorphisms between the twisted de Rham and Betti cohomologies in their local and global versions. In the present paper, we will mainly concentrate on the case where the exponential integral is defined starting from a holomorphic function on a complex manifold $X$, $f:X\rightarrow \mathbb C$, so that
\begin{align*}
 I_\Gamma(f,\gamma)=\int_{\Gamma} e^{-\gamma f} \mu, 
\end{align*}
where we assume $f$ to have a finite number of isolated critical points, $\gamma$ is a non-vanishing complex parameter, $\mu$ is a holomorphic volume form over $X$, and $\Gamma$ is an open integration chain. This kind of integrals are met in several physical applications, where one evaluates them asymptotically for large values of $\gamma$, by using the saddle point approximations. The associated thimbles indeed represent selected basis of integration cycles \cite{Witten:2010cx}, relative to some subset $D_0\subset X$ of positive codimension, and allow to understand the cohomology structures underlying the integral. The twisted de Rham cohomology is determined by the exact form $df$, the twisting being given by the covariant differential $\nabla_f=d+df\wedge$. To the triple $(X,D_0,f)$, one can construct four Local Systems, given by the local and global twisted de Rham and Betti cohomologies, all deeply related. In these terms, exponential integrals can be interpreted as periods of such cohomologies. In proving this, a key role is played by the Wall Crossing Structures associated to the integrals, which appear when the parameter $\gamma$ meets Stokes lines in the complex plane. The proof of these facts, given in \cite{Kontsevich:2024mks}, relates on a generalization of the Riemann-Hilbert problem along the Stokes rays.\\
We will review these facts in Section \ref{EIHF}, adapting the notation and formalism to the application to physics we have in mind. In a sequence of recent papers, it has been shown that the Skyrme model admits exact analytic solutions describing nuclear matter in different pasta states \cite{Alvarez:2020zui,Cacciatori:2021neu,Cacciatori:2022kag}. In particular, for the gauged Skyrme model one finds solutions representing baryonic layers at finite baryon density in the presence of a constant magnetic field \cite{Cacciatori:2024ccm}. The grand-canonical partition function of such system is expressed in terms of a  Pearcey integral. Its importance relies on the fact that it provides low energy non-perturbative effective description of chromodynamics and gives the occasion to replace cumbersome numerical analyses with analytical studies. The phase space of this system strongly depends on the Stokes lines of the partition function, which also determine critical curves in the $\mu_B-B_{ext}$ plane, $\mu_B$ being the (complexified) chemical potential of the external magnetic field $B_{ext}$. In Section \ref{PI} we will interpret the Pearcey integral as an exponential integral and apply the general theory to it in order to analyze the Stokes phenomenon.  Interestingly, the Stokes phenomenon also plays a significant role in analogue gravity and in the interpretation of the Hawking effect, \cite{Belgiorno:2020xlv,Belgiorno:2020ehk,Belgiorno:2024xpr,Trevisan:2024jvn}. \\
In this work, we further consider extensions of the previous analysis to include exponential integrals involving multivalued functions, which serve as a tool for computing Feynman integrals.
In Section \ref{EICFFI},  we briefly review how the constructions related to the triple $\left( X, D_0, f \right)$ can be generalized to the triple $\left( X, D_0, \alpha \right)$, where the $1$-form $\alpha$, representing the twisting of the covariant differential, is now a generic closed holomorphic form rather than necessarily an exact one, as developed in \cite{Kontsevich:2024mks}. Our motivation is that for Feynman integrals the function $f$ is replaced by the logarithm of a polynomial function, $f=\log \mathcal{B}$, which is thus not well defined globally, while its differential is. We will elaborate on a general strategy for applying such tools to Feynman integrals. Our strategy will provide an interpretation that is not restricted by specific underlying geometries or special assumptions on the spacetime dimensions. This will be illustrated through a concrete example in Section \ref{SI}.\\
A more extended analysis of the multivalued case in relation to Feynman integral will be presented in a companion paper \cite{Angius:2025acm}.\\
Finally, we will conclude with a discussion of our results and future perspectives in Section \ref{C}.

\section{Exponential integrals for holomorphic functions}\label{EIHF}

Exponential integrals are ubiquitous in physics, particularly in path integral computations across any quantum field theory, including conformal field theory correlators and non-perturbative analyses in string theory. In this section, we provide a concise overview of the mathematical techniques developed to handle these integrals, while referring the reader to \cite{Kontsevich:2024mks} for more detailed information.

Let $X$ be a smooth $n-$dimensional complex affine algebraic variety, $\Ol_X$ its structure sheaf and $\Omega^k_X$ the sheaf of differential $k-$forms on $X$. Given a holomorphic function $f\in \Ol(X)=\Gamma(X,\Ol_X)$, a Borel-Moore $n-$chain $\Gamma$ (locally compact) and an algebraic volume form $\mu\in\Gamma(X,\Omega^n_X)$, one defines the exponential integral of $f$ over $\Gamma$ with respect to $\mu$ as:

\begin{equation}
    I \equiv I_{\Gamma}(f)= \int_{\Gamma} e^{-f} \mu.
    \label{ExpIntegral1}
\end{equation}
\nn
Since we are working with smooth algebraic varieties, we can assume the support of $\Gamma$ to be an integer linear combination of closed oriented submanifolds. However, differently from ordinary homology, $\Gamma$ may have a nonempty boundary $\de \Gamma$. We will assume that the boundaries of the integration cycles are contained in a closed algebraic subset
$D_0\subset X$ of strictly positive codimension, ($Supp(\de\Gamma)\subset D_0$). Therefore, if the integration chain $\Gamma$ is such that the map
\begin{equation}
    Re(f)|_{Supp(\Gamma)}:Supp(\Gamma)\rightarrow\R
\end{equation}
\nn
is proper\footnote{The pre-image of any compact is compact.} and bounded from below, the exponential integral \eqref{ExpIntegral1} is absolutely convergent.\\
Furthermore, we can even generalize the notion of exponential integral by rescaling the function $f \mapsto \gamma f$ and studying how the structure of the resulting integral 

\begin{equation}
    I(\gamma) = I_{\Gamma} (f, \gamma) = \int_{\Gamma} e^{-\gamma f} \mu.
    \label{ExpIntegralgamma}
\end{equation}
\nn
depends on the complex parameter $\gamma \in \C^*=\C\backslash \{0\}$\footnote{In order to emphasize the variable we will often use the notation $\C_\gamma$ to mean the copy of $\C$ where $\gamma$ takes values.}.\\
For generic $\gamma$, the integral $I(\gamma)$ can be expressed as a linear combination of exponential integrals over special integration cycles, called \textit{thimbles}. These are real, non-compact cycles formed by the gradient flow lines of $Re(\gamma f)$ with respect to an auxiliary Hermitian metric on $X$.   In general, these gradient flow lines, which originate from a critical point, do not cross any other critical point along their trajectory. However, as the argument $arg(\gamma)$ varies, there exist special values of $\gamma \in \mathbb{C}^{\ast}$ at which this condition fails, leading to a change in the number of independent gradient flow lines. When one of these special loci, known as \textit{Stokes lines}, is crossed in $\mathbb{C}^{\ast}$, the linear combination of thimbles undergoes a discontinuous change (jump) described by a Stokes automorphism. \\
The collection of Stokes automorphisms along the plane $\mathbb{C}^{\ast}_{\gamma}$ forms the wall crossing structure associated to the integrals \eqref{ExpIntegralgamma}, which coincides with the one arising from the holomorphic version of Morse theory, see \cite{Nicolaescu2007} for an introduction to complex Morse theory. Exponential integrals can also be placed within the framework of exponential Hodge theory and interpreted as periods. In particular, they can be embedded into a generalized Riemann-Hilbert correspondence to study the relationship between de Rham and Betti cohomologies, both at the local and global levels.  In the global setting, the isomorphism between these two cohomologies associated with the triple $(X,D_0,f)$ is precisely realized through the exponential integral.
In the next subsections, we will study in detail the four cohomologies associated with this triple, each of which defines a vector bundle over $\mathbb{C}^{\ast}_{\gamma}$, and we will discuss their mutual relations. 

Let us  define the \textit{bifurcation set} $S^*\subset\mathbb{C}$ as the minimal finite set of points such that for any $t \in \mathbb{C} \setminus S^*$ there exists an open neighborhood $U$ (in analytic topology\footnote{Algebraic geometry makes use of the Zariski topology. However, since we have invoked smoothness, we can always view $X$ as a complex manifold and use the corresponding topology. This is called analytification of the topology}) and a homeomorphism $f^{-1}(U) \simeq U \times f^{-1}(t)$ which is compatible with the natural projections on both spaces to $U$ and such that it induces a homeomorphism:
\begin{equation}
    f^{-1}(U) \cap D_0 \, \simeq \, U \times \left( f^{-1}(t) \cap D_0\right).
\end{equation}
\nn
Smoothness implies $X$ is a complex manifold, i.e. it locally looks like $\C^n$. Since $f$ is continuous, and $U\subset \C$ is open, $f^{-1}(U)$ can be identified with an open set of $\C^n$. If $U$ does not contain bifurcation points, the above definition states that $f:\C^n \supset f^{-1}(U)\rightarrow U$, defines a local fibration on $\C$ (indeed a fibration on $U$), whose fiber is $f^{-1}(t)\cong Spec[\C[x_1,\dots,x_n]/\langle f\rangle]$,\footnote{With $\langle f \rangle$ we mean the ideal generated by $f$ in $\C[x_1,\dots,x_n]$, e.g. the elements of the form $gf$, $g\in \C[x_1,\dots,x_n]$. Therefore, $R:=\C[x_1,\dots,x_n]/\langle f\rangle$ is a ring. On the other hand, $f=0$ determines an affine subvariety $Y$ of $\C^n$. If $x\in Y$ and $h\in R$, we have an evaluation map $ev_x:R\rightarrow \C$ which is a homomorphism. The map $ev: Y\rightarrow R,\ x\rightarrow ev_x$, gives a bijection between $Y$ as a subset of $\C^n$, and the set of homomorphisms $R\rightarrow \C$. The latter is called $Spec[\C[x_1,\dots,x_n]/\langle f\rangle]$.} with $t\in U$, up to constant deformations (that are deformations depending trivially on $t$). Such fiber is smooth whenever the Milnor algebra\footnote{Also called local Jacobian ring} (Chiral ring)

\begin{equation}
    \mathcal{M}_f= \frac{\Ol_{\C^n,\mathbf{z}}}{\langle \de_{x_1} f, \ldots, \de_{x_n} f\rangle}
\end{equation}
\nn
is trivial, that is, if its dimension $\mu$, called (local) \textit{Milnor number}, vanishes \cite{Porteous_1971}. Here $\Ol_{\C^n,\mathbf{z}}$ is the stalk at $\mathbf{z}$ of $\Ol(X)$, that is the ring of germs of power series converging in some neighborhood of $\mathbf{z}$, $f(\mathbf{z})=t$.
If $\mu=0$ for any $t'\in U(f(\mathbf{z}))$, Ehresmann's lemma \cite{Dundas} implies that $f$ is a locally trivial fibration over $U$. On the other hand, if non zero Milnor numbers arise, the transition functions are constrained by elements of the Jacobian and the fibration cannot be trivial: the bifurcation set contains at least the set $S\equiv \{t_i\equiv f(\sigma_i)\}_i$ of critical values of $f$.\\
In the latter case, one can still define a locally trivial fibration $M$ on $U^\odot_i\equiv U(t_i)\backslash\{t_i\}$, called \textit{Milnor fibration} \cite{Ebeling,Ebeling1,ebeling2005monodromy,Gabrielov1979, Kulikov_1998}, whose fiber $M_{\mathbf{z}_i}$ is a CW complex homotopy equivalent to a bouquet of $\mu_{i}\equiv \mu|_{t_i}$ copies of $(n-1)-$spheres. Each of such spheres, or equivalently each element $\Delta_i$ of $H_{n-1}(M_{\mathbf{z}_i},\C)$\footnote{$H_k(S^{n-1},\C)=\C$, for $k=0,n-1$ and vanishes otherwise.}, is called (algebraic)\textit{vanishing cycle}. Their denomination follows from the fact that they shrink to zero when approaching the critical point.  
The importance of the role they play here derives from Brieskorn and Malgrange \cite{Brieskorn1970,Malgrange1994} proof of the isomorphism between the homology generated by the vanishing cycles and the hypercohomology of the De-Rham complex twisted by middle extended Gauss-Manin connection:

\begin{equation}
    H_{n-1}(M_z,\C)\cong \mathbb{H}(\Omega^\bullet_{X},\nabla^{mid}_{GM}), 
\end{equation}

\nn
which means that, in some sense, the homology of the full space is determined by the homology of the fiber.\footnote{One can get an idea as follows. In the above local fibration all fibers are isomorphic so have equal (co)homology. This determines a vector bundle over the complement of the bifurcation points, with fibers the (co)homology groups, and whose transition functions are nontrivial only around the bifurcation points, so they are locally constant. In this sense, one can think of (co)homology classes as functions of the base point $t$. The Gauss-Manin connection is essentially a flat connection telling how the (co)homology classes change along the basis, t. i. how to take their covariant derivative in $t$.} The \textit{Picard-Lefschetz theory} \cite{Lefschetz:1924,HuseinSade} (see Appendix \ref{Appendix} for a concise review) provides a concrete tool for determining and studying these vanishing cycles.\\
As we will see explicitly in Section \ref{sec:the relative homology}, the set of $n-$dimensional(real) manifolds $\mathcal{T}_i \simeq\Delta_i\times \R^+\subset X$ corresponding to the traces of the vanishing cycles along the vanishing directions in the base space, called Picard-Lefschetz thimbles, provide a basis of thimbles for the global Betti cohomology associated with the triple $\left( X, D_0, f \right)$. \\
Suppose now $X$ can be compactified to a smooth projective variety $\overline{X}$ such that $f$ extends to a regular map (t.i. everywhere defined):
\begin{equation}
    \overline{f} \, \, : \, \, \overline{X} \, \longmapsto \, \mathbb{P}^1.
\end{equation}
\nn
We can decompose $\overline{X}-X = D_h \sqcup D_v$, where the \textit{vertical divisor} $D_v = \overline{f}^{-1}(\infty)$ is the locus at infinity where $f$ diverges, and the \textit{horizontal divisor} $D_h$ is the locus at infinity where $f$ has finite limit.
In the following we will assume the set $D_0 \cup D_v \cup D_h$ is a normal crossing divisor and that no critical points lie at infinity nor at $D_0$. With this, we mean that the restriction of $f$ to $D_0$ or to the infinity locus,\footnote{In general, the infinity locus is not a submanifold but rather a stratifold. Thus, one has to check that the restrictions of $f$ to each open stratum has no critical points.} the restricted function has no critical points, which implies $S^*=S$. 
Finally, suppose that no degeneration of critical points occurs.

\subsection{Twisted de Rham Cohomology}
In the previous paragraph we introduced the triple $\left( X, D_0, f \right)$ and the exponential integral \eqref{ExpIntegral1} where $\mu$ is a holomorphic top form on $X$. From a cohomological perspective $\mu$ is closed with respect to the differential
\begin{equation}
    \nabla_f \, = \, d \, \, - \, \, df \, \wedge.
\end{equation}
Since the connection is flat, this differential gives rise to a complex of sheaves (in Zariski topology):
\begin{equation}
    \Omega_X^{\bullet} \, = \, \Omega_X^0 \, \xrightarrow{\nabla_f} \, \Omega^1_X \, \xrightarrow{\nabla_f}\, \dots \, \xrightarrow{\nabla_f} \, \Omega^n_X.
\end{equation}
\nn
In order to incorporate the boundary divisor $D_0$, we restrict to the subcomplex $\Omega^{\bullet}_{X,D_0}$ of $\Omega^{\bullet}_X$ of forms with support on $X \setminus D_0$. This subcomplex is the basis for the following definition of global cohomology. \\
\nn
\textbf{Definition:} \textsc{[global twisted de Rham, \cite{Kontsevich:2024mks}, Def.2.2.1]}\\
\textit{The global twisted de Rham cohomology is the graded abelian group:}
\begin{equation}
    H^{\bullet}_{dR, global} \left( (X,D_0),f \right)= \mathbb{H}^{\bullet} \left( X_{Zar}, (\Omega^{\bullet}_{X, D_0}), \nabla_f \right)
    \label{GlobalDerham}
\end{equation}
\textit{of equivalence classes of forms on $X \setminus D_0$ with respect to the differential $\nabla_f$.}\\\\
Notice, for instance, that any $1$-form $\alpha$ closed  with respect to the standard de Rham differential, yields a $\nabla_f$-closed $1$-form $e^{f} \alpha$:
\begin{align}
        \nabla_f (e^f \alpha) \, & = d (e^{f} \alpha) - df \wedge e^f \alpha
        = e^f d \alpha + e^f df\wedge \alpha - df \wedge e^f \alpha = 0 .
\end{align}
\noindent
If we now fix $\gamma \in \mathbb{C}^{\ast}_{\gamma}$, and replace $f \mapsto \gamma f$, we obtain the graded $\mathbb{C}-$vector space
\begin{equation}
H^{\bullet}_{dR,global,\gamma} \left( (X, D_0), f\right)=H^{\bullet}_{dR,global} \left( (X, D_0), \gamma f \right).
\end{equation}

\noindent
In addition to this global version of cohomology, one may also study the cohomology localized near each critical point of $f$.

\noindent
\textbf{Definition:} \textsc{[local twisted de Rham, \cite{Kontsevich:2024mks}, Def.2.3.3]}\\
\textit{Let $\Sigma = \left\lbrace \sigma_i \in X \setminus D_0 \vert df(\sigma_i)=0 \right\rbrace$ be the set of critical loci of $f$ in $X \setminus D_0$. The local twisted de Rham cohomology associated to the triple $(X,D_0,f)$ is the $\mathbb{C} [[1/\gamma]]$-module:\footnote{ Notation: $\mathbb{C}[[1/\gamma]]$ is the ring of formal power series of $\frac{1}{\gamma}$ with coefficients in $\mathbb{C}$. A $\mathbb{C}[[1/\gamma]]-$module is an abelian group equipped with an induced action of $\mathbb{C}[[1/\gamma]]$.}}
\begin{equation}
    H^{\bullet}_{dR, local} \left( (X, D_0), f \right) = \bigoplus_{i \in \Sigma} \mathbb{H}^{\bullet} \left(  U_{form} (\sigma_i), (\Omega^{\bullet}_{X,D_0} \left[ \left[\gamma \right] \right], (1/\gamma) d - df \wedge )\right)
\end{equation}
\textit{where $U_{form}(\sigma_i)$ is the formal neighborhood of the critical locus $\sigma_i \in X$. Each summand is called local de Rham cohomology associated with $\sigma_i$ (or $t_i=f(\sigma_i) \in S$) and it is denoted with $H^{\bullet}_{dR,loc,\sigma_i} (X, D_0, f)$. }

\ 

\nn With the above definitions in mind, the following proposition (\cite{Kontsevich:2024mks}, Prop. 2.3.4) summarizes the key structural properties  of the global de Rham cohomology.

\ 

\nn\textbf{Proposition:}\\
\textit{Assume that $f$ is proper, and set $\tau=1/\gamma$. Then
\begin{itemize}
\item[(i)] The coherent sheaf $\mathcal{H}^{\bullet}_{dR, global}(X,f)$ on $\mathbb{C}$, defined as:
    \begin{equation}
        \mathcal{H}^{\bullet}_{dR,global} (X,f) = \mathbb{H}^{\bullet}_{Zar} \left(  X \times \mathbb{C}_{\tau} , \left(pr^{\ast}_X ( \Omega^{\bullet}_X, \tau d_X - df \wedge ) \right)\right)
    \end{equation}
    gives rise to a graded vector bundle over $\mathbb{C}_{\tau}$.\footnote{Notation: $pr^{\ast}_X$ is the map induced from the projection of $X \times \mathbb{C}_{t}$ onto $X$.} Its restriction to $\mathbb{C}^{\ast}_{\tau}$ carries a flat connection $\nabla_{\tau}$, encoding how cohomology varies with the parameter $\tau$.
     \item[(ii)] The connection $\nabla_{\tau}$ has a regular singularity at $\tau = \infty$ (i.e. $\gamma=0$) and a second order pole at $\tau=0$ (i.e. $\gamma = \infty$). As we approach the point $\tau=0$, the global connection $\nabla_{\tau}$ splits into a direct sum of blocks, each of which is the tensor product of an exponential factor $e^{t_i \gamma} $ (rank 1 irregular D-module on $\mathbb{C}$) and a regular connection:
    \begin{equation}
        \left( \mathcal{H}^{\bullet}_{dR,global}(X,f), \nabla_{\tau} \right) \simeq \bigoplus_{i \in S} e^{t_i \gamma} \otimes \left( E_i, \nabla_i \right).
    \end{equation}
    In physical parlance this is the statement that, as $\tau \mapsto 0 $ (i.e. $\gamma \mapsto \infty$), the integral localizes around each critical point $\sigma_i$ giving an irregular contribution $e^{t_i \gamma}$ times a regular contribution solution of the system $\left( E_i, \nabla_i \right)$.
        \item[(iii)] The fiber of $\mathcal{H}^{\bullet}_{dR,global} (X,f)$ at $\tau = 0$ (i.e. $\gamma=\infty$) is isomorphic to the sum
    \begin{equation}
        \bigoplus_{i \in S} \mathbb{H}^{\bullet} \left(  U_{form}(\sigma_i), \left( \Omega^{\bullet}, -df \wedge   \right)\right).
    \end{equation} 
        \item[(iv)] Formally near $\tau=0$, the global twisted de Rham cohomology can be reconstructed using the local pieces around each critical point via the following global-to-local isomorphism:
    \begin{equation}
        \varphi_{dR} \, \, : \, H^{\bullet}_{dR, global} ((X,D_0),f) \otimes_{\mathbb{C}[\gamma]}  \mathbb{C[[\gamma]]} \, \simeq \, H^{\bullet}_{dR,loc} ((X,D_0),f)
        \label{dR:global_to_local}
    \end{equation}
        \item[(v)] For any $\gamma \in \mathbb{C}^{\ast}$ there is a non-degenerate pairing 
    \begin{equation}
        H^{\bullet}_{dR,global,- \gamma} (X,f) \otimes H^{\bullet}_{dR,global, \gamma} (X,f) \quad \longmapsto \quad \mathbb{C} \left[ -2\,dim_{\mathbb{C}} X\right]
    \end{equation}
    which extends to a non-degenerate pairing at $\gamma = \infty$ (i.e. $\tau=0$). This is the twisted Poincaré duality, shifting cohomological degree by $-2\,dim_{\mathbb{C}} X=-2n$.
\end{itemize}
}
\nn
We will show how to compute it concretely in section (\ref{PI}).

\subsection{Betti (Co-)Homology}\label{sec:betti}
To complement the twisted de Rham picture, we now introduce the corresponding Betti (co)-homology groups, which capture the topology of chains on $X$ relative to the level sets of $f$ at infinity. We begin by fixing a real constant $c >0$ and considering the singular relative homology
\begin{equation}
    H_{\bullet} \left( X, D_0 \cup f^{-1} (Re(z) \geq c), \mathbb{Z} \right) \simeq H_{\bullet} \left(  X, D_0 \cup f^{-1} (c), \mathbb{Z}\right).
\end{equation}
Once $c > max_{\sigma \in \Sigma} Re(f(\sigma))$, the critical points do not lie on the boundary, and then the relative homology stabilizes (i.e. it is the same replacing $c$ with any $c'>c$).\\\\
\noindent
\textbf{Definition:} \textsc{[global Betti, \cite{Kontsevich:2024mks}, Def.2.4.1]}\\
\textit{The global Betti homology of $(X,D_0,f)$ is}
\begin{equation}
    H_{\bullet}^{Betti,global} \left( (X,D_0), f, \mathbb{Z} \right) := H_{\bullet} \left( (X,D_0), f^{-1} (\infty), \mathbb{Z}\right)
\end{equation}
\textit{and, similarly, the global Betti cohomology is}
\begin{equation}
    H^{\bullet}_{Betti,global}  \left( (X,D_0), f, \mathbb{Z} \right) \equiv H^{\bullet} \left( (X,D_0), f^{-1} (\infty), \mathbb{Z}\right),
\end{equation}
{\it where the infinity means selecting the stabilized (co)homology.}

\

As in the case of de Rham cohomology, we consider the rescaling of the function $f \mapsto \gamma f$, which allows to define the global Betti cohomology at any point in the plane $\mathbb{C}^{\ast}_{\gamma}$.

\

\noindent
\textbf{Definition:} \textsc{[\cite{Kontsevich:2024mks}, Def.2.4.2]} \\
\textit{Let $\gamma \in \mathbb{C}^{\ast}_{\gamma}$. For each fixed $\gamma$ we define the graded abelian group:}
\begin{equation}
    H^{\bullet}_{Betti,global,\gamma} \left( (X, D_0),f , \mathbb{Z} \right) \equiv H^{\bullet} \left( (X, D_0), (\gamma f)^{-1} (\infty), \mathbb{Z} \right)
    \label{Def:Betti_global_gamma}
\end{equation}

\

By extending these groups from $\mathbb{Z}$ to $\mathbb{Q}$, one obtains a definition for the Poincaré duality in this global Betti setting.

\

\noindent
\textbf{Proposition:} \textsc{[Poincaré duality, \cite{Kontsevich:2024mks}, Prop.2.4.3]}\\
\textit{Let $X'=\overline{X}-D_v- \overline{D}_0$ and $D_0' = D_h - (D_h \cap D_v)$. We have the following isomorphism:}
\begin{equation}
    H_{\bullet}^{Betti,global} \left( (X,D_0),f \right) \simeq H^{\bullet}_{Betti,global} \left( (X',D'_0),-f\right) \left[ 2\,dim_{\mathbb{C}} X \right].
    \label{Betti:Poincare_duality}
\end{equation}
The family of abelian groups \eqref{Def:Betti_global_gamma} over the whole space $\mathbb{C}^{\ast}_{\gamma}$ defines a \textit{Local System} (a locally constant sheaf) over $\mathbb{C}^{\ast}_{\gamma}$ denoted as $\mathcal{H}^{\bullet}_{Betti,global} \left( (X,D_0), f\right)$.\\
Now, we want to relate these groups to local data. In order to do this, we look at the codomain $\C_t$ of the function $f$ as the real plane $\R^2$ and choose an open region $U$, whose closure $B=\bar U$ is a submanifold of $\R^2$ isomorphic to a unit disc. Assuming that the boundary $\partial B$ does not intersect the critical locus of $f$, we fix an arbitrary point $t_0 \in \partial B$. The idea is the following. For each $k \in \mathbb{Z}_{\geq 0}$ and $k<n$, we associate to the pair $(B, t_0)$ the abelian group
\begin{equation}
    V^k(B,t_0) := H^{k} \left( f^{-1} (B), (D_0 \cap f^{-1} (B)) \cup f^{-1}(t_0), \mathbb{Z}\right)
\end{equation}
and we look at it as a vector space. We now assume that all finite, non-degenerate critical values of $f$ lie in the interior of $B$. Since they are isolated points, we can find a finite number of subsets $B_i=\bar U_i \subset B$, each containing exactly one critical value and such that they have vanishing intersection in $B$ and intersect $\partial B$ precisely in the same marked point $b$. By retracting $B$ to the bouquet of $B_i$, one gets the isomorphism 

\begin{equation}
    V(B, b) \simeq \bigoplus_j V(B_j,b).
\end{equation}
This allows us to explore each component separately and study cohomologies with a single critical point. This leads one to introduce the following definitions.
Let us assume $D_0 = \emptyset$ ($X$ is projective) to lighten notation.  

\

\noindent
\textbf{Definition:} \textsc{[local Betti cohomology, \cite{Kontsevich:2024mks}, Def.2.5.1]}\\
\textit{For each critical value $t_i$, a small positive $\epsilon$ and $\gamma \in \mathbb{C}^{\ast}$, we define the local Betti cohomology associated with the pair $(t_i,\gamma)$ as the graded abelian group}
\begin{equation}
    H^{\bullet}_{Betti,local,t_i, \gamma} (X,f) = V\left( D( \gamma t_i, \epsilon), t_{\theta_\gamma}\right) = H^{\bullet} \left( (\gamma f)^{-1}(D(\gamma t_i, \epsilon)) , f^{-1} (t_{\theta_\gamma}), \mathbb{Z}\right)
    \label{Betti_local:ti}
\end{equation}
\nn
\textit{where $D( \gamma t_i, \epsilon)$ is a closed disc in $\mathbb{C}$ of radius $\epsilon$ centered in $\gamma t_i$ and $t_{\theta_\gamma}$ is the point on the boundary of the disc  such that $t_{\theta_\gamma}= \gamma t_i+ \epsilon e^{i \theta_\gamma}$ with $\theta_\gamma = \pi-\arg (\gamma)$.\\
At fixed $\gamma$, the direct sum of these cohomology groups for each $t_i \in S$ form the local Betti cohomology $H^{\bullet}_{Betti,local, \gamma} (X,f)$:}
\begin{equation}
    H^{\bullet}_{Betti,local, \gamma} (X,f) = \bigoplus_{t_i \in S} H^{\bullet}_{Betti,local, t_i, \gamma} (X, f).
\end{equation}
\nn
Similar to the global case, the family of local Betti cohomologies over the space $\mathbb{C}^{\ast}_{\gamma}$ forms a Local System denoted as $\mathcal{H}^{\bullet}_{Betti, local}(X,f)$.

\begin{figure}[h!]
    \centering
    \includegraphics[width=0.8\linewidth]{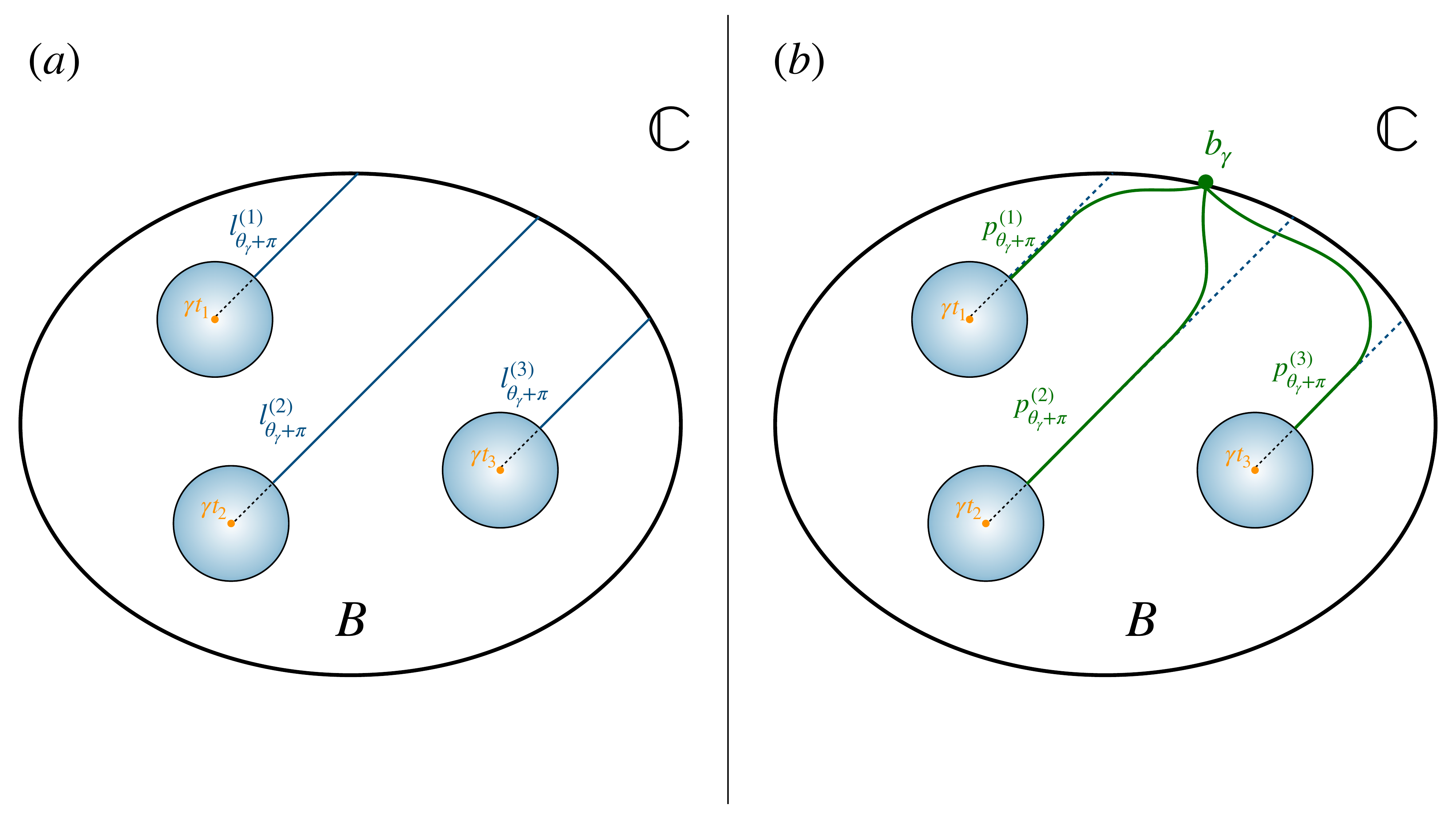}
    \caption{Pictorial representation of the homotopy deformation implementing the boundary retraction that induces the isomorphism \eqref{Betti:global_to_local}.}
    \label{fig:Betti_local_to_global}
\end{figure}

 Now we will make use of the description of the local and global cohomology groups as vector spaces $V(B_j,t_j)$ in order to relate them to each other. First, let us construct a sufficiently large disc $B \subset \mathbb{C}$ containing all the critical values of $S$, and for each critical value in $t_i \in S$ let us construct its proper disc $D(\gamma t_i, \epsilon)$ with marked point $t^{(i)}_{\theta_{\gamma}}$ on its boundary. From each of these points $t^{(i)}_{\theta_{\gamma}}$ let us construct a ray $l^{(i)}_{\theta_{\gamma}+\pi}$ in the direction $\theta_{\gamma}+\pi$. The resulting configuration consists of a set of parallel lines originating from the small discs and terminating at the boundary of the large disc, as depicted in figure \ref{fig:Betti_local_to_global}-(a). At this point, we can construct a homotopy of the large disc such that the deformed rays $l^{(i)}_{\theta_{\gamma}+\pi} \mapsto p^{(i)}_{\theta_{\gamma}+\pi}$ intersect at a unique point $b_{\gamma}$ on the boundary of the large disc (figure \ref{fig:Betti_local_to_global}-(b)). \\
For all the $\gamma \in \mathbb{C}^{\ast}$ that do not belong to the Stokes rays, defined below, the retraction of the complement of the big disc with respect to the union of the small discs and the paths $p^{(i)}_{\theta_{\gamma}+\pi}$ gives rise to the Betti local to global isomorphism:
\begin{equation}
    \varphi_{Betti} \, \, : \, \,  \,  \mathcal{H}^{\bullet}_{Betti,local} (X,f) \stackrel{\sim}{\longrightarrow} \mathcal{H}^{\bullet}_{Betti, global} (X,f).
    \label{Betti:global_to_local}
\end{equation}

\noindent
We call $\varphi_{\theta}$ the restriction of this isomorphism along a specific ray $\theta$ in $\mathbb{C}^{\ast}_{\gamma}$. We can think of it as generated by the embedding of a neighborhood of the ray (intersected with $B$) in $B$.

\

\noindent
\textbf{Definition:} \textsc{[Stokes ray, \cite{Kontsevich:2024mks} Def.2.5.3]}\\
\textit{We call the ray $s_{\theta}=\left\lbrace \gamma \vert \arg(\gamma)= \pi-\theta_{ij} = \theta \right\rbrace= \mathbb{R}_{\geq 0} \cdot (t_i-t_j)^{-1} \subset \mathbb{C}_{\gamma}$ with $\theta_{ij}= \arg(t_j-t_i)$ a Stokes ray.\\
Rays with vertex at the origin that are not Stokes rays are called generic rays. }

\

\nn Notice that there can be more copies $t_a,t_b\in \C_t$ of critical points such that $\arg(t_a-t_b)=\arg(t_i-t_j)$. All these copies give the same Stokes ray.
 Whenever $\gamma$ lies on the Stokes ray of slope $\pi-\arg(t_i-t_j)$ in the plane $\C_\gamma$, the corresponding line $l^{(i)}_{arg(t_i-t_j)}$ in the $\C_{\gamma t}$-plane, used to construct the Betti isomorphism $\varphi_{Betti}$, passes through both points $\gamma t_i$ and $\gamma t_j$ before reaching the boundary of the large disc. For all other points nothing special happens.  
Therefore, we see that for each Stokes ray $s_{\theta}$ (namely a Stokes ray with slope $\theta$), we have an isomorphism $T_{\theta}$ among the graded abelian groups $H^{\bullet}_{Betti, local, \gamma}(X,f)$ with $\arg(\gamma)$ sufficiently close to $\theta$.  Concretely, choose a sector in the $\gamma$-plane with boundary rays at angles  $\theta_{\pm}= \theta \pm \epsilon$. These two rays lift to sectors in the $\mathbb{C}_{\gamma t}$-plane whose retraction paths avoid all but one critical value, so for each of those values the two deformations give homotopic maps and hence the same identification of local Betti groups.

However, when $\theta= \pi - \arg(t_i-t_j)$, the corresponding rays in $\mathbb{C}_{\gamma t}$ pass through both $\gamma t_i$ and $\gamma t_j$. In that case the edges cannot be deformed into one another without crossing the line $l^{(i)}_{\theta_{ij}}$.  The required isomorphism for the jump, given by $\varphi_{\theta^-}^{-1} \circ \varphi_{\theta^+}$, is implemented by the operator:
\begin{equation}
    T_{\theta} = \mathbb{1} + \sum_{\begin{array}{c}
       i \neq j   \\ \arg(t_i-t_j)= -\theta 
    \end{array}} T_{ij},\label{Tij}
\end{equation}
where
\begin{equation}
    T_{ij} \, \, : \,\ \, H^{\bullet} \left( D (t_i, \epsilon), t_{\theta_{ij}}; \mathbb{Z} \right) \quad \longmapsto \quad H^{\bullet} \left( D(t_j, \epsilon), t_{\theta_{ij}}; \mathbb{Z}\right).
\end{equation}
With this isomorphism we can glue the Local System $\mathcal{H}_{Betti,local} (X,f)$ across the Stokes rays. Equipped with these Stokes automorphisms, the local system $\mathcal{H}_{Betti,local} (X,f)$ over the circle $S^1 = \left\lbrace \vert \gamma \vert = const \right\rbrace$ provides a concrete example of an analytic wall-crossing structure.

\ 
To this point we have introduced four cohomology theories, de Rham and Betti in their global and local versions, each pair related by its own isomorphism. In the statements that follow, we will establish the comparison isomorphisms between the de Rham and Betti frameworks.

\nn\textbf{Definition:} \textsc{[exponential period map]}\\
\textit{The integration over cycles defines a non-degenerate pairing
\begin{equation}
   \int \, \, : \,\,  H_{\bullet}^{Betti,global} \left( (X,D_0), f \right) \, \otimes  \, H^{\bullet}_{dR,global} \left( (X, D_0), f \right) \quad \longmapsto \quad \mathbb{C},
   \label{GlobalPairing}
\end{equation}
called Exponential Period Map.}

\

\nn From this pairing, for each $\gamma \in \mathbb{C}^{\ast}_{\gamma}$, we can construct the following isomorphism, \cite{Kontsevich:2024mks}, Prop.2.7.1:
\begin{equation}
    \varphi_{\gamma} \, \, : \, \, H^{\bullet}_{dR,global, \gamma} (X,f) \, \simeq \, H^{\bullet}_{Betti, global, \gamma} (X,f) \otimes \mathbb{C}.
    \label{isomorphism:dR_Betti_global}
\end{equation}

\

\noindent
Then we can refer to the integrals \eqref{ExpIntegralgamma} as exponential periods of de Rham cocycles over Betti cycles. \\
Finally, by promoting the Local Systems over $\mathbb{C}_{\gamma}^{\ast}$ to vector bundles with connection $\nabla_{\tau=1/\gamma}$, we have the following local version of the isomorphism \eqref{isomorphism:dR_Betti_global} (\cite{Kontsevich:2024mks}, Prop.2.7.2)
\begin{equation}
    RH^{-1}_{loc} \left( \mathcal{H}^{\bullet}_{Betti,local} ((X,D_0),f) \otimes \mathbb{C} \right) \, \simeq \, \mathcal{H}^{\bullet}_{dR,local} ((X,D_0), f),
\end{equation}
where $RH^{-1}_{loc}$ is the Riemann-Hilbert inverse functor from the category of Local Systems of complex vector spaces to the category of regular singular connections of vector spaces over $\C[[t]]$. 

\subsection{WCS for Exponential Integrals}
One of the main consequences of our initial generalization of the exponential integral \eqref{ExpIntegral1} to the one-parameter family \eqref{ExpIntegralgamma},  achieved by rescaling the function $f \mapsto \gamma f$, is the emergence of Stokes phenomena for specific values of the parameter $\gamma \in \mathbb{C}^{\ast}$. For these special values $\gamma^{\ast}$, the number of independent lines $l^{(i)}_{\theta_{\gamma^{\ast}+ \pi}}$ used to construct the Betti local to global isomorphism decreases, leading to discrete changes in the graded abelian groups $H^{\bullet}_{Betti,local, \gamma} (X,f)$. These changes are controlled by wall crossing formulas. In this section we discuss the wall crossing structure for exponential integrals \cite{Kontsevich:2022ana, Kontsevich:2024mks}, which provides a generalization of the $2d$ version used by Cecotti and Vafa in \cite{Cecotti:1992ccv}.

\begin{figure}[h!]
    \centering
    \includegraphics[width=0.9\linewidth]{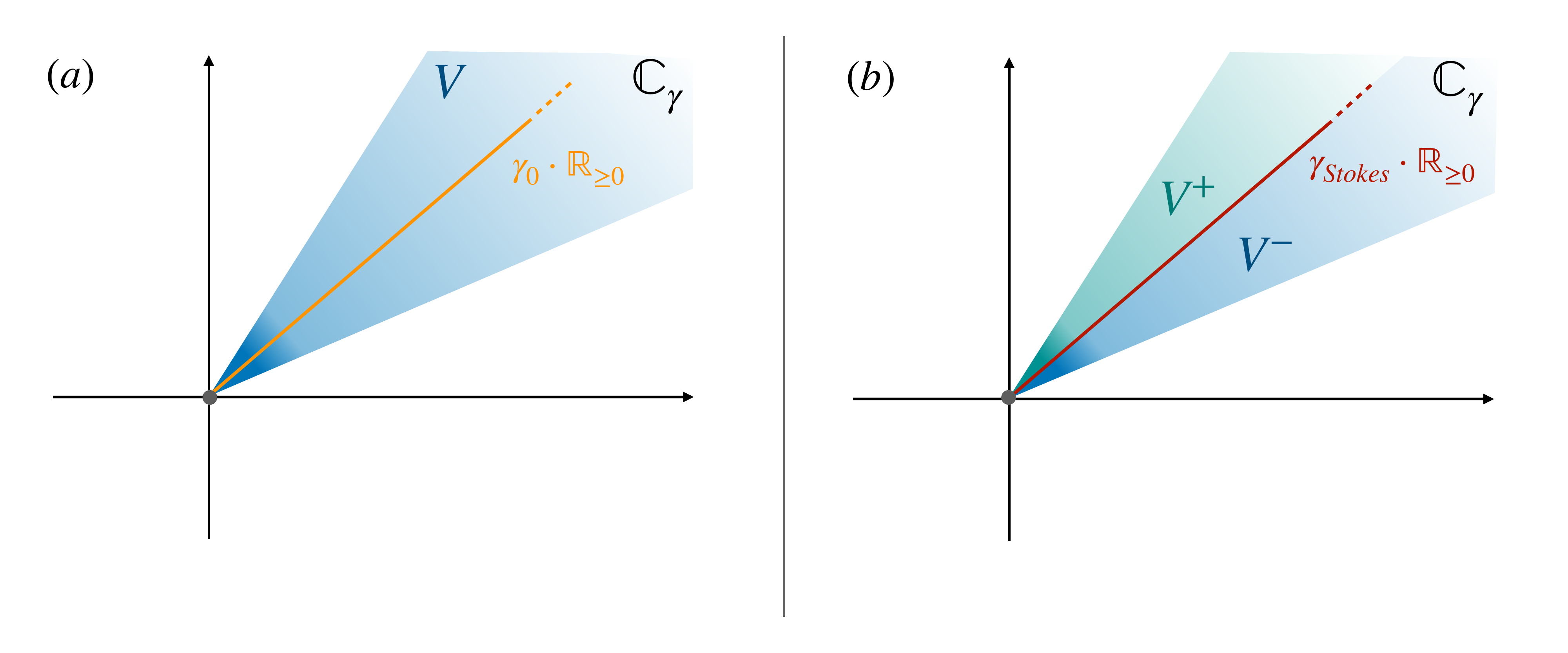}
    \caption{(a) Sector $V$ in the complex plane $\mathbb{C}_{\gamma}$ containing the generic ray $\gamma_0 \cdot \mathbb{R}_{\geq 0}$. (b) Sectors $V^+$ and $V^-$ in the complex plane $\mathbb{C}_{\gamma}$ separeted by the Stokes ray $\gamma_{Stokes} \cdot \mathbb{R}_{\geq0}$}
    \label{fig:sectors}
\end{figure}

Let us fix a region $\mathcal{R}\subset\mathbb{C}_{\gamma}$, such that for any $\gamma\in \mathcal{R} $, the exponential integral \eqref{ExpIntegralgamma} is an analytic function of $\gamma$, depending on $\Gamma$. Notice that if $Supp(\Gamma)$ is compact, then the region is unrestricted. Otherwise, it is necessary to ensure that $\gamma f$ is bounded from below. In general, if $\gamma_0\in \mathcal R$, then $\R_{\geq 0}\gamma_0 \subset \mathcal R$. \\
If we do not fix the integration cycle but we keep the volume form fixed, we can interpret $I(\gamma)$ as a morphism of sheaves of abelian groups on $\mathbb{C}^{\ast}_{\gamma}$
\begin{equation}
    \begin{split}
        \mathcal{H}_{\bullet}^{Betti,global} \left( (X,D_0),f\right) \, \, \, & \longmapsto \, \, \, \mathcal{O}^{an}_{\mathbb{C}_{\gamma}^{\ast}}\\
        \Gamma \, \, \, & \longmapsto \, \, \, \int_{\Gamma} \, e^{-\gamma f} \, \mu. \\
    \end{split}
    \label{morphism:1}
\end{equation}
\nn
If we choose $\gamma_0$ lying on a generic ray in $\mathbb{C}^{\ast}_{\gamma}$, then, for any $\gamma$ in a small sector $V\subset \mathcal R$ containing the ray $\mathbb{R}_{\geq 0} \cdot \gamma_0$ (see figure \ref{fig:sectors}-(a)), the canonical isomorphism between global and local Betti homologies induced by \eqref{Betti:global_to_local} is well defined and it gives rise to the following morphism among sheaves:
\begin{equation}
    \bigoplus_{t_i \in S} \mathcal{H}^{Betti,local, \gamma, t_i} \left( f^{-1} (V), f \right) \, \, \, \longmapsto \, \, \, \mathcal{O}^{an}_{\mathbb{C}^{\ast}_{\gamma}}(V).
\end{equation}
\noindent
Let us now choose a Stokes ray $s_{\theta}$ and consider a new small sector $V= V^+ \cup V^-$ in the plane $\mathbb{C}^{\ast}_{\gamma}$ containing the ray (see figure \ref{fig:sectors}-(b)). We choose two bases $\left\lbrace \Gamma^+_{(i)} \right\rbrace$ and $\left\lbrace \Gamma^-_{(i)} \right\rbrace$  for the local Betti homology in the sectors $V^+$ and $V^-$, respectively, corresponding to the angles $\theta^+=\arg(\gamma^+ )= \theta + \epsilon$ and $\theta^-=\arg(\gamma^- )= \theta - \epsilon$. With these choices, we can define two vector valued analytic functions:
\begin{equation}
    \overline{I}^+ (\gamma) =\left( \begin{matrix} \int_{\Gamma_{(1)}^+} e^{-\gamma f} \mu \\ \int_{\Gamma_{(2)}^+} e^{-\gamma f} \mu \\ \dots \\ \dots \\ \int_{\Gamma_{(k)}^+} e^{-\gamma f} \mu \end{matrix} \right) \quad \quad \quad \quad \quad \overline{I}^- (\gamma) = \left( \begin{matrix} \int_{\Gamma_{(1)}^-} e^{-\gamma f} \mu \\ \int_{\Gamma_{(2)}^-} e^{-\gamma f} \mu \\ \dots \\ \dots \\ \int_{\Gamma_{(k)}^-} e^{-\gamma f} \mu \end{matrix} \right)
\end{equation}
related by the following wall crossing formulas
\begin{equation}
    \int_{\Gamma^-_{(i)}} e^{-\gamma f} \mu= \int_{\varphi^{\ast}_{\theta_+}(\Gamma^+_{(i)})} e^{-\gamma f} \mu + \sum_{\substack{j \neq i \\ \arg(t_i - t_j) = \theta}}\int_{(\varphi^{\ast}_{\theta^+} \circ T_{ij}) \Gamma^+_{(i)}} e^{-\gamma f} \mu,
    \label{wcf:general}
\end{equation}
where $\varphi^{\ast}$ denotes the dual isomorphism to the one in \eqref{Betti:global_to_local}. Roughly speaking, these formulas describe the analytic continuation of the function $\overline{I}^- (\gamma)$ from the sector $V^-$ to the adjacent sector $V^+$ across the Stokes ray.

In the special case in which $f$ is a Morse function with $k$ different critical points, there is a special basis for the local and global Betti homologies for each direction $\theta=\pi-\arg(\gamma)$ which is the one of Lefschetz thimbles $th_{i, \theta + \pi}$. By definition, $th_{i, \theta + \pi}$ is the union of gradient lines of the function $Re (e^{-i \theta} f)$ emerging from the critical point $\sigma_i$, while $f(th_{i, \theta + \pi})$ is the line with direction $\theta + \pi$ emerging from the critical value $f(\sigma_i)=t_i$. Using this basis of thimbles, we can define the following collection of integrals for any generic direction $\gamma \in \mathbb{C}^{\ast}$, with $\theta= \pi-\arg(\gamma)$, such that $\gamma$ does not lie on a Stokes ray:
\begin{equation}
    I_i (\gamma)= \int_{th_{i, \theta + \pi}} e^{-\gamma f} \mu.
    \label{expInt:thimble}
\end{equation}
\nn Let us suppose to have defined them along the direction $\theta_-=\theta-\epsilon$, where $\theta$ identifies now a Stokes ray. Then, when we move toward $\theta_+=\theta+\epsilon$ through the Stokes line, the integrals undergo a discontinuous jump according to \eqref{wcf:general}
\begin{align}
    I_i\mapsto I_i+n_{ij}I_j,
\end{align}
where $n_{ij}$ are integers counting the number of gradient trajectories of $Re(e^{i \theta_{ij}}f)$ joining the critical values $\sigma_i$ and $\sigma_j$. Equivalently, $n_{ij}$ is the intersection index of the opposite thimbles $th_{i,\theta_++\pi}$ and $th_{j, \theta_--\pi}$ emerging from the critical points $\sigma_i$ and $\sigma_j$.

As $\gamma \mapsto \infty$, the integrals \eqref{expInt:thimble} admit a power expansion:
\begin{equation}
    I_i (\gamma) = e^{-\gamma t_i} \sum_{\lambda} c_{i,\lambda} \gamma^{- \lambda -1},
    \label{series:Exp_integral_gamma}
\end{equation}
for some $c_{i,\lambda} \neq 0$. \\
In order to analyze this series let us start isolating the exponential dependence at the critical point
\begin{equation}
   I_i(\gamma)= e^{-\gamma t_i} I_i^{mod} (\gamma)
   \label{Exp_integral:mod}
\end{equation}
and define the new variable
\begin{equation}
    s= f (\mathbf{z}) -t_i.
    \label{s:definition}
\end{equation}
Since the function $\rm{Im}(e^{-i \theta} f)$  remains constant along the cycle $th_{i, \theta + \pi}$, the variable $s$ ranges over the real interval from zero to infinity. Let us denote by $\Delta_i(s)$ the $(n-1)$-dimensional closed hypersurfaces defined by the level equations $f(\mathbf{z})=s=const$. These level sets are known as vanishing cycles of the homology group $H_{n-1} \left( f^{-1}(s), \mathbb{Z} \right)$ \cite{Pham:1967} (see Appendix \ref{Appendix} for a concise introduction to the topic). When $\gamma$ does not lie on a Stokes ray, the trace of these vanishing cycles along the variation of $s$ in the range $\left[ 0; + \infty \right[$ span the thimble $th_{i, \theta + \pi}$ 
\begin{equation}
    th_{i, \theta + \pi} = \bigcup_{s \geq 0} \Delta_i (s).
    \label{family:vanish_cycles}
\end{equation}
Using the Gelfand-Leray form $\frac{\mu}{ds} \Big\vert_{\Delta_i (s)}$, the exponential integral \eqref{Exp_integral:mod} can be rewritten as
\begin{equation}
    I_i^{mod}(\gamma) = \int_{0}^{\infty} ds  e^{- \gamma s} vol_{\Delta_i} (s),
\end{equation}
where 
\begin{equation}
    vol_{\Delta_i} (s) = \int_{\Delta_i (s)} \frac{\mu}{ds} \Big\vert_{\Delta_i (s)}
    \label{volume:vanishing_cycle}
\end{equation}
denotes the volume of the $(n-1)-$dimensional vanishing cycles $\Delta_i(s)$ in the family \eqref{family:vanish_cycles}. Note that the modified integral $I^{mod}_i(\gamma)$ can be interpreted as the Laplace transform of $vol_{\Delta_i}(s)$. On the other hand, the function $vol_{\Delta_i}(s)$ can be read as the pairing between the holomorphic cohomology class $\left[ \frac{\mu}{ds}\right] \in H^{n-1} \left( f^{-1}(s), \mathbb{C}\right)$ and the homology class $\left[ \Delta_i \right] \in H_{n-1} \left( f^{-1}(s), \mathbb{Z} \right)$. According to the resolution of singularities theorem (see for example \cite{Arnold}), this function admits an absolutely convergent power series expansion for $0 < s \ll \varepsilon$ of the form:
\begin{equation}
    vol_{\Delta_i}(s) = \sum_{\lambda } \sum_{ 0 \leq k \leq k_{max}} a_{\lambda,k} s^{\lambda} \log (s)^k.
     \label{expansion:vol_Delta}
\end{equation}
The numbers $\lambda$ correspond to the eigenvalues $e^{2 \pi i \lambda}$ of the monodromy operator $M_i$ acting on $H_{n-1}(f^{-1}(s), \mathbb{Z})$ when we turn around the singularity $s=0$, while the integer $k_{max}+1$ determines the size of the largest Jordan block associated with that eigenvalue. 
Taking the total differential of the definition \eqref{s:definition}, we obtain
\begin{equation}
    ds = \frac{\partial f}{\partial z^1} dz^1 + \frac{\partial f}{\partial z^2} dz^2 + \dots + \frac{\partial f}{\partial z^n} dz^n.
\end{equation}
This expression shows that the function $vol_{\Delta_i}(s)$ develops potential singularities in the complex $s$-plane whenever all partial derivatives of $f$ vanish simultaneously, that is, at the critical points of $f$. Consequently, a series expansion of $vol_{\Delta_i} (s)$ in powers of $s$ will have a radius of convergence determined by the distance to the nearest singularity on the same Riemann sheet. \\
Substituting the expansion \eqref{expansion:vol_Delta} into the exponential integral \eqref{Exp_integral:mod}, and using the following identity:
\begin{equation}
    \int_{0}^{\infty} e^{- \gamma s} s^{\lambda} (\log s)^k ds = \frac{d^k}{d \lambda^k} \left[ \gamma^{-(\lambda +1)} \Gamma \left( \lambda + 1\right) \right]
\end{equation}
we obtain
\begin{equation}
     I_{i} (\gamma) \,= e^{- \gamma t_i} \sum_{\lambda} \sum_k a_{\lambda, k} \int_{0}^{\infty} ds e^{- \gamma s} s^{\lambda} (\log s)^k = e^{- \gamma t_i} \sum_{\lambda} \sum_k a_{\lambda, k} \frac{d^k}{d \lambda^k} \left[ \gamma^{-(\lambda +1)} \Gamma \left( \lambda + 1\right) \right]. 
\end{equation}
Comparing this result with the power series in \eqref{series:Exp_integral_gamma}, we have
\begin{equation}
    c_{i, \lambda} = \frac{1}{\gamma^{-(\lambda+1)}}\sum_k a_{\lambda,k} \frac{d^k}{d \lambda^k} \left[ \gamma^{- (\lambda +1)} \Gamma \left( \lambda +1 \right) \right].
\end{equation}


\section{Holomorphic Morse theory }\label{PI}

In this section, we present a concrete application of the formalism developed in the previous paragraphs, accompanied by a discussion of its connection to holomorphic Morse theory. This connection not only provides deeper geometric insight into the structure of the theory but also clarifies the role of the abstract objects we introduced in the construction of Betti cohomology.

\begin{figure}[h!]
    \centering
    \includegraphics[width=0.6\linewidth]{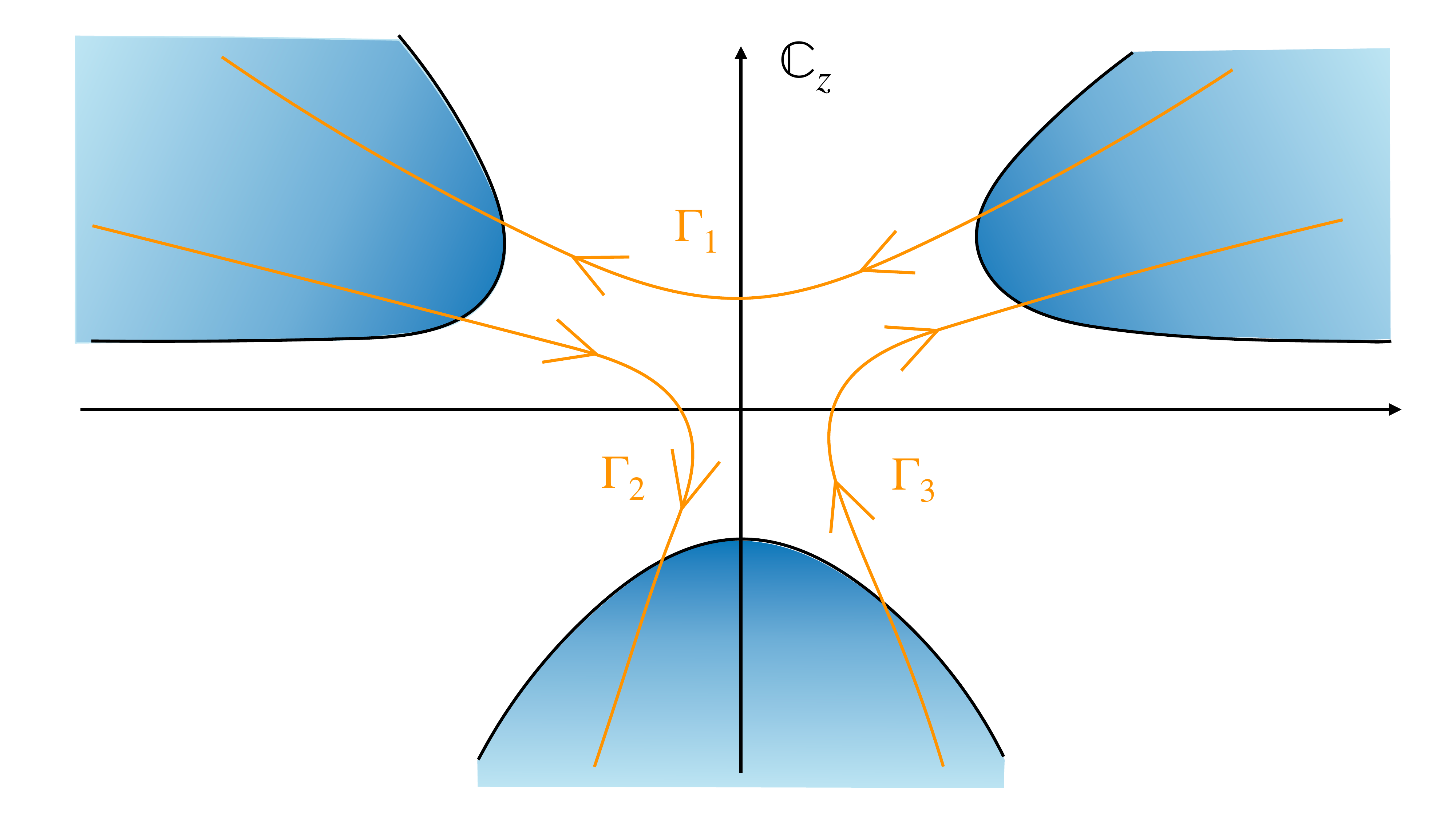}
    \caption{The blue areas in the complex plane represent the regions $D_N$ for $f(\mathbf{z}=Ai(\mathbf{z}))$ and large $N$. Despite $X= \mathbb{C}$ is contractible, $D_N$ can be the union of different disjoint pieces. The $1-$cycles $\Gamma_i \in H_1 (X, D_N , \mathbb{Z})$ must connect distinct components of $D_N$.  }
    \label{fig_th1}
\end{figure}

In order to build geometric intuition and develop familiarity with the setup, we now focus on the case $X\cong \C^n$ and consider exponential integrals of the form
\begin{equation}
    I(f,\gamma)=\int_\Gamma e^{-\gamma f(\mathbf{z})}g(\mathbf{z})d^n\mathbf{z},
    \label{ExpIntegral3}
\end{equation}
where $f(\mathbf{z})$ is a holomorphic function and $g (\mathbf{z}) d^n \mathbf{z}$ is a holomorphic $n$-form on $X$.
The aim of the procedure is the one to provide a basis for the integration contours, for any $\gamma \in \mathbb{C} \setminus \left\lbrace 0 \right\rbrace$, such that the integral \eqref{ExpIntegral3} converges. As $\gamma$ varies over its domain, the admissible integration contours $\Gamma$ must be deformed accordingly to ensure convergence.\\
Let us define the set $D_N$ in $X$ as

\begin{equation}
    D_N = \left\lbrace \mathbf{z} \in \mathbb{C}^n \vert Re(\gamma f(\mathbf{z})) \geq N \right\rbrace,
\end{equation}
\nn
for $N \in \R$, with $ |N|>>1$. This subset of $X$ consists in general of different disconnected components (an illustrative example is given by the blue regions in Figure \ref{fig_th1}). Any reasonable cycle $\Gamma$ for \eqref{ExpIntegral3} should connect two distinct regions of this subset, namely it should be a non-compact $n-$cycle of $X$ with boundaries in $D_N$, i.e. an element of the relative homology $H_n \left( X, D_N , \mathbb{Z} \right)$.
\nn
The condition on the boundaries is just part of the requirements that our integration contours have to satisfy. To ensure that the integrals are well-behaved, we must also impose conditions on the portions of the cycles extending into the complementary region $X \setminus D_N$. In particular, the cycles must avoid regions of $X$ where $Re(\gamma f (\mathbf{z}))\mapsto - \infty$,  as such behavior would lead to divergence. Furthermore, to prevent oscillations, we must impose the condition that $\rm{Im}(\gamma f (\mathbf{z}))$ remains constant along $\Gamma$, ensuring that we can factor out the phase $e^{i c}$ and reduce the problem to a real-valued integral. 

The techniques described in Section \ref{EIHF} provide a systematic method for analyzing the cycles in this relative homology, constructing a basis for them, and defining a well-behaved intersection pairing. 

\subsection{Relative Homology}\label{sec:the relative homology}
Let $H_k \left( X, D_N, \mathbb{Z}\right)$ be the $k-$homology group of $X$, on $\Z$, relative to $D_N$. The elements of this group, called relative cycles, are equivalence classes of $k$-chains in $X$ whose boundaries lie in $D_N$, modulo those chains that are homologous to chains entirely contained in $D_N$. Notice that, in the limit $N \mapsto + \infty$, this homology group corresponds, up to Poincaré duality as given in \eqref{Betti:Poincare_duality}, to the Betti homology group defined in \eqref{Def:Betti_global_gamma}, with $D_0= \emptyset$.  
By applying the constructions outlined in Section \ref{EIHF} we can determine the dimension of this relative homology group and construct an explicit basis for it.  In doing so, we recover the same geometric objects that arise in Morse theory, which analyzes the topology of $X$ by studying the properties of the differential functions defined on it. \\
In the present case, the function we will use to carry out the analysis is the height function 
\begin{equation}
    h= Re(\gamma f(\mathbf{z})).
\end{equation}
\nn
The set $\Sigma$ of critical points of this function $h$ coincides with the one of $f$ because of the Cauchy-Riemann equations. A critical point is said to be non-degenerate if the Hessian matrix associated to $h$ in that point is invertible. If all critical points are non-degenerate, the height function is a well-defined Morse function.
 The number of negative eigenvalues of the Hessian equips critical points of an index, called Morse index. For non-degenerate holomorphic functions on complex manifolds of complex dimension $n$, the Morse indices are all equal to $n$. Consequently, the Betti inequalities, which provide lower bounds on the dimensions of the homology groups, are saturated
\begin{equation}
       rank[H_k (X,D_N,\R)]= \begin{cases} 0,\quad k<n,\\ \texttt{\#}\Sigma,\quad k=n.
   \end{cases}
   \label{BettiRelation}
\end{equation}
 This provides a direct way to compute the dimension of $H_n(X,D_N, \mathbb{Z})$. Let us now determine a basis for this group.
If the function $h$ is perfect\footnote{The differences between the indices of distinct critical points of $h$ are never equal to $\pm 1$.}, Morse theory provides a way to construct a relative $n-$cycle $\Gamma_i$ for each critical point in $\Sigma$. Let us make the simplifying assumption that all the critical points $\sigma_i \in \Sigma$ are isolated points in $X$, and the corresponding distinct critical values form the set:
\begin{equation}
 S=\left\lbrace t_i \in \mathbb{C} \vert \, f(\sigma_i)=t_i\right\rbrace.
\end{equation}

\begin{figure}[h!]
    \centering
    \includegraphics[width=0.8\linewidth]{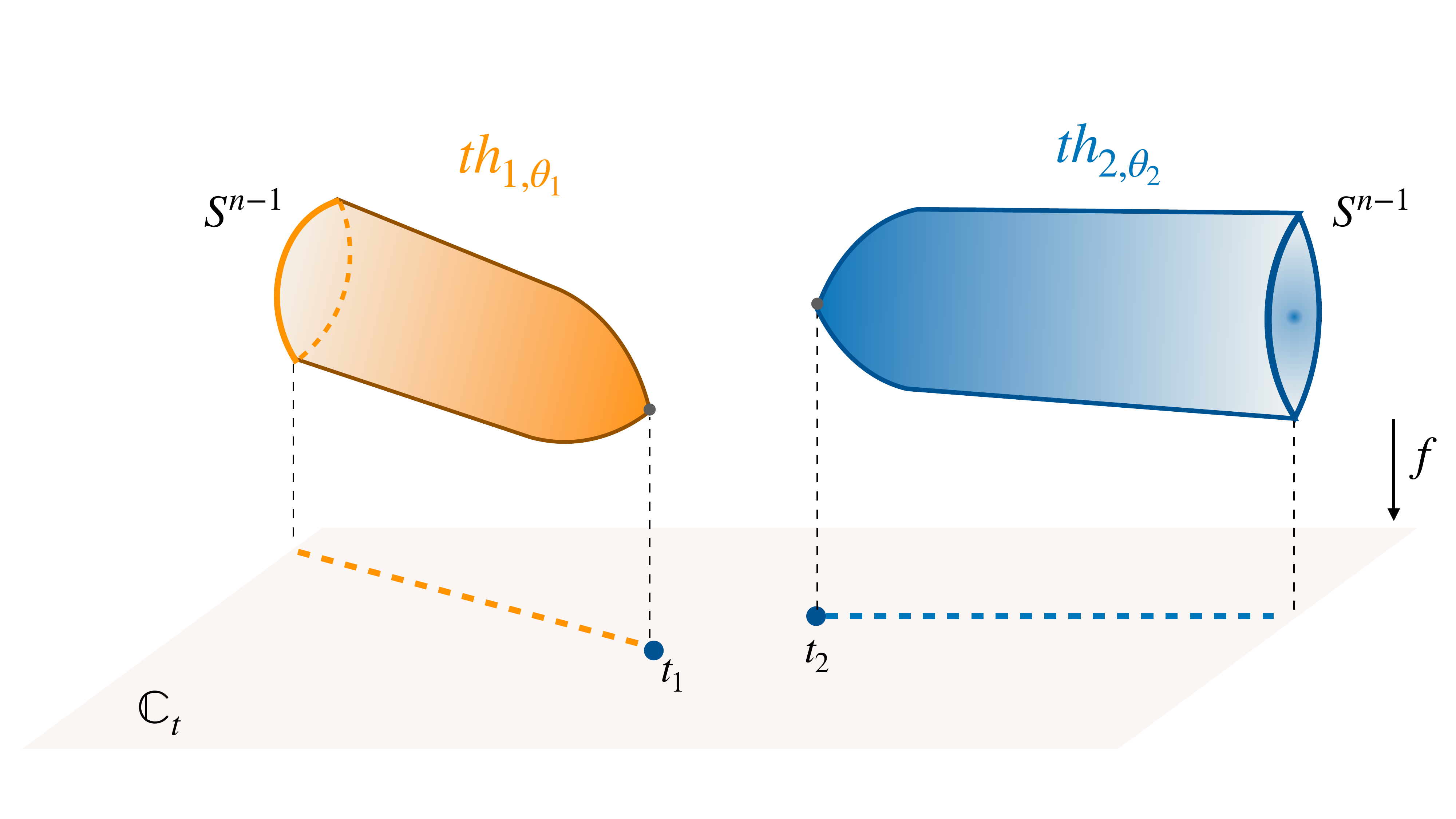}
    \caption{Pictorial representation of two thimbles: $th_{1,\theta_1}$ and $th_{2, \theta_2}$ constructed as continuations of the vanishing cycles $\Delta_i$ and $\Delta_j$, diffeomorphic to $S^{n-1}$, along the paths $t_1 + e^{i \theta_1} \mathbb{R}_{\geq 0}$ and $t_2 + e^{i \theta_2} \mathbb{R}_{\geq 0}$ in $\mathbb{C}_t$.}
    \label{fig_th2}
\end{figure}

For each critical point $\sigma_i \in X$, there is a unique vanishing cycle $\Delta_i \subset f^{-1}(t_i)$ diffeomorphic to $S^{n-1}$. An explicit method to construct these cycles is provided in the Appendix \ref{Appendix}.  Moreover, for each critical point $\sigma_i$ and a generic direction $\theta \in \mathbb{R} / 2 \pi \mathbb{Z}$, we can construct the Lefschetz thimble $th_{i, \theta} \sim \mathbb{R}^n \subset X$ as the continuation of the vanishing cycle along the path $t_i + e^{i \theta} \mathbb{R}_{\geq 0} \subset \mathbb{C}_t$ (see Figure \ref{fig_th2} for a conceptual visualization), ill-defined only if $\theta = \arg (t_j-t_i)$ for some $t_j \neq t_i$.\\
Among these thimbles, we aim to select a basis for the relative homology. This is achieved by considering the continuations of vanishing cycles along special paths of the form $t_i + e^{i \theta_i} \mathbb{R}_{\geq 0}$ in $ \mathbb{C}_t$ which start from the critical points $t_i$ and reach $t= \infty$ while maintaining a constant phase $\theta_i = \rm{Im}(\gamma f(t_i))$. These paths are solutions of the gradient flow equations:
\begin{equation}
\frac{du^i}{dt} =+ g^{ij} \frac{\partial h}{\partial u^j},
\label{eq:gradient_flow_step-up}
\end{equation}
where $u^i$ are real coordinates on $X$ and $g_{ij}$ is a Riemannian metric on $X$. These paths define the steepest ascent Lefschetz thimbles $\Gamma_i^+$, which have the key properties that the function $h$  increases monotonically along them, and, if $h=Re(\gamma f)$ for a holomorphic function $f$, then the imaginary part $\rm{Im}(\gamma f)$ remains constant along the thimble. If $\Gamma_i^+$ contains exactly one critical point, it corresponds to a \textit{good} Lefschetz thimble. Otherwise, it is referred to as a \textit{Stokes line}\footnote{Note that we use the term "\textit{Stokes rays}" to refer to the semi-infinite lines in the complex plane $\mathbb{C}_{\gamma}$  as established in the definition from Section \ref{sec:betti}. Instead, we use the term "\textit{Stokes lines}" to denote Lefschetz thimbles in $\mathbb{C}^n$ that contain more than one critical point. Each time $\gamma \in \mathbb{C}_{\gamma}$ lies on a Stokes ray, Stokes lines appear in $\mathbb{C}^n$ as a manifestation of the associated Stokes phenomenon.}. Assuming no Stokes lines are present in our set of thimbles, the number of thimbles exactly matches the rank of the relative homology group. To prove that they indeed generate the homology group, namely that they are independent cycles, we need to establish a method for uniquely decomposing any element $\Gamma \in H_n (X, D_{N}, \mathbb{Z})$  as a linear combination of the form
\begin{equation}
    \Gamma = \sum_{\sigma \in \Sigma} n_{\sigma} \Gamma^+_{\sigma}.
    \label{path_decomposition}
\end{equation}
Such a decomposition exists, and the coefficients $n_{\sigma}$ that appear on it are integer numbers representing the intersection between the cycle $\Gamma$ and the basis of steepest descent Lefschetz thimbles $\Gamma_i^-$ which belong to the dual homology group $H_n(X,D_{-N}, \mathbb{Z})$. Here,  $D_{-N}= \left\lbrace  \mathbf{z} \in X \vert Re(\gamma f(z)) \leq -N\right\rbrace$ with $N \in \mathbb{R}$ taken to be sufficiently large.
The new thimbles $\Gamma_i^-$ are solutions of the gradient flow equations with opposite sign:
\begin{equation}
\frac{du^i}{dt} =- g^{ij} \frac{\partial h}{\partial u^j}.
\end{equation}
They represent the downward-flowing cycles associated with each critical point $\sigma_i \in \Sigma$. These cycles retain the property that  $ \rm{Im}(\gamma f(z))$ remains constant along them, while the function $h$ decreases monotonically.  In the absence of Stokes rays, the intersection pairing is given by:
\begin{equation}
    \langle \Gamma^+_i , \Gamma^-_j \rangle = \delta_{ij}.
    \label{pairing:up_down_thimbles}
\end{equation}
It is straightforward to evaluate this formula for perfect Morse functions with no flows between distinct critical points. Indeed, if there are no flows between two distinct points $ \sigma_i, \sigma_j \in \Sigma$, the corresponding Lefschetz thimbles do not intersect. This is because they are associated with different constant values of the phase $\rm{Im}(\gamma f)$, which remains constant along each thimble. Conversely, the thimbles $\Gamma^{\pm}_{\sigma}$ follow paths along which the function $h$ is monotonically increasing or decreasing. As a result, they intersect exactly once, at the critical point  $\sigma$ itself. 

Therefore, a generic integration contour $\Gamma \in H_n(X,D_{N}, \mathbb{Z})$  can be decomposed in terms of the paths $\Gamma^+_i$ for generic $\gamma \in \mathbb{C}^{\ast}$, away from a Stokes ray, as in \eqref{path_decomposition}, with coefficients uniquely determined by
\begin{equation}
    n_{\sigma} = \langle \Gamma, \Gamma^-_{\sigma} \rangle.
\end{equation}
They count, with appropriate orientation, the number of downward flows from each critical point $\sigma$ to $\Gamma$. \\
Let us consider the example where $\gamma$ is purely imaginary, and the function $f$ is given by the quotient $P_1 (\mathbf{z})/P_2(\mathbf{z})$, where $P_1$ and $P_2$ are polynomials with real coefficients. We take the integration cycle $\Gamma=\Gamma_{\mathbb{R}}$ to be the product of $n$ real lines in $\mathbb{C}^n$. Let us partition the set of critical points $\Sigma$ into three subsets:
\begin{equation}
    \Sigma = \Sigma_{\mathbb{R}} + \Sigma_{\leq 0} + \Sigma_{> 0},
\end{equation}
where $\Sigma_{\mathbb{R}}$ denotes the set of critical points lying on the real axis, $\Sigma_{\leq 0}$ consists of critical points off the real axis for which the associated value of $h$ satisfies $h \leq0$, and $\Sigma_{> 0}$ includes those off the real axis for which $h > 0$. \\
For critical points lying on the real line, since $\gamma \in Im$, the function $h = Re (\gamma f)$, vanishes identically. In particular, we have $h_{\sigma}=0$ for all $\sigma \in \Sigma_{\mathbb{R}}$. Because $h$ strictly decreases along downward gradient flows, there can be no such flows starting at 
$\sigma \in \Sigma_{\mathbb{R}}$ that remain on the real line. Consequently, the only intersection between the downward Lefschetz thimble $\Gamma^-_{\sigma}$ and the real cycle $\Gamma_{\mathbb{R}}$ is the point $\sigma$ itself:
\begin{equation}
    \sigma \, \in \, \Sigma_{\mathbb{R}} \quad \quad \Longrightarrow \quad \quad n_{\sigma}= \langle \Gamma_{\mathbb{R}}, \Gamma^-_{\sigma} \rangle = 1.
\end{equation}
If $\sigma \in \Sigma_{\leq 0}$,  no downward flows originating from $\sigma$ intersect $\Gamma_{\mathbb{R}}$. This follows from the fact that $h$ is strictly decreasing along downward flows, and by definition, $h \leq 0$ for points in $\Sigma_{\leq}$. Thus, we have: 
\begin{equation}
    \sigma \, \in \, \Sigma_{\leq 0} \quad \quad \Longrightarrow \quad \quad n_{\sigma}=\langle \Gamma_{\mathbb{R}}, \Gamma^-_{\sigma} \rangle =0. 
\end{equation}
Finally, if $\sigma \in \Sigma_{>0}$,  it is in principle possible for downward flows originating from $\sigma$ to intersect the real section $\Gamma_{\mathbb{R}}$. The precise number of such intersections depends on the specific geometry of the function and must be determined case by case. Altogether, we obtain the decomposition:
\begin{equation}
    \Gamma_{\mathbb{R}} = \sum_{\sigma \in \Sigma_{\mathbb{R}}} \Gamma^+_{\sigma} + \sum_{\sigma\in \Sigma_{>0}} n_{\sigma} \Gamma^+_{\sigma}.
\end{equation}
\subsection{Stokes rays}
In the previous paragraph, we described the relative homology $H_n (X,D_{N}, \mathbb{Z})$, constructed a basis of thimbles for it, and defined an intersection pairing with the dual homology $H_n (X, D_{-N}, \mathbb{Z})$ to express a generic cycle $\mathcal{C} \in H_n (X,D_{N}, \mathbb{Z})$ in terms of this basis. However, as we explained in Section \ref{sec:betti}, the presence of Stokes rays affects the well-definedness of certain Lefschetz thimbles, making the previous construction insufficient. In this section, we explain why some Lefschetz thimbles become ill-defined in the presence of Stokes phenomena and how the framework introduced earlier can be adapted to restore consistency.

The key perspective we adopt is to construct a description of the homology group $H_n \left( X, D_{N}, \mathbb{Z}\right)$ that ensures a well-defined pairing \eqref{pairing:up_down_thimbles} for any value of $\gamma \in \mathbb{C}_{\ast}$. This is achieved by first establishing the structure for a specific $\gamma$ where no Stokes rays appear, as we did in the previous section, and then extending it across the entire $\mathbb{C}_{\ast}$. 

\begin{figure}
    \centering
    \includegraphics[width=0.5\linewidth]{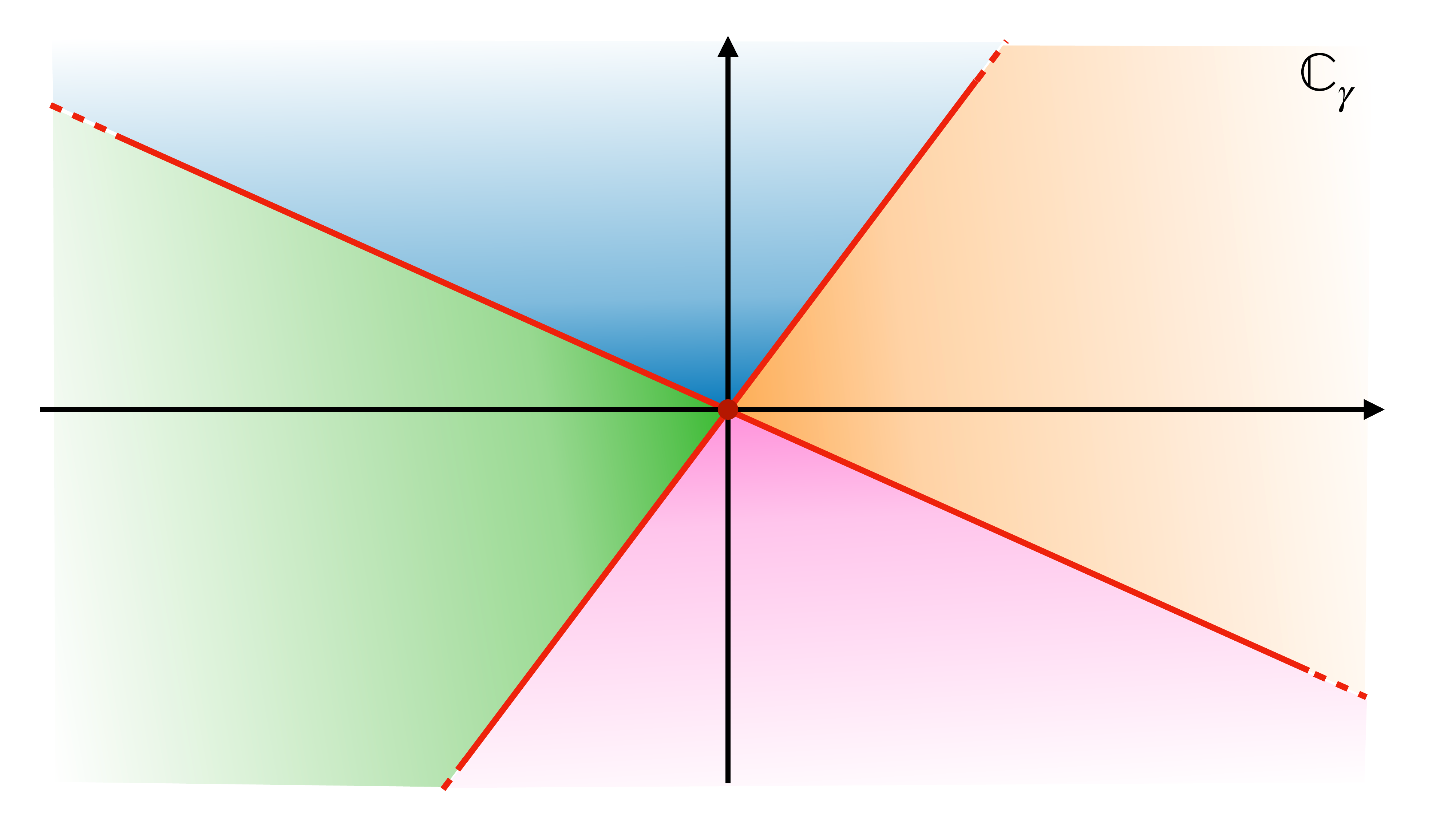}
    \caption{ The $\mathbb{C}_{\gamma}$ plane: The red lines indicate the Stokes lines, across which discontinuities arise in the definition of the Lefschetz thimble structure for the corresponding exponential integrals. The differently colored regions represent the distinct petals of the fan.}
    \label{fig_petals}
\end{figure}

Stokes lines are solutions of the gradient flow equation \eqref{eq:gradient_flow_step-up} that cross at least two critical points of the function $\gamma f(\mathbf{z})$. Since the imaginary part of this function is preserved along the flows, we have
\begin{equation}
    \rm{Im}(\gamma f(\mathbf{z})) \, \vert_{\sigma_i} = \rm{Im}(\gamma f(\mathbf{z})) \, \vert_{\sigma_j} 
    \label{StokesRaysFormula}
\end{equation}
for any point in the thimbles $\Gamma^+_{i}$ and  $\Gamma^+_{j}$. Moreover, since the number of critical points is finite, there can only be a finite number of Stokes lines. By assumption, $\gamma f(\mathbf{z})$ takes distinct values at different critical points, meaning that Stokes lines appear only for specific values of  $\gamma$. For  $\sigma_i \neq \sigma_j \, \in \, \Sigma$, the loci 
\begin{equation}
    l= \left\lbrace  \gamma \,  \in \, \mathbb{C}_{\ast} \,\ \vert\ \, \rm{Im}(\gamma f(\mathbf{z})) \, \vert_{\sigma_i} = \rm{Im}(\gamma f(\mathbf{z})) \, \vert_{\sigma_j} \right\rbrace
\end{equation}
in the complex plane $\mathbb{C}_{\gamma}$, define regions where the Lefschetz thimble structure undergoes discontinuities: they are the Stokes rays discussed in Section \ref{sec:betti}. These rays always pass through the origin; however, since $\left\lbrace 0 \right\rbrace \notin \mathbb{C}_{\gamma}$, they remain disconnected and form straight lines radiating outward from the center. As a result, the complex $\gamma$-plane  is divided into a fan-like structure composed of distinct sectors—referred to as petals in Figure \ref{fig_petals}. 
\nn
At this point, the procedure is to fix $\gamma$ within a specific petal of the fan, say the zeroth region $(0)$, away from any Stokes rays, and define the Lefschetz thimble structure for the corresponding integral. We can then vary $\gamma$ along the complex plane. As we cross a Stokes ray $s_{\theta}$, associated with a Stokes line between the critical points $\sigma_i $ and $\sigma_j$, for which $h_{\sigma_i} < h_{\sigma_j}$, the corresponding thimbles $\Gamma^+_{i}$ and $\Gamma^+_{j}$ undergo a discontinuous jump to the adjacent region $(1)$ of the form: 
\begin{equation}
    \left( \begin{matrix}   \Gamma^{+(1)}_{i} \\ \Gamma^{+(1)}_{j} \end{matrix} \right) = \left( \begin{matrix} 1 & \Delta_{ij} \\ 0 & 1 \end{matrix} \right) \, \left( \begin{matrix}   \Gamma^{+(0)}_{i} \\ \Gamma^{+(0)}_{j} \end{matrix} \right) \quad ,\quad \text{for } \, \, h_{\sigma_i} < h_{\sigma_j}
    \label{equation:jump}
\end{equation}
\nn
where the integers $\Delta_{ij}$ receive a contribution $\pm 1$ for each upflow line from $\sigma_i$ to $\sigma_j$, with the sign depending on cycles orientation and on the direction from which $\gamma$ crosses the Stokes line.  This is nothing but the intersection number of the corresponding vanishing cycles expressed by the Picard-Lefschetz formula \eqref{PicardLefschetzFormula}, up to a sign depending on the relative orientation of the cycles, when we cross the cut line starting from $t_j=f(\sigma_j)$ in the plane $\mathbb{C}_t$:
\begin{equation}
    \Delta_{ij} = (\pm 1) \Delta_i \circ \Delta_j.
     \label{TotalIntersection}
\end{equation}
This means that the new thimble $\Gamma_i^{+(1)}$ in the region $(1)$ is associated to the new vanishing cycle
\begin{equation}
    \Delta_i^{(1)}= \Delta_i^{(0)} \pm \left( \Delta_i \circ \Delta_j \right) \Delta_j^{(0)}.
\end{equation}
\nn
In order for the decomposition \eqref{path_decomposition} to be continuous, the coefficients $n_{\sigma_i}$ and $n_{\sigma_j}$ transform across the ray by 

\begin{equation}
    \left( \begin{matrix}   n_{\sigma_i} \\ n_{\sigma_j} \end{matrix} \right) \quad \longmapsto \quad  \, \left( \begin{matrix} 1 & 0 \\ - \Delta_{ij} & 1 \end{matrix} \right) \, \left( \begin{matrix}   n_{\sigma_i}  \\ n_{\sigma_j} \end{matrix} \right) \, .
\end{equation}

\begin{figure}[h!]
    \centering
    \includegraphics[width=0.8 \linewidth]{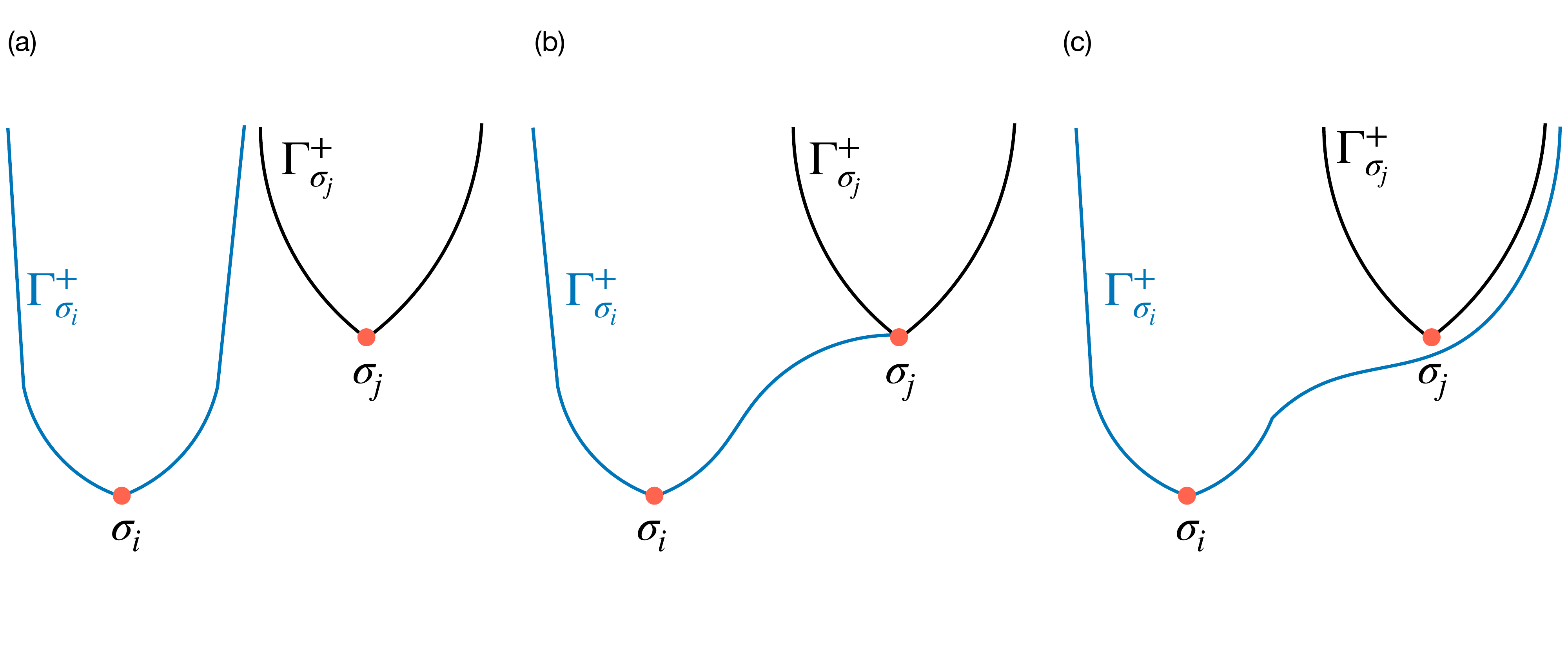}
    \caption{Jump of the thimble $\Gamma^+_{i}$ across a Stokes ray. From left to right, $\gamma$ crosses a Stokes ray: (b), (c).}
    \label{fig_jump}
\end{figure}

To understand the reason for these jumps and the meaning of the integer coefficients 
$\Delta_{ij}$ appearing in the jump matrix in \eqref{equation:jump} let us consider a simple one-dimensional example. Suppose that for a suitable 
$\gamma$, away from any Stokes line, we have two critical points $\sigma_i$ and $\sigma_j$ with distinct values of $h_{\sigma_i} < h_{\sigma_j}$ and distinct imaginary parts $\rm{Im}(\gamma f(\mathbf{z}))$, for which we can define two distinct thimbles without any intersection (see Figure \ref{fig_jump} (a)). As we move $\gamma$ towards a Stokes line, the thimble $\Gamma^+_{i}$ is continuously deformed until it crosses the thimble 
 $\Gamma^+_{j}$ at the critical point $\sigma_j$ (see Figure \ref{fig_jump} (b)). At this point, the first thimble is no longer well defined. As we continue moving 
$\gamma$ across the Stokes line, the support of the thimble $\Gamma^+_{i}$ continues to be deformed on the other side of the thimble $\Gamma^+_{j}$, as shown in Figure \ref{fig_jump} (c). The comparison between the representations (a) and (c) in Figure \ref{fig_jump} illustrates the jump.

The number of upward flows from $\sigma_i$ that intersect the point $\sigma_j$ along a Stokes ray, counted with appropriate sign based on the orientation, gives the number $\Delta_{ij}$ in the matrix \eqref{equation:jump}.

{\bf Remark:} An interesting relation between the total monodromy acting on the thimbles after a transformation $\gamma \mapsto e^{2 \pi i } \gamma$ and the transformation of a basis of $(n-1)-$forms dual to the vanishing cycles has been pointed out in \cite{Cecotti:1992ccv}. Let us start from a regular point $\gamma$ and let us transport it along a circle in a clockwise direction with $\vert \gamma \vert$ fixed. Each time that $\gamma$ crosses a Stokes ray $l_{ij}$, corresponding to the crossing of the Stokes line connecting the critical points $\sigma_i$ and $\sigma_j$, we have a change on the thimble basis given by \eqref{equation:jump}. Let us indicate the matrix giving the jump as
\begin{equation}
    M_{ij} = \mathbb{1} + A_{ij} ,
\end{equation}
where the only non-zero entry in the matrix $A_{ij}$ is $(ij)=\Delta_{ij}$ counting the intersection number among the vanishing cycles $\Delta_i$ and $\Delta_j$. After a tour of $\pi$ around the origin, the total change on the basis of thimbles is
\begin{equation}
    S= \prod_{l_{ij}} M_{ij}.
\end{equation}
In the second half sector, beyond $\pi$, each time $\lambda$ crosses a Stokes line we have a jump given by:
\begin{equation}
    M_{ji}^{-t} = \mathbb{1} - A_{ji}.
\end{equation}
The total jump along this second half circle is represented by the matrix:
\begin{equation}
    S^{-t}=\prod_{l_{ij}} M_{ji}^{-t}.
\end{equation}
The full monodromy is defined via
\begin{equation}
    H=S^{-t}S.
\end{equation}
This matrix is invariant under deformations of the function $\gamma f({z})$ and it is quasi-unipotent.\footnote{This means that some power of it is unipotent: the sum of the identity plus a nilpotent matrix} Then, its eigenvalues are always roots of the unity
\begin{equation*}
    \text{Eigenvalues of H} \quad \quad \longrightarrow \quad \quad \left\lbrace e^{2 \pi i q_k}, \qquad q_k\in\Q \right\rbrace.
\end{equation*}

\subsection{Twisted de Rham Cohomology}
Let us now move to the cohomological side of the exponential pairing, explicitly showing its construction in the one variable case $X=\C$, considering as holomorphic function the polynomial $f=\mathcal{P}_\ell\in \C[z]$ of degree $\ell$. Adding to $\C$ the divisor $D_v=p=\{\infty\} $, we end up with the good normalization $\overline{X}=\Pro^1$, on which $\mathcal{P}_\ell$ naturally extends to $\overline{\mathcal{P}}_\ell:\Pro^1\rightarrow\Pro^1$. Denoting the twisted differential as $\nabla\equiv (\gamma^ {-1} d+d\mathcal{P}_\ell\wedge)$, let consider the complex of sheaves: 

\begin{equation}
    (\Omega^\bullet_{\Pro^1,p},\nabla): 0\rightarrow \Ol_{\Pro^1}(*p) \xrightarrow{\nabla}\Omega^1_{\Pro^1}(*p)\rightarrow 0.
\end{equation}
\nn
On $\Pro^1$, meromorphic functions with only allowed pole at infinity are in fact polynomials on $\C$, thus we have:
\begin{equation}
\begin{split}
    &\Ol_{\Pro^1}(*p)\cong \C[z],\\
    &\Omega^1_{\Pro^1}(*p)\cong \{P(z)dz|P(z)\in \C[z]\}.
    \end{split}
\end{equation}
\nn
By definition \eqref{GlobalDerham}, the global twisted de Rham cohomology is
\begin{equation}
    H^\bullet_{dR,t}(\C,d\mathcal{P}_n)=\mathbb{H}(\Pro^1,(\Omega^\bullet_{\Pro^1,p},\nabla))=H^\bullet(R\Gamma(\Pro^1,(\Omega^\bullet_{\Pro^1,p},\nabla)))
\end{equation}
\nn
We can compute the hypercohomology by means of the Grothendieck spectral sequence with second page

 \begin{equation}
     E_2^{p,q}=R^p\Gamma(\Pro^1,H^q(\Omega^\bullet_{\Pro^1,p},\nabla)).
 \end{equation}
 \nn
 Let us firstly determine the cohomology of $(\Omega^\bullet_{\Pro^1,p},\nabla)$:
 \begin{equation}
     H^\bullet(\Omega^\bullet_{\Pro^1,p},\nabla)=\ker\nabla\oplus {\rm coker}\nabla.
 \end{equation}
 \nn
 Thus, the computation reduces to the calculation of the kernel and the cokernel of the twisted differential. For $g(z)\in \C[z]$, the stalk of $\ker\nabla$ is 

\begin{equation}
    \ker\nabla=\{g(z)\in\C[z]\ |\ \gamma^{-1} g'(z)+\mathcal{P}'_\ell(z)g(z)=0\},
\end{equation}
\nn
for $\gamma \in\C^*$, the constrain on $g(z)$ becomes:

\begin{equation}
    g(z)=e^{-\gamma\mathcal{P}_\ell(z)},
\end{equation}
\nn
that cannot be a polynomial unless $\ell=0$. Therefore:

\begin{equation}
    \ker\nabla=\emptyset.
\end{equation}
\nn
The cokernel of $\nabla$ measures the failure of $\nabla$ to be surjective: away from the critical points of $\mathcal{P}_\ell$, $\nabla$ is locally surjective and its cokernel vanishes. On the other hand, near each critical point the equation $\nabla g=\eta$ for $\eta\in \Omega^1_{\Pro^1}(*p)$ may not have a polynomial solution for $g$. This generates a one dimensional obstruction to surjectivity, thus the stalk ${\rm coker}(\nabla)_\sigma\cong \C$. The cokernel of $\nabla$ is a direct sum of skyscraper sheaves with support on critical points

\begin{equation}
    {\rm coker}\nabla=\bigoplus_{\sigma\in\Sigma}\C.
\end{equation}
\nn
An alternatively and, for our purposes, more interesting way to determine it consists to notice that the cokernel of $\nabla$ is isomorphic to the Jacobian ring associated to $\mathcal{P}_\ell$ 
\begin{equation}
    {\rm coker}\nabla \cong \frac{\C[z]}{\rm{Im}(\nabla)}\cong\frac{\C[z]}{(\mathcal{P}'_\ell(z))}=J_{\mathcal{P}_\ell}.
\end{equation}
\nn
One can indeed prove the image of $\nabla$ is isomorphic to the ideal generate by $\mathcal{P}'_\ell$. 
The Jacobian ring, as a $\C$ vector space, is:
\begin{equation}
    J_f=span\{1,z,\dots,z^{\ell-2}\}\cong \C^{\mu},
    \label{CokerJacobian}
\end{equation}
\nn
with $\mu=\ell-1$ the total Milnor number.\\
Because of the vanishing of the higher cohomology groups of $(\Omega^\bullet_{P^1}(*p),\nabla)$, the terms $E_2^{p,q}$ vanish for $q>1$ and because of the acyclicality of skyscraper sheaves $E_2^{p,q}=0$ for $p>0$. Thus, the only possibly non zero ``turning page'' differential could be $d_2:E_2^{0,1}\rightarrow E_2^{2,0}$, but $E_2^{2,0}$ is also zero. Hence, the spectral sequence degenerates at page $E_2$. Therefore,
\begin{equation}
\begin{split}
    H^n(R\Gamma(\Pro^1,(\Omega^\bullet_{\Pro^1,p},\nabla)))&=\bigoplus_{p+q=n}E_\infty^{p,q}=\bigoplus_{p+q=n}E_2^{p,q}\\&=\bigoplus_{p+q=n}R^p\Gamma(\Pro^1,H^q(\Omega^\bullet_{\Pro^1,p},\nabla)).
    \end{split}
\end{equation}
\nn
That is:
\begin{equation}
\begin{alignedat}{2} 
    &\mathbb{H}^0(\Pro^1,(\Omega^\bullet_{\Pro^1,p},\nabla))\cong H^0(\Pro^1,\ker\nabla)\cong 0,\\ &\mathbb{H}^1(\Pro^1,(\Omega^\bullet_{\Pro^1,p},\nabla))\cong H^1(\Pro^1,\ker\nabla)\oplus H^0(\Pro^1,{\rm coker}\nabla)\\& \hspace{2.6cm} \cong H^0(\Pro^1,\C^{\ell-1})\cong \C^{\ell-1}\otimes H^0(\Pro^1,\Z)\cong \C^{\ell-1},\\
    \end{alignedat}
\end{equation}
\nn
where the first isomorphism in the last line is given by the universal coefficient theorem.
Finally, the global de Rham cohomology is
\begin{equation}
\begin{split}
    &H^0_{dR,t}(\C,d\mathcal{P}_\ell)\cong 0,\\
    &H^1_{dR,t}(\C,d\mathcal{P}_\ell)\cong \C^{\ell-1}.
    \end{split}
    \label{1DTwistedPoly}
\end{equation}
\nn
As we can see, the global twisted de Rham cohomology is independent on the possible coalescence of critical points. The point is that the global Jacobian ring has dimension equal to the total Milnor number \eqref{CokerJacobian}, which takes into account any possible multiplicity $m_i$. So, although the support of ${\rm coker}(\nabla)$ and its local structure changes

\begin{equation}
    {\rm coker}(\nabla)=\bigoplus_{\Sigma}\C_{\sigma_i}^{m_i},
\end{equation}
\nn
its global sections remain always $\Gamma({\rm coker}(\nabla))\cong \C^{\ell-1}$.
The sheaf ${\rm coker}\nabla$ is not sensitive to the coalescence of critical points, due to its naively construction as a direct sum of skyscraper shaves, which loses information about the local structure. In order to recover such information, we need to turn it into a perverse sheaf and to consider a suitable extension of the twisted de Rham complex, i.e. of $\nabla$.\\
Given a connection defined on a open dense subset $U\subset X$ (smooth locus), its extension along a divisor $D=X\backslash U$ come with a prescription about its behaviour near $D$. The ones we are possibly interested in are the so called Middle and Logarithmic extensions: the first one, arising in the context of perverse sheaves, avoids the addition of unnecessary singularities while preserving key invariants; on the contrary, the second one allows for the connection to have logarithmic singularities near the divisors. Although $\nabla$ is defined on the whole complex plane $\C$, it does not define a local system $\La$ on it, because locally constancy fails on $\Sigma$, due to the obstructions arising in solving $\nabla s=0$: flat sections do not freely generate the cohomology. In fact, $\nabla$ defines a local system on $\C\backslash (D\cup\Sigma)$. Such obstructions arise as a consequence of the non trivial monodromy around critical points (and around branch points for a multivalued function).
Indeed, near a critical point $\sigma_i$, the expansion $\mathcal{P}_\ell(z)\sim \mathcal{P}_\ell(\sigma_i)+c(z-\sigma_i)^{m_i+1}$ shows that the monodromy has a Jordan block of size $m_i$, and it becomes unipotent in full degenerate case. Thus, the number of critical points influences the rank of $\rm coker\nabla$ by reducing it by the size of the Jordan blocks of the monodromy matrices.
Explicitly, if $\mathcal{P}_\ell$ has $\ell-1$ distinct critical points, the monodromy acts on flat sections via distinct eigenvalues, thus $\La$ ha no invariant subspaces and ${\rm coker}\nabla\cong\C^{\ell-1}$: each critical point contributes with independent obstructions. Instead, if $\mathcal{P}_\ell$ has only one critical point with multiplicity $\ell-1$, the monodromy matrix becomes unipotent ($\ell-1$ equal eigenvalues), introducing $\ell-2$ relations among obstructions and thus ${\rm coker}\nabla=\C$. The solutions to $\nabla s=0$ is $s(z)=e^{-\gamma\mathcal{P}_\ell(z)}(\sum_{i=0}^{\ell-1} (c_i\log^i(z-\sigma))$. 
Thus:

\begin{equation}
    ({\rm coker}\nabla)_{mid} = \bigoplus_{\sigma\in\Sigma} \C=\C^{n_c},
\end{equation}
\nn
with $n_c$, the number of distinct critical points. Taking into account the monodromy in this way, equivalently, means to restrict on sections with moderate growth, that is to consider the middle (mid) extension $(\Omega\bullet_{\C,D})_{mid}$. Notice that no prescription along the divisor $D$ is added. In particular, the mid extension is independent on $D$.\\
We want just to add a comment on perversity, without dwelling too much on the subject;\footnote{Readers interested in a deeper understanding of perversity are referred to the lecture notes \cite{goresky2021lecturenotessheavesperverse}.} in this context, we could forget about it, since its consideration is necessary only for categorical reasons, but totally irrelevant for the purposes of the present calculations. Consider, for instance, a skyscraper sheaf $\delta$. It fails to be perverse because it does not satisfy co-support conditions, however we can easily make it into a perverse sheaf by just shifting it by $[-1]$, meaning now $H^{i-1}(\delta[-1])=H^i(\delta)$. The ``perversification'' of ${\rm coker}\nabla$ then just imply a unit shift to the left of the spectral sequence, leaving, in fact, hypercohomology unchanged.\\
Finally, supposing only one critical point has multiplicity $m$ 

\begin{equation}
\begin{split}
    &H^0_{dR,t,local}(\C,d\mathcal{P}_\ell)_{(m)}\cong 0,\\
    &H^1_{dR,t,local}(\C,d\mathcal{P}_\ell)_{(m)}\cong \C^{\ell-m},
    \end{split}
    \label{1DDeRham}
\end{equation}
We will need this refinement in the next section when considering the case of degenerate points.




\subsection{Pearcey’s integral}
As a first application of the Lefschetz thimble decomposition discussed above, we examine a Pearcey's integral\cite{Olver}, appearing in \cite{Cacciatori:2024ccm} as the grand-canonical partition function of the gauged Skyrme model, describing baryonic layers living at finite baryon density within a constant magnetic field.
We want to study the integral
\begin{equation}
    \mathcal{P}(a)= \int_{- \infty}^{+\infty} dz \,e^{-a \left( z^4 +bz^2+cz+d\right)},
    \label{Integral:Pearcey_real}
\end{equation}

\nn
for generic values of the real parameters $a,b,c,d$.

\nn
Following the prescription of the previous section, let proceed extending the polynomial argument of the exponential to a holomorphic function over $\C_z$, by complexifying both the variable $z$ and the parameters. In particular, the real parameter $a$ is promoted to the complex parameter $\gamma$, over which we will build the wall crossing structure. We then analyze the integral over a generic contour $\Gamma$ in $\C_z$
\begin{equation}
    \mathcal{P}(\gamma) = \int_{\Gamma} dz\,e^{-\gamma \left( z^4 +bz^2+cz+d\right)},
\end{equation}
and seek a basis of integration cycles along which the integral remains convergent. Once a basis and a intersection product (in homology) are identified, the real integration contour can be decomposed, with integral coefficients, in terms of such basis. As a result, the integral in \eqref{Integral:Pearcey_real} becomes a linear combination of integrals evaluated over the basis. For large values of the parameters, these basis integrals admit an asymptotic expansion, which is then transferred to the initial integral. The expectation is that for different values of the parameters, both the basis for the integration contours and the decomposition of the real line in terms of them will be modified.\\
The set of critical points of the holomorphic function
\begin{equation}
    f(z)\equiv z^4+bz^2+cz+d,
\end{equation}
 i.e. the solutions of the cubic equation $f'(z)=4z^3+2bz+c=0$, can be compactly written as

\begin{equation}
    \Sigma = \left \{-\frac{b}{\sqrt[3]{3(-9c+\sqrt{3\Delta})}}+\frac{\sqrt[3]{-9c+\sqrt{3\Delta}}}{2\sqrt[3]{9}},\frac{(1\pm i\sqrt{3})b}{2\sqrt[3]{3(-9c+\sqrt{3\Delta})}}-\frac{(1\pm i\sqrt{3})\sqrt[3]{-9c+\sqrt{3\Delta}}}{4\sqrt[3]{9}}\right\},
\end{equation}
\nn
where according to the sign of the discriminant $\Delta$, three different situation arise: 
\begin{equation}
    \Delta\equiv 8b^3+27c^2 \quad \quad \quad \begin{cases}
     > 0  \quad \quad 1 \, \text{ real and } 2 \text{ complex conjugate solutions,}\\
    < 0  \quad \quad 3 \, \text{ real different solutions,}\\
      =0  \quad \quad 3 \,\text{ real solutions with at least a multiple root.}\\
    
    \end{cases}
\end{equation}
We will analyze these three cases separately, since each one defines a different connected region in the parameter space $U=\left\lbrace (b,c) \in \mathbb{C}^2 \right\rbrace$, and on each region we can define a distinct local system of Betti homologies $H^{\bullet}_{Betti,glob,\gamma}(\mathbb{C}, f_{(a,b)}, \mathbb{Z})$ equipped with its own wall-crossing structure.
\paragraph{Positive discriminant.}
Let us firstly consider the case where $\Delta >0$. \\
For concreteness, we fix the parameters to $(b,c,d)=(3/2,7,-1)$ and carry out the explicit computations for this choice. The critical points are computed to be

\begin{equation}
    \sigma_i\in \Sigma=\left \{-1,\frac{1}{2}(1+i\sqrt{6}),\frac{1}{2}(1-i\sqrt{6})\right \},
\end{equation}

\begin{figure}[!htb]
\centering
\hspace{-1cm}
\begin{minipage}{0.5\textwidth}
\centering
  \includegraphics[width=0.98\linewidth]{figure/StokeslinesPearcey.pdf}
\end{minipage}
\begin{minipage}{0.5\textwidth}
\centering
  \includegraphics[width=1.05\linewidth]{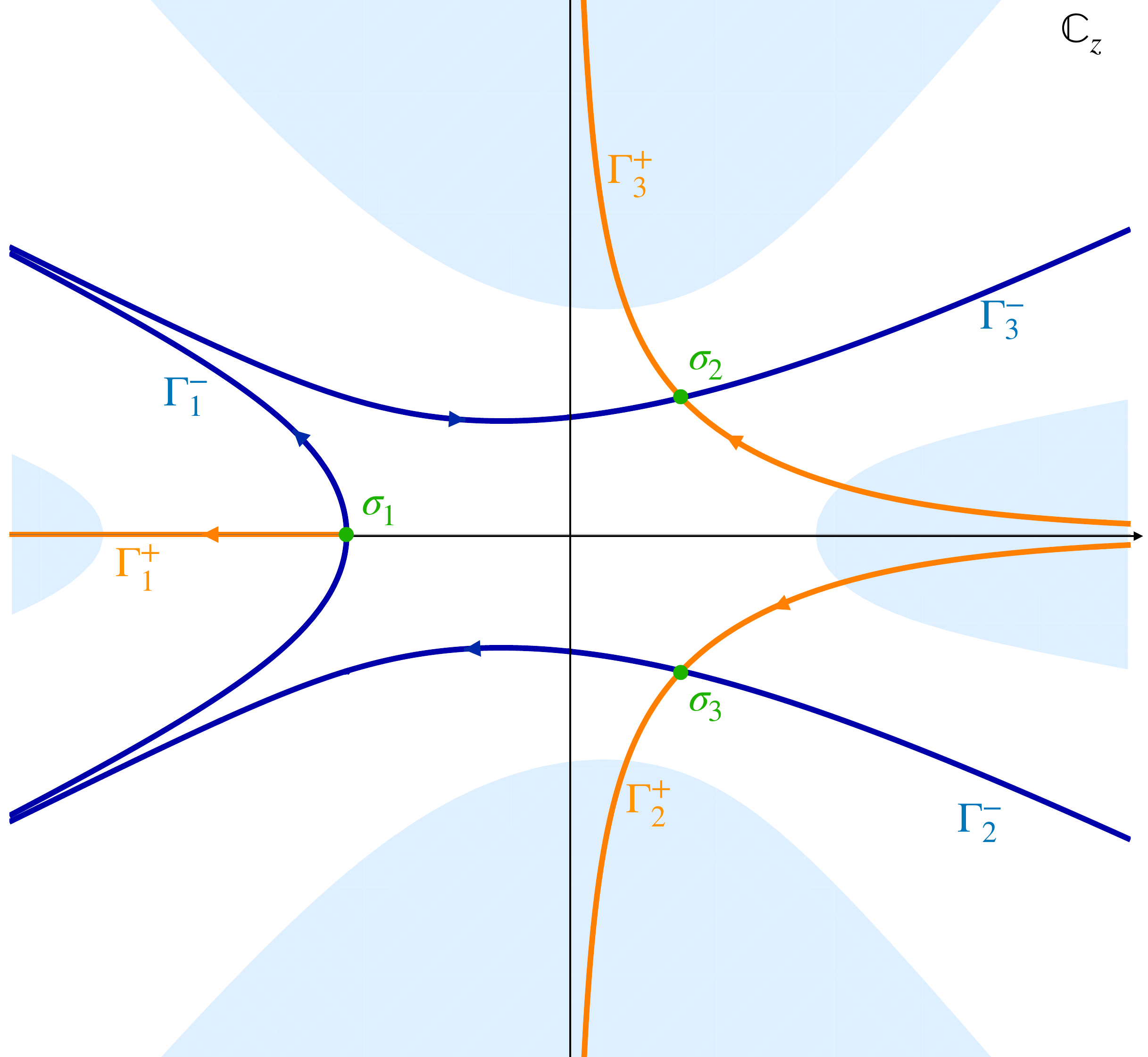}
\end{minipage}
\caption{Positive discriminant case $(\gamma,b,c,d)=(1,3/2,7,-1)$. (Left) Stokes' lines on the $\C_\gamma$ plane. (Right) Ascendant paths $\Gamma_i^+$ spanning $H_1(\C_z,D_N,\Z)$ (in orange), descendant paths $\Gamma_i^-$ spanning $H_1(\C_z,X_{-N},\Z)$ (in blue).}
\label{PearcyPositiveDiscriminant}
\end{figure}

\nn
where $f(z)$ takes (respectively) the critical values 

\begin{equation}
    t_i \equiv f(\sigma_i)\in \mathcal{S}=\left \{-\frac{11}{2},\left (\frac{11}{16}+3i\sqrt{6}\right ),\left (\frac{11}{16}-3i\sqrt{6}\right)\right\}.
\end{equation}
\nn
The non-degeneracy of the Hessian at each critical point, together with the fact that all Morse indices equal one guarantee the saturation of \eqref{BettiRelation}, thus $dim H_1(X,D_N,\Z)=3$. Here $D_N\subset \C$ denotes the union of the four connected regions in the complex $z=(u,v)$ (shaded blue in Figure \ref{PearcyPositiveDiscriminant}) where the Morse function $h(u,v)=Re( \gamma f(u,v))>N$.\\
Using \eqref{StokesRaysFormula}, we find that the Stokes' rays are the three lines

\begin{equation}
\begin{alignedat}{3}
    &l_0:Re(\gamma)=-\frac{11}{16}\sqrt{\frac{3}{2}}\rm{Im}(\gamma), \quad \quad &&\mbox{where}\quad \rm{Im}(\gamma f(z))|_{\sigma_1}=\rm{Im}(\gamma f(z))|_{\sigma_2},\\
    &l_1:Re(\gamma)=0, \quad \quad &&\mbox{where}\quad \rm{Im}(\gamma f(z))|_{\sigma_2}=\rm{Im}(\gamma f(z))|_{\sigma_3},\\
    &l_2:Re(\gamma)=\frac{11}{16}\sqrt{\frac{3}{2}}\rm{Im}(\gamma),\quad \quad &&\mbox{where}\quad \rm{Im}(\gamma f(z))|_{\sigma_1}=\rm{Im}(\gamma f(z))|_{\sigma_3},
\end{alignedat}
\end{equation}
\nn
resulting in a splitting of the $\C_{\gamma}$ plane in three regions with different thimbles structures. Let us fix $\gamma=1$, lying in the first petal of the fan (orange region labeled with $(0)$ on the right side of Figure \ref{PearcyPositiveDiscriminant}).
We identify the three thimbles $\Gamma_i\equiv \Gamma_{\sigma_i}$ as the paths passing through a critical point and keeping constant the imaginary part of the Morse function (Figure \ref{PearcyPositiveDiscriminant}):

\begin{equation}
\begin{split}
    &H_1(X,D_N,\Z)= span\{\Gamma_1^+,\Gamma_2^+,\Gamma_3^+\}\cong \Z^3,\\
    &H_1(X,D_N,\Z)^\vee= span\{\Gamma_1^-,\Gamma_2^-,\Gamma_3^-\}\cong \Z^3.
    \end{split}
\end{equation}
\nn
Let us set $f(z)=t$ and look at the preimage 

\begin{equation}
    f^{-1}(t)=\begin{pmatrix} z_1(t)\\z_2(t)\\z_3(t)\\z_4(t)\end{pmatrix}.
\end{equation}
\nn
When approaching a critical point $\sigma_i$, the four point fiber degenerates to a three point set, identifying a vanishing cycle $\Delta_i$. We have

\begin{equation}
    \Delta_1=\{z_3\}-\{z_4\} ,\quad \quad \Delta_2=\{z_1\}-\{z_4\}\quad \mbox{and}\quad \Delta_3=\{z_1\}-\{z_4\}.
    \label{VanishingPearcey}
\end{equation}
\nn
The monodromy matrices acting on this base of vanishing cycles are computed to be 

\begin{equation}
    M_1=\begin{pmatrix}-1&0&0\\-1&1&0\\-1&0&1\end{pmatrix} \quad \mbox{and}\quad M_2=M_3=\begin{pmatrix}1&-1&0\\0&-1&0\\0&0&-1\end{pmatrix}.
\end{equation}
\nn
By using \eqref{PicardLefschetzFormula} and \eqref{TotalIntersection} we compute the intersection numbers

\begin{equation}
    \Delta_{12}= 1 \quad, \quad \Delta_{13}=1 \quad\mbox{and}\quad \Delta_{23}= -2.
\end{equation}
\nn
Note that the intersection numbers $\Delta_{ij}$ are defined in \eqref{TotalIntersection} up to a sign depending on the orientation of the cycles. 
When crossing a Stokes' line, the base of thimbles undergoes a change of the type \eqref{equation:jump}. Let $\Gamma_i^{+(k)}$ be the vector of thimbles in the $(k)$-th sector of the fan. With the clockwise ordering showed in Figure \ref{PearcyPositiveDiscriminant} and $T^{\theta_{(k)}}$\footnote{Matrix representation of the operator dual to $T_{\theta}$ in \eqref{Tij}.} the jump matrix associated to the Stokes line $l_{(k)}\to s_{\theta_k}$ connecting the $(k)$-th and the $(k+1)$-th sectors of the fan, 

\begin{equation}
     \Gamma_i^{+(k+1)}=T_{ij}^{\theta_{(k)}}\Gamma_j^{+(k)},
\end{equation}
\nn
we have\\
\begin{equation}
\begin{alignedat}{3}
    &Re(\gamma f(z))|_{\sigma_1}<Re(\gamma f(z))_{\sigma_2} \quad \quad\quad \quad &&\mbox{along}\,\, l_0,\\
    &Re(\gamma f(z))|_{\sigma_2}>Re(\gamma f(z))_{\sigma_3} \quad \quad\quad \quad &&\mbox{along}\,\, l_1,\\
    &Re(\gamma f(z))|_{\sigma_1}>Re(\gamma f(z))_{\sigma_3} \quad \quad\quad \quad &&\mbox{along}\,\, l_2,\\ 
\end{alignedat}
\end{equation}
\nn
so that

\begin{equation}
    T^{\theta_{(0)}}=\begin{pmatrix}1&1&0\\0&1&0\\0&0&1\end{pmatrix} \quad,\quad T^{\theta_{(1)}}=\begin{pmatrix}1&0&0\\0&1&0\\0&-2&1\end{pmatrix}\quad,\quad T^{\theta_{(2)}}=\begin{pmatrix}1&0&0\\0&1&0\\1&0&1\end{pmatrix}.
\end{equation}
\noindent
These matrices define the wall crossing structure in $\mathbb{C}_{\gamma} \times U_{b,c}^{\Delta > 0}$, in which the walls are defined on subregions  of this space where exactly two critical values are aligned
\begin{equation}
    W_{ij}= \left\lbrace (\gamma, u) \in \mathbb{C}^{\ast} \times U_{b,c}^{\Delta > 0} \vert \rm{Im} \left( \gamma (t_i(u)-t_j(u))\right)=0 \right\rbrace.
\end{equation}
They correspond to walls of the second type in the sense of \cite{kontsevich:2008kos}.\\
After a complete round of $2\pi$, we get the full monodromy matrix
\begin{equation}
    S=\begin{pmatrix}
        0&2&-1\\1&-2&2\\1&-1&1
    \end{pmatrix},
\end{equation}
\nn
with eigenvalues $-1,-1,1$.

\nn
Let us now consider the case of non positive discriminant. Then, singular points lie on the real axes of $\C_z$, this meaning that a Stokes' line appears along $\rm{Im}(\gamma)=0$, splitting the $\C_\gamma$ plane into two regions, corresponding to the upper and lower half planes. It is worthwhile to emphasize that the a priori naive choice of a real $\gamma$, in this case, would give rise to a wrong description, since we would have set precisely on the Stokes' line. In order to proceed, let us thus set $\gamma=i$.

\begin{figure}[!htb]
\centering
\hspace{-1cm}
\begin{minipage}{0.5\textwidth}
\centering
  \includegraphics[width=1\linewidth]{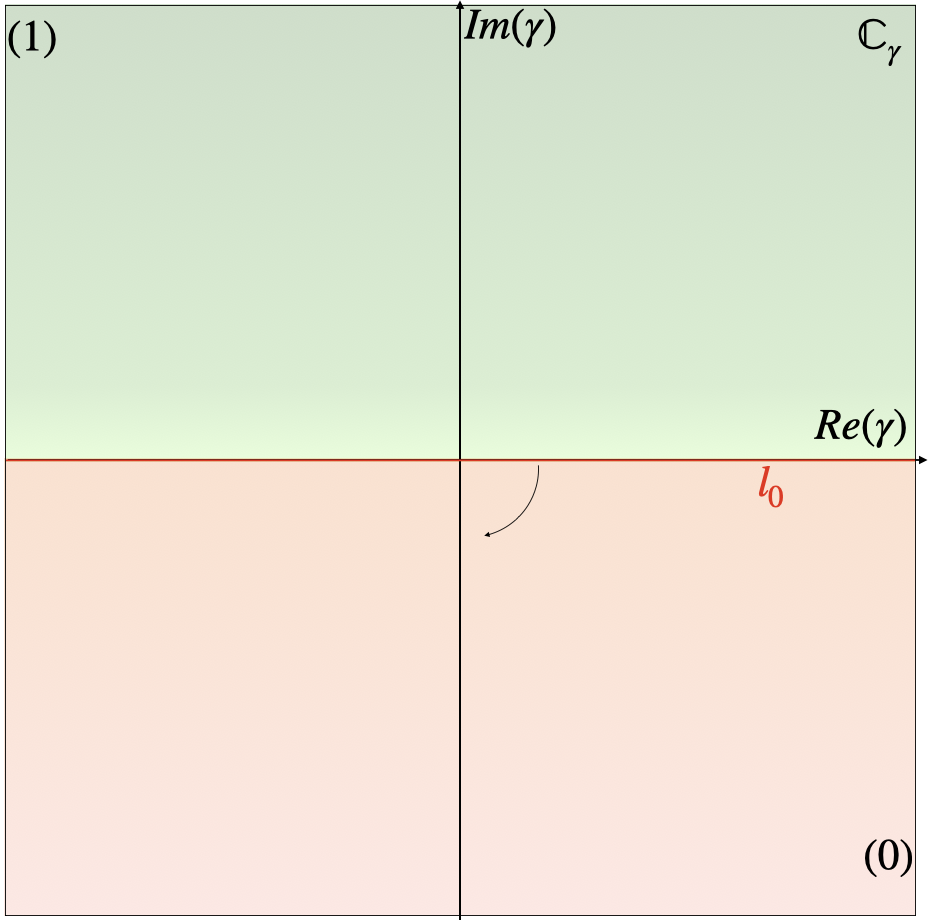}
\end{minipage}
\begin{minipage}{0.5\textwidth}
\centering
  \includegraphics[width=0.99\linewidth]{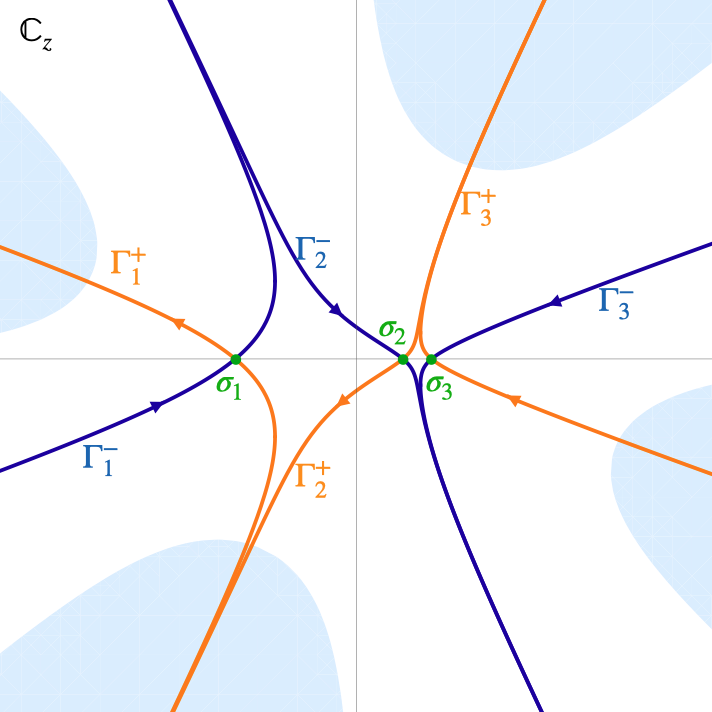}
\end{minipage}
\caption{Negative discriminant case $(\gamma,b,c,d)=(i,-1,1/2,-1)$. (Left) Stokes' lines on the $\C_\gamma$ plane. (Right) Ascendant paths $\Gamma_i^+$ spanning $H_1(\C_z,D_N,\Z)$ (in orange), descendant paths $\Gamma_i^-$ spanning $H_1(\C_z,X_{-N},\Z)$ (in blue).}
\label{PearcyNegativeDiscriminant}
\end{figure}

\paragraph{Negative discriminant.} Firstly, let us consider the case $\Delta<0$, characterized by three different real critical points. The local Betti homology generated by local thimbles, shown in Figure \ref{PearcyNegativeDiscriminant}(right), is pretty much the same as in the positive discriminant case, being it of course unaffected by the reality of critical points. However, as early emphasized, the relevant peculiarity appears in the thimbles structure, due to a Stokes' line $l_0$ on the real axis of the $\C_\gamma$ plane, (see Figure \ref{PearcyNegativeDiscriminant}(left)). Setting $(b,c,d)=(-1,1/2,-1)$, we get

\begin{equation}
    \Sigma=\left \{\frac{1}{2},\frac{1}{4}(1\pm \sqrt{5})\right \} \quad \mbox{and} \quad \mathcal{S}=\left \{-\frac{15}{16},\frac{1}{32}(-41\pm 5 \sqrt{5})\right \}.
\end{equation}
\nn
Proceeding as above, we identify three vanishing cycles

\begin{equation}
    \Delta_1=\{z_1\}-\{z_2\}, \quad \Delta_2=\{z_3\}-\{z_4\}, \quad \Delta_3=\{z_1\}-\{z_4\}
\end{equation}
\nn
and the corresponding monodromy matrices

\begin{equation}
    M_1=\begin{pmatrix}-1&0&0\\0&1&0\\-1&0&1\end{pmatrix},
    \,M_2=\begin{pmatrix}1&0&0\\0&-1&0\\0&-1&1\end{pmatrix} \quad \mbox{and}\quad M_3=\begin{pmatrix}1&0&-1\\0&1&-1\\0&0&1\end{pmatrix}.
\end{equation}
\nn
In order to determine the jump matrices, we have to compute the intersection numbers among thimbles. However, the Morse function vanishes in all critical points. In order to avoid it, we slightly move away from the imaginary axis setting $\gamma=i+\delta$. We get 

\begin{equation}
    Re(\gamma f(z))|_{\sigma_3}>Re(\gamma f(z))|_{\sigma_1}>Re(\gamma f(z))_{\sigma_2}, \quad \quad\quad \quad \mbox{along}\,\, l_0.
\end{equation}
\nn
Note that in this case the ray $l_0$ corresponds to a Stokes line intersecting three distinct critical values. The natural generalization of the jump matrix \eqref{equation:jump}  in this case accounts for the double jump of $\Gamma^{+(0)}_2$:  the first caused by crossing the branch cut emanating from $t_1$ in the $\mathbb{C}_t$ plane, and the second by crossing the cut associated with $t_3 \in \mathbb{C}$. Then, when we cross the line $l_0$ in the clockwise direction, the corresponding jump matrix is given by the following upper triangular matrix:

\begin{equation}
      \begin{pmatrix}\Gamma^{+(1)}_2\\\Gamma^{+(1)}_1\\\Gamma^{+(1)}_3\end{pmatrix}=\begin{pmatrix}1&\Delta_{21}&\Delta_{23}-\Delta_{21}\Delta_{13}\\0&1&\Delta_{13}\\0&0&1\end{pmatrix}\begin{pmatrix}\Gamma^{+(0)}_2\\\Gamma^{+(0)}_1\\\Gamma^{+(0)}_3\end{pmatrix},  
\end{equation}

\nn
where intersection numbers among vanishing cycles $\Delta_{ij}$ are computed with \eqref{PicardLefschetzFormula}. Reordering the base of thimbles, we get

\begin{equation}
   T^{(0)}=\begin{pmatrix}1&0&-1\\0&1&-1\\0&0&1\end{pmatrix},
   \label{jump:Negative_discriminant}
\end{equation}
\nn
and 
\begin{equation}
    S=\begin{pmatrix}1&0&-1\\0&1&-1\\1&1&-1\end{pmatrix},
\end{equation}
\nn
with eigenvalues $i,-i,1$.
The matrix \eqref{jump:Negative_discriminant} define the wall crossing structure in $\mathbb{C}_{\gamma}^{\ast} \times U_{(b,c)}^{\Delta<0}$, in which the walls are defined by subregions of this space where three critical values are aligned. They correspond to walls of the first type in the sense of \cite{kontsevich:2008kos}.

\begin{figure}[!h]
\centering
\hspace{-1cm}
\begin{minipage}{0.5\textwidth}
\centering
  \includegraphics[width=1\linewidth]{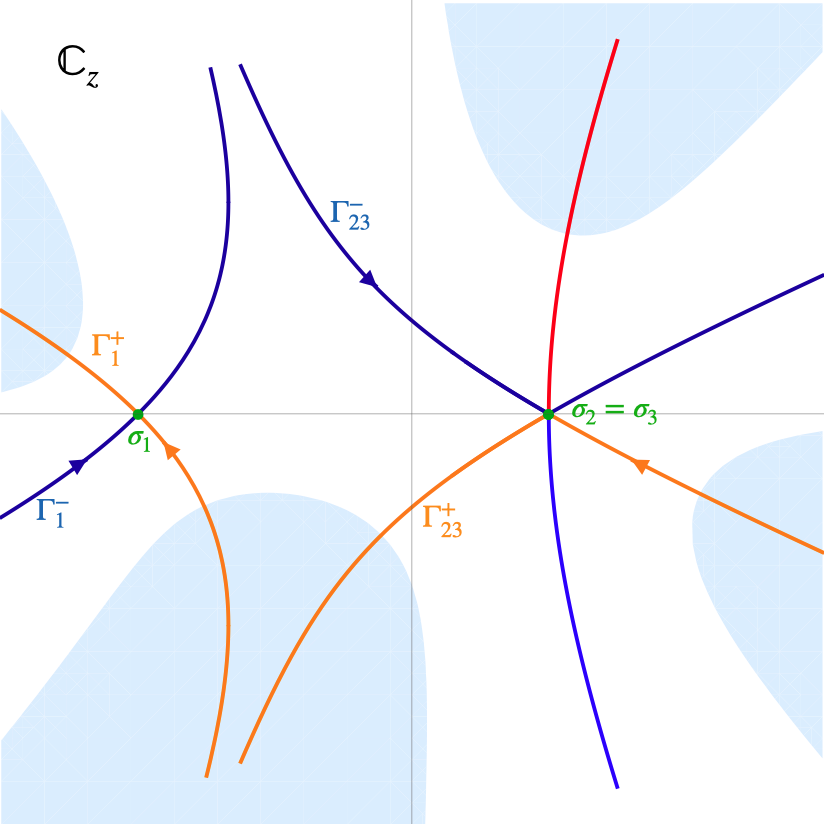}
\end{minipage}
\begin{minipage}{0.5\textwidth}
\centering
  \includegraphics[width=0.99\linewidth]{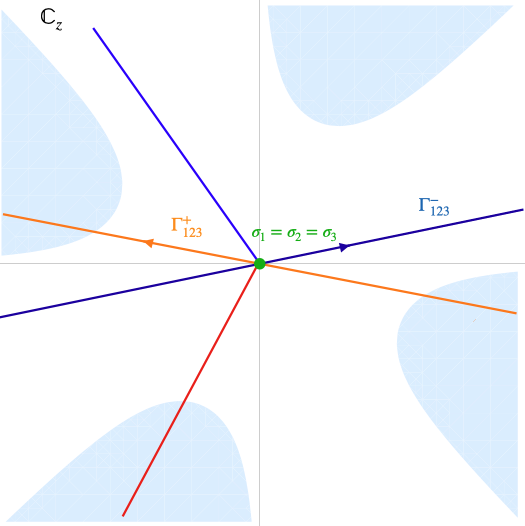}
\end{minipage}
\caption{Vanishing discriminant case. \\(Left)Double Degeneration, $(\gamma,b,c,d)=(i,-6,8,-1)$. Ascendant paths $\Gamma_i^+$ spanning $H_1(\C_z,D_N,\Z)$ (in orange), descendant paths $\Gamma_i^-$ spanning $H_1(\C_z,X_{-N},\Z)$ (in blue).\\(Right)Triple Degeneration, $(\gamma,b,c,d)=(i,0,0,-1)$. Ascendant paths $\Gamma_i^+$ spanning $H_1(\C_z,D_N,\Z)$ (in orange), descendant paths $\Gamma_i^-$ spanning $H_1(\C_z,X_{-N},\Z)$ (in blue).(Both) Coalescing ascendant (red) and descendant (light blue) paths.}
\label{PearcyVanishingDiscriminant}
\end{figure}

\paragraph{Vanishing discriminant.}
The last case we want to discuss is $\Delta=0$, where two or all three critical points coalesce. Comparing Figures \ref{PearcyNegativeDiscriminant}(right) and \ref{PearcyVanishingDiscriminant}(left) we observe that as $\sigma_2$ and $\sigma_3$ coalesce, the upward branches of $\Gamma_2^+$ and $\Gamma_3^+$, as well as the downward branches of $\Gamma_2^-$ and $\Gamma_3^-$,  begin to overlap with opposite orientations. As a result, the combination of these four paths yields only two independent thimbles:

\begin{equation}
    \Gamma_2^++\Gamma_3^+=\Gamma_{23}^+ \quad \mbox{and}\quad  \Gamma_2^-+\Gamma_3^-=\Gamma_{23}^-, 
\end{equation}
\nn
This is a consequence of the fact the vanishing cycles $\Delta_2$ and $\Delta_3$, appearing in \eqref{VanishingPearcey}, identify the same homology class when the associated critical points coalesce.\\
Similarly, when the triple degeneration occurs\footnote{For $b=c=0$.}, shown in Figure \ref{PearcyVanishingDiscriminant}(right), four branches overlap, leading to a further reduction in the number of independent thimbles. We are left with:

\begin{equation}
    \Gamma_1^++\Gamma_{23}^+=\Gamma_{123}^+ \quad \mbox{and}\quad  \Gamma_1^-+\Gamma_{23}^-=\Gamma_{123}^-.
\end{equation}
\nn
Therefore, the Betti homology turn out to be

\begin{equation}
\begin{alignedat}{2}
    &H_1(X,D_N,\Z)_{(2)}=span\{\Gamma_1^+,\Gamma_{23}^+\} \quad \quad &&\mbox{and}\quad H_1(X,D_N,\Z)_{(2)}^\vee=span\{\Gamma_1^-,\Gamma_{23}^-\},\\
    &H_1(X,D_N,\Z)_{(3)}=span\{\Gamma_{123}^+\}, \quad \quad &&\mbox{and}\quad H_1(X,D_N,\Z)_{(3)}^\vee=span\{\Gamma_{123}^-\},\\
\end{alignedat}
\end{equation}
\nn
where bracket subscripts denote the multiplicity of coalescent critical points. 
In the cohomology side, we can compute the relative cohomology using (\ref{1DDeRham}). We obtain

\begin{equation}
\begin{alignedat}{2}
    &H^1(X,D_N,\C)_{(2)}=span\{1,z\} \cong \C^2,\\
    &H^1(X,D_N,\C)_{(3)}=span\{1\}\cong \C.
\end{alignedat}
\end{equation}
\nn
Notice that the universal coefficient theorem for cohomology explicitly shows the duality \footnote{$Ext(H_0(X,D_N,\Z)_{(i)},\C)=0.$}
\begin{equation}
    H^1(X,D_N,\C)_{(i)}\cong Hom(H_1(X,D_N,\Z),\C)_{(i)}.
\end{equation}

\section{Exponential integrals for closed forms and Feynman integrals}\label{EICFFI}

In the previous section, we associated to each triple $\left( X, D_0,f \right)$ four Local Systems over $\mathbb{C}_{\gamma}^{\ast}$ corresponding to the global and local de Rham and Betti cohomologies. Moreover, we showed that, using this language, exponential integrals  can be naturally interpreted as periods pairing these two types of (co)homologies. 
Here, still following \cite{Kontsevich:2024mks}, we extend the discussion to the more general setting of a triple $(X, D_0, \alpha)$ where $X$ is a complex smooth algebraic variety, $D_0 \subset X$ is a normal crossing divisor, and $\alpha$ is a closed algebraic 1-form. This generalized setup allows us to consider exponential integrals defined by multivalued functions, which represent ubiquitous objects in physics. Understanding such integrals and developing general methods to evaluate them remains a challenging and often open problem across various domains of theoretical physics.
The aim of this section is to outline a general framework in which exponential integrals of this type can be systematically studied. Our primary motivation comes from the study of Feynman integrals. In particular, we introduce here the study of Feynman integrals in the Baikov representation \cite{Baikov:1996iu} (see also \cite{Grozin:2011mt} for a review) using the language of exponential integrals. A more detailed analysis will be provided in \cite{Angius:2025acm}.

Any $D$ dimensional, $L$-loops Feynman integral with $E$ external legs, can be rewritten, using denominators as integration variables, in the so called Baikov representation 

\begin{equation}
    I(\{\nu_i\})=\int_\Gamma \mathcal{B}^{-\gamma} \prod_{i=1}^N \frac{dz_i}{z^{\nu_i}},
    \label{Baikov}
\end{equation}
\nn
where $\mathcal{B}$, known as the Baikov polynomial, is a polynomial on the integration variables which coefficients depends on the masses of the internal particles and the external momenta, and $\gamma=-(D-E-L-1)/2$. If $\gamma \in \mathbb{Q}$, the integral is multivalued but its underlying geometry\footnote{Here, by geometry we mean the geometric space where the integrand becomes single-valued.} is a computable Riemann foliation obtained by gluing a finite number of sheets. In such cases, the integral \eqref{Baikov} can still be interpreted as a standard pairing between the de Rham cohomology and the singular homology associated to this geometry. However, in the context of dimensional regularization, $\gamma\notin \Q$. As a result, we lose our geometric intuition behind the integral, and the previous interpretation of \eqref{Baikov} as a period no longer applies. Reformulating the integrand as an exponential
\begin{equation}
    I(\left\lbrace \nu_i \right\rbrace) = \int_{\Gamma} e^{-\gamma \log \mathcal{B}} \prod_{i=1}^N \frac{dz_i}{z^{\nu_i}}
\end{equation}
solves the issue on $\gamma$ since it allows to interpret the integral as an exponential period for any $\gamma \in \mathbb{C}^{\ast}$ at the cost of dealing with a non-holomorphic function in the argument of the exponential. 

\subsection{Twisted de Rham Cohomology}

The first important consequence of dealing with a multivalued function in the exponential is that the twisted de Rham cohomology side of the pairing is defined with respect to the differential
\begin{equation}
    \nabla_{\alpha} \, = \, d \, - \,  \alpha \,  \wedge,
\end{equation}
where $\alpha$ is a closed $1$-form on the complex algebraic variety $X$ that is not necessarily exact. This means that, in general, $\alpha$ cannot be written globally as $\alpha = d f$ for some function $f$, but additional contributions may enter into its definition. To identify the various potential contributions, one needs to choose a suitable compactification $\overline{X}$ of $X$. However, the final result should ultimately be independent of the specific choice. \\
This ``good'' compactification for $X$ must satisfy the properties stated in Proposition $3.1.1$ of \cite{Kontsevich:2024mks}. In particular it must be constructed using a set of normal crossing divisors $D_h, D_v$ and $D_{\log}$
\begin{equation}
    \overline{X} - X = D_h \cup D_v \cup D_{\log},
\end{equation}
such that
\begin{itemize}
    \item[\textit{(i)}] Among all the normal crossing divisors $\overline{D}_0, D_h, D_v$ and $D_{\log}$ only $D_v$ and $D_{\log}$ can have common irreducible components;
    \item[\textit{(ii)}] For any point $x \in \overline{X}$ there exists a small analytic neighborhood $U$ and a closed meromorphic $1-$form $\alpha$ locally given by the expression
\begin{equation}
    \alpha = \alpha_{reg} + \alpha_{log} + \alpha_{\infty}.
\end{equation}
The first contribution represent a regular form on $U$. The second contribution, $\alpha_{\log}$, can be expressed in local coordinates near a divisor $D_{log}$ as
\begin{equation}
    \alpha_{log} = \sum_i c_i d \log z_i,
\end{equation}
where $z_i$ are local coordinates in which $D_{log}$ is given by $\prod_i z_i=0$, and $c_i \in \mathbb{C} \setminus \left\lbrace 0 \right\rbrace$ represent the periods of $\alpha$ around the loci $z_i=0$, computed as
\begin{equation}
    c_i = \frac{1}{2 \pi i} \oint_{S^1_i} \alpha,
\end{equation}
the circle $S^1_i$ encircling $D_{log}$ in a smooth point.
The final contribution admits the form $\alpha_{\infty}=dF$ in local coordinates near $D_v$, where $F$ is an analytic function of the form
\begin{equation}
    F = \frac{c}{\prod_j z_j^{k_j}} \left( 1 + o (1) \right) \, , \quad k_j \geq 1
\end{equation}
and $D_v$ is locally defined by $\prod_j z_j=0$ in those coordinates.
\end{itemize}
\nn
Once we have such a compactification, we can construct the sheaf $\Omega^{\bullet}_{\overline{X},D_{\log}}$ of differential forms on $\overline{X}$ with possible logarithmic poles along the divisors $D_{\log}$ and use this sheaf to construct the global de Rham cohomology at any $\gamma \in \mathbb{C}^{\ast}$ using the following definition.\\[0.5cm]
\textbf{Definition:} \textsc{[global twisted de Rham, \cite{Kontsevich:2024mks}, Def. 3.2.1]}\\
\textit{Let $\gamma \in \mathbb{C}_{\gamma}^{\ast}$. The (twisted) de Rham cohomology is the graded abelian group constructed by the hypercohomology
\begin{equation}
   H^{\bullet}_{dR,glob,\gamma} \left( X, D_0, \alpha \right) \, = \, \mathbb{H}^{\bullet} \left( \overline{X}, (\Omega^{\bullet}_{\overline{X},D}, \gamma^{-1}d + \alpha \wedge \, ) \right).
   \label{GlobaldR:forms}
\end{equation}
}\\
Also in the case of closed forms, one can define a local version of the cohomology and establish a global-to-local isomorphism, analogous to equation \eqref{dR:global_to_local} for exact differentials. To provide this definition we need to introduce the subsheaf $\Omega^{\bullet}_{\overline{X}, \alpha}$ of $\Omega^{\bullet}_{\overline{X}}(\log (\overline{D}_0 +D_h+D_v+D_{\log}))(- \overline{D}_0)$ consisting of forms $\eta \in \Omega^{\bullet}_{\overline{X}}(\log (\overline{D}_0 +D_h+D_v+D_{\log}))(- \overline{D})$ such that $\alpha \wedge \eta \in \Omega^{\bullet}_{\overline{X}}(\log (\overline{D}_0 +D_h+D_v+D_{\log}))(- \overline{D}_0)$, which means that the $1$-form $\alpha$ can have poles of order one along the divisors $D_v \cup D_h \cup D_{\log}$. 
\\
[0.5cm]
\textbf{Definition:} \textsc{[local twisted de Rham, \cite{Kontsevich:2024mks}, Def. 3.2.1]}\\
\textit{Let us indicate with $\mathcal{Z} (\alpha)$ the set of zeros of $\alpha$ on $X$ and the zeros of the restriction of $\alpha$ on $D_0 \cup D_h$. We define the local (twisted) de Rham cohomology as
\begin{equation}
   H^{\bullet}_{dR,loc,\gamma} \left( X, D_0, \alpha \right) \, = \bigoplus_{i \in I} \, \mathbb{H}^{\bullet} \left( U_{form } (z_i), (\Omega^{\bullet}_{U_{form(z_i),\alpha}} \left[ \left[ \gamma^{-1} \right] \right], \gamma^{-1}d + \alpha \wedge \, ) \right),
   \label{LocaldR:forms}
\end{equation}
where $U_{form}(z_i)$ is the formal neighborhood of the component $z_i$ of $\mathcal{Z} (\alpha)$.}\\[0.5cm]
\noindent
The following isomorphism holds:
\begin{equation}
    H_{dR,glob, \gamma}^{\bullet} \left( X, \alpha \right) \, \simeq \, H_{dR,loc, \gamma}^{\bullet} \left( X, \alpha \right).
\end{equation}

\subsection{Betti Cohomology}\label{SubsectionBetti}
The direct construction of the global Betti cohomology is extremely technical, we just briefly recall it here for fixing notations, referring to \cite{Kontsevich:2024mks} for details. 
Let $\widetilde{X}$ be the real oriented blow-up of $\overline{X}$ along $D=D_h\cup D_v\cup D_{log}$, that is the manifold with boundary, and possibly corners, obtained from $\overline{X}$ replacing $D$ with the $S^1-$ bundle of its normal bundle. Let $\pi:\widetilde{X}\rightarrow\overline{X}$ and $\Pi:X\rightarrow\widetilde{X}$ be the natural projection and embedding, respectively. Let $\La_{\alpha,\gamma}$ be the local system on $X$ of flat sections of the trivial vector bundle on $X$ with respect to $\nabla=(d+\gamma\alpha)$, i.e.

\begin{equation}
    \La_{\alpha,\gamma}(U)=\ker \nabla|_{U}, \quad \quad U\subset X.
\end{equation}
\nn
Let $D^{\R}_k=\{y \in \de \widetilde{X}|\pi(y)\in D_k\}$, $k=h, v, log$, be the portion of the boundary of $\widetilde{X}$ whose points are projected onto $D_k$, and let us split each of them according to the growth behavior of $F$,

\begin{equation}
\begin{split}
    &D^{\R\,\pm}_v=\{\tilde{z}\in D^{\R}_v| \pm Re(\gamma F(\pi(\tilde{z}))\geq 0)\},\\
    &D^{\R\,\pm}_{log,i}=\{\tilde{z}\in D^{\R}_{log}| \mp Re( \gamma c_i)> 0)\},
    \end{split}
    \label{PlusMinusDivisors}
\end{equation}
\nn
where $c_i$ is the residue of $\alpha$ in the $i-th$ irreducible component of $D^\R_{log}$.
Next, let us define $\widetilde{X}^-$ as the subset of $\widetilde{X}$ obtained by removing the normal directions where $F$ has a ``bad'' behavior: 
\begin{equation}
    \widetilde{X}^-=\widetilde{X}\backslash \left ( (D^\R_{log}-D^{\R\,+}_{log})\cup(D^\R_{v}-D^{\R\,-}_{v})\cup D_h \right ),
\end{equation}
\nn
with $D^{\R\,\pm}_{log}=\overline{\cup_i D^{\R\,\pm}_{log,i}}$.
Let $i:\widetilde{X}^-\rightarrow \widetilde{X}$ its open inclusion. \\[0.5cm]
\textbf{Definition:} \textsc{[Global Betti cohomology, \cite{Kontsevich:2024mks}, Def.3.4.3]}\\
\textit{The global Betti cohomology is defined as}
\begin{equation}
\begin{split}
    H^{\bullet}_{Betti,glob, \gamma} \left( X, \alpha \right)& \equiv H^{\bullet} \left( \widetilde{X}, \Pi_\ast (\La_{\alpha,\gamma})\otimes i_!(\underline{\Z}_{\widetilde{X}^-})\right)\\ &\cong H^{\bullet} \left( \widetilde{X}, i_!(\underline{\Z}_{\widetilde{X}^-}\otimes i^\ast \Pi_\ast (\La_{\alpha,\gamma})\right)\cong H^{\bullet} \left( \widetilde{X}, i_!i^\ast \Pi_\ast (\La_{\alpha,\gamma})\right) \\&\cong H^{\bullet} \left( \widetilde{X}, D^{\R\,+}_v\cup D^{\R\,-}_{log},\Pi_\ast (\La_{\alpha,\gamma})\right).
    \label{GlobalBetti1form}
    \end{split}
\end{equation}
\vspace{0.3cm}

\nn
Let us describe the construction of the \textit{local} Betti cohomology in a way similar to that presented in section  \ref{sec:betti} for the case of holomorphic functions. 
Fixing a Riemannian metric on $X$, for each $z_i \in \mathcal{Z}(\alpha)$, we can always choose a sufficiently small $\varepsilon-$neighborhood $U_{\varepsilon,i} \left( z_i  \right) \subset X$ and a holomorphic function $f_i$ defined on it such that, locally in this neighborhood,
\begin{equation}
    \alpha = d f_i.
\end{equation}
For each $\theta \in \mathbb{R} / 2 \pi \mathbb{Z}$, with $\theta = \arg \left( \gamma \right)$, we can define the graded $\mathbb{Z}-$module $H^{\bullet}_{Betti,local,z_i,\gamma}$, analogous of \eqref{Betti_local:ti}, via the relative cohomology with respect to preimage of the point $t_i +\varepsilon e^{i \theta}$ in the boundary of $\overline{U}_{\varepsilon,i} \left( z_i  \right)$:
\begin{equation}
    H^{\bullet}_{Betti,local,z_i, \gamma} \left( X, \alpha \right) = H^{\bullet} \left(U_{\varepsilon,i}(z_i), \overline U_{\varepsilon,i}(z_i) \cap f_i^{-1} \left( t_i+\varepsilon e^{i \theta} \right) \right),
\end{equation}
with $t_i=f_i \left( z_i \right)$.\\[0.5cm]
\textbf{Definition:} \textsc{[local Betti cohomology, \cite{Kontsevich:2024mks}, Def.3.4.6]}\\
\textit{For fixed $\gamma$, the direct sum}
\begin{equation}
    H^{\bullet}_{Betti,local, \gamma} \left( X, \alpha \right) = \bigoplus_{z_i \in \mathcal{Z}(\alpha)} H^{\bullet}_{Betti,local,z_i,\gamma} \left( X, \alpha \right)
\end{equation}
\textit{is called the local Betti cohomology.}
\vspace{0.3cm}

\nn
 Since the divisor $D_h$ is empty, as we will see below, the only zeros of $\alpha$ contributing to $\mathcal{Z}(\alpha)$ arise from the domain $X= \mathbb{C}^n \setminus \left\lbrace \mathcal{B}=0 \right\rbrace$. 
If the roots of $d \mathcal{B}$ do not lie on the hypersurface $\mathcal{B}=0$, the set $\mathcal{Z}(\alpha)$ coincides with the set $\Sigma$ of critical points of the polynomial $\mathcal{B}$ within $X$.  Since these are the only points contributing to the construction of the local Betti (co)homology, we can proceed in these cases analogously to the approach described for holomorphic functions.

\nn
Let us now discuss the global-to-local isomorphism for these Betti cohomologies. First, let us fix a new definition of Stokes rays in terms of zeros of the $1$-form $\alpha$ rather than in terms of critical points as done in the case of holomorphic functions.\\[0.5cm]
\textbf{Definition:} [\textsc{Stokes ray, \cite{Kontsevich:2024mks}, Def. 3.9.1}]\\
\textit{We call the ray $s_{\theta} = \left\lbrace \gamma \, \vert \, \arg{\gamma}= \pi - \theta_{ij} = \theta \right\rbrace = \mathbb{R}_{\geq 0} \cdot e^{i \theta} \subset \mathbb{C}_{\gamma}$ with $\theta= \arg \left( \int_{\Gamma_{ij} } \alpha \right)$, where $\Gamma_{ij}$ is the homotopy class of paths in $X$ joining the two points $z_i$ and $z_j$ in $\mathcal Z(\alpha)$, a Stokes ray. \\ Rays with vertex at the origin that are not Stokes rays are called generic rays.}\\[0.3cm]
If $\gamma$ does not lie on a Stokes ray, given the local system $\mathcal{L}_{\alpha, \gamma}$ associated with the holomorphic $1$-form $\gamma \alpha$, we always have a well defined isomorphism
\begin{equation}
    \varphi_{\arg{\gamma}} \, : \, H^{\bullet}_{Betti,glob, \gamma} (X, \alpha) \, \simeq  \, H^{\bullet}_{Betti,local, \gamma} (X, \alpha).
\end{equation}
Close to a Stokes ray $s_{\theta}$, there exist two isomorphisms, $\varphi_{\theta^+}$ and $\varphi_{\theta^-}$,  corresponding to angles immediately adjacent to the ray.  The discrepancy between these isomorphisms is captured by the Stokes automorphism
\begin{equation}
    \varphi_{\theta^-}^{-1} \circ \varphi_{\theta^+} \, : \, H^{\bullet}_{Betti,local, \gamma} (X, \alpha) \, \mapsto \, H^{\bullet}_{Betti,local, \gamma} (X, \alpha). 
    \label{Stokes_automorphism:closed_forms}
\end{equation}
Just as for holomorphic functions, we can also associate a wall-crossing structure to the pair $\left( X, \alpha \right)$ by using the maps \eqref{Stokes_automorphism:closed_forms} as $\gamma$ varies along $S^1_{\theta}$ in $\mathbb{C}^{\ast}_{\gamma}$. For a continuous family of pairs $\left( X, \alpha \right)$, as far as $\pi_0 \left( \mathcal{Z}(\alpha)\right)$ is locally constant, the corresponding wall crossing structures form a continuous family of WCS. 

Now, let us assume that the zeros of $\alpha$ are isolated and simple. Then, for each $z_i \in \mathcal{Z}(\alpha)$, and its associated holomorphic function $\gamma f_i$, we can construct the thimble $th_{i , \theta_{\gamma}}$ emanating from $z_i \in X$ by tracing the vanishing cycles $\Delta_i (s)$ on the level sets $f_i - f_i(z_i)=s$ for $s \in \left[0; + \infty\right)$ along the direction $\theta_{\gamma}$. We say that $th_{i , \theta_{\gamma}}$ is compatible with $\alpha$ in the direction $\theta_{\gamma}$ if, for any point $x \in \Delta_i(s)$, as $s \mapsto + \infty$, we have $\gamma f_i(x) \mapsto + \infty$. In this case, we can define the pairing between the de Rham cohomology and the Betti homology class represented by $th_{i , \theta_{\gamma}}$ via the exponential integral
\begin{equation}
    I_i (\gamma)= \int_{th_{i , \theta_{\gamma}}} e^{- \gamma f_i} \, \mu,
    \label{Pairing:dR_Betti_forms}
\end{equation}
which is well defined only when the integral is convergent. Following a construction analogous to the one described in Section \ref{EIHF}, the integral \eqref{Pairing:dR_Betti_forms} can rewritten as
\begin{equation}
    I_i (\gamma) = e^{- \gamma f_i (z_i)} \int_{0}^{+ \infty} ds\ e^{- \gamma s} vol_{\Delta_i} (s).
\end{equation}
where $vol_{\Delta_i} (s)$ denotes the volume of the vanishing cycle $\Delta_i (s)$ on the level set $f_i - f_i(z_i)=s$. Since $vol_{\Delta_i} (s)$ typically increases as $s \mapsto + \infty$, the converge of the integral \eqref{Pairing:dR_Betti_forms} is ensured if and only if this volume growth is at most exponential.  However, this condition may fail when $\dim_{\mathbb{C}} X \geq 3$.

We finally get to the point. Feynman integrals in the Baikov representation \eqref{Baikov}, involve multivalued logarithmic functions of the form
\begin{equation}
    f(z_1,z_2,\dots,z_n)= \log \mathcal{B} \left( z_1, z_2,\dots,z_n \right).
\end{equation}
The domain of definition of $f$ is the complex manifold  $X=\mathbb{C}^n \setminus \left\lbrace \mathcal{B}=0 \right\rbrace$, which excludes the zero locus of $\mathcal{B}$. A natural compactification of this space, satisfying the conditions outlined in Proposition 3.1.1 of \cite{Kontsevich:2024mks}, is the complex projective space $\overline{X}=\mathbb{P}^n= \mathbb{C}^n  \cup \mathbb{P}^{n-1}$, obtained introducing the additional divisors corresponding to the hyperplane at infinity and the zeros of the homogenized Baikov polynomial:

\begin{equation}
    D_{\overline{\mathcal{B}}}\equiv\{[z_1,\dots,z_n,\zeta]\in \Pro^n\ |\ \overline{\mathcal{B}}(z_1,\dots,z_n,\zeta)=0\},
\end{equation}
where $\overline {\mathcal B}$ is the extension of $\mathcal B$ to the compactification, with values in $\mathbb P^1$.  Notice that $f$ does not extend, but this is totally irrelevant.
The two added divisors intersect non trivially in $D_{\overline{\mathcal{B}}}\cap \Pro^{n-1} = \{[z_1,\dots,z_n,0]\in \Pro^n\ |\ \overline{\mathcal{B}}(z_1,\dots,z_n,0)=0\}$, i.e. the points at infinity of the compactification in $\Pro^n$ of the variety defined by the Baikov polynomial. 
Let us call the hyperplane at infinity
\begin{equation}
    D_\infty \equiv \{[z_1,\dots,z_n,0]\in \Pro^n\ \},
\end{equation}
\nn
and finally write
\begin{equation}
    \overline{X}- X= D_{\overline{\mathcal{B}}}\cup D_\infty.
\end{equation}
\nn
Consider now the globally-defined 1-form 

 \begin{equation}
    \alpha = d \log \overline{\mathcal{B}} = \frac{d \overline{\mathcal{B}}}{\overline{\mathcal{B}}},
\end{equation}
\nn
which clearly shows logarithmic poles along $D_{\overline{\mathcal{B}}}$ with residue $c_{\overline{\mathcal{B}}}=+1$, thus $D_{\overline{\mathcal{B}}}\subset D_{log}$. In order to study the behavior at infinity, let choose the coordinate $\eta^{-1} = \overline{\mathcal{B}}$: as $\eta$ approaches zero, $\alpha=-d\eta/\eta$ exhibits again a logarithmic singularity, but with opposite residue: $c_\infty=-1$. We then finally conclude:

\begin{equation}
    D_{log}=D_{\overline{\mathcal{B}}}\cup D_\infty \quad \quad \mbox{and}\quad\quad  D_h=D_v = \emptyset.
\end{equation}
\nn
The $S^1-$bundle with Euler class $1$ over $D_\infty\cong S^{n}$ is given by the Hopf fibration: $D^\R_\infty\cong S^{2n-1}$. On the other hand $D^\R_{\overline{\mathcal{B}}}$ has Euler class $e=d c_1$, with $d$ the degree of $\mathcal{B}$ and $c_1=c_1(\Ol_{D_{\overline{\mathcal{B}}}})$ the first Chern class of the normal bundle of $D_{\overline{\mathcal{B}}}$ in $\C\Pro^n$.
Applying definitions \eqref{PlusMinusDivisors} we get 

\begin{equation}
    D^{\R+}_{log} =\begin{cases}D^\R_\infty, \quad\quad \mbox{if}\quad Re(\gamma)>0,\\ D^\R_{\overline{\mathcal{B}}},\quad\quad\, \mbox{if}\quad Re(\gamma)<0 ,\end{cases} \quad \quad \mbox{and}\quad \quad D^{\R-}_{log} =\begin{cases}D^\R_\infty, \quad\quad \mbox{if}\quad Re(\gamma)<0,\\ D^\R_{\overline{\mathcal{B}}},\quad\quad\, \mbox{if}\quad Re(\gamma)>0,\end{cases}
\end{equation}
\nn
and the global Betti cohomology \eqref{GlobalBetti1form} becomes
\begin{equation}
    H^{\bullet}_{Betti,glob, \gamma} \left( X, \alpha \right)\cong \begin{cases}H^{\bullet} \left( \widetilde{X}, D^\R_\infty ,\Pi_\ast (\La_{\alpha,\gamma})\right), \quad\quad \mbox{if}\quad Re(\gamma)>0,\\
    H^{\bullet} \left( \widetilde{X}, D^\R_{\overline{\mathcal{B}}}\,, \Pi_\ast (\La_{\alpha,\gamma})\right),\quad\quad\, \mbox{if}\quad Re(\gamma)<0 .\end{cases}
    \label{GlobalBettiLog}
\end{equation}
\nn
For better readability, we now simplify the notation by setting $\La \equiv \La_{\alpha,\gamma}$.
The key of the whole discussion lies in the behavior of the direct image $\Pi_\ast (\La)$ that, by definition, is the extension of $\La$ to $\widetilde{X}$ by zero, 
i.e. the sheaf associating to $U\subset \widetilde{X}$ the group of sections of $\La$ on $U\cap X$. Note that no nonzero sections are entirely supported on the boundary, and since $X$ is a deformation retract of $\widetilde{X}$, we get 

\begin{equation}
    H^k(\widetilde{X},\Pi_\ast \La)\cong   H^k(X,\La).
\end{equation}
\nn
Therefore, the long exact sequence for the pair $(\widetilde{X},D^\R_k)$ reduces to 

\begin{equation}
    \cdots\rightarrow H^k(\widetilde{X},D^\R_k,\Pi_\ast \La) \rightarrow H^k(X,\La)\rightarrow H^k(D^\R_k, \Pi_\ast \La) \rightarrow H^{k+1}(\widetilde{X},D^\R_k,\Pi_\ast \La) \rightarrow \cdots .
    \label{LES}
\end{equation}
\nn
In order to compute \eqref{GlobalBettiLog}, we have thus to determine $H^k(D^\R_k, \Pi_\ast \La)$ and $H^k(X,\La)$. 
Now, let $M_k\in GL(r,\C)$ be the monodromy matrix around $D_k$, with $r$ the rank of $\La$. In order for a global section on $X$ to be flatly extendable to the boundary $D^\R_k$, it must belong to $\ker(M_k-I)$. Indeed, subspaces that are invariant under the action of monodromies determine directions towards the boundary, along which global sections can be analytically continued. If $M_k$ is semisimple  with no eigenvalues $1$ and no product relations (like, e.g., $\det\,(M_1 M_2)=1$) occur, no nonzero section of $\La$ extends along the boundary $\de\widetilde{X}$. In the opposite situation, when all $M_k$ are the identity matrix, all global sections flatly extend along the boundary. In the general case, whenever $M_k$ has some unitary eigeinvalue and/or relations among monodromies appear, some global sections extend, whereas others do not. 
Intuitively, the cohomology relative to a portion of the boundary kills all global sections that can be flatly extended to that boundary. Thus it depends on the possibility to extend global sections, encoded in  $H^k(X,\La)$, and on possible obstructions due to the topology of the boundary, encoded in  $H^k(D^\R_k, \Pi_\ast \La)$. 
Moreover, unless the local system is trivial (e.g. a constant sheaf), the isomorphism $H^k(-,\La)\cong H^k(-,\C)\otimes\La$ fails to be true, due to the torsion of $\La$ on $\C$\footnote{Measured by the $Ext(\C,\La)$.}, and the cohomologies with coefficients in $\La$, significantly depends on the behaviour of the local system, encoded in the monodromy matrices defining it. This contribution must be computed case by case, however, we can compute, in fully generality, the contribution coming from the cohomologies with constant coefficients.  
We assume $D_{\overline{\mathcal{B}}}$ is smooth. In particular, $D_{\mathcal B}\equiv\{z\in\mathbb C^n|\mathcal B(z)=0 \}$ is a smooth affine variety. Therefore, we can proceed as follows. First, the Alexander duality theorem ensures that (here smoothness is irrelevant)
\begin{align}
    H^{k}(\mathbb C^n\backslash D_{\mathcal B}, \mathbb C)\simeq \tilde H_{2n-k-1}(D_{\mathcal B},\mathbb C),
\end{align}
where $\tilde H_*$ is the reduced homology. Next, since $D_{\mathcal B}$ is smooth and has complex codimension $1$, despite being noncompact, we can use the Poincar\'e duality to get (for $k\geq 1$)
\begin{align}
    H^{k}(\mathbb C^n\backslash D_{\mathcal B}, \mathbb C)\simeq \tilde H^{k-1}(D_{\mathcal B},\mathbb C).
\end{align}
Finally, we have just to remember that for $q\geq 1$ one has $\tilde H^q=H^q$, while $\tilde H^0=H^0/\mathbb C$ to get
\begin{equation}
    H^k(X,\C) \cong \begin{cases}\C,\hspace{5cm} k=0\\ 0,\hspace{5.05cm} k=1  \\ H^{k-1}(D_{\mathcal B},\mathbb C)\cong \C^{m_k} , \hspace{2.1cm} k=2,\ldots,n
    \end{cases},
    \label{Primi}
\end{equation}
\nn
with $m_k\equiv \dim H^{k-1}(D_\mathcal{B},\mathbb C)$. The cohomologies of the boundaries components are computed via the Serre spectral sequence applied to the circle bundles $(S^1\rightarrow D^\R_j\rightarrow D_j)$. The second page $E_2^{p,q}=H^p(D_j,H^q(S^1))$ is $E_2^{p,q}=H^p(D_j,\C)$ for $q=0,1$ and it vanishes otherwise. Both spectral sequences degenerate at page $E_2$. The case of $D_{\overline {\mathcal B}}$ clearly depends on $\mathcal B$ and must be computed case by case.\footnote{Notice $D^\R_{D_\mathcal{B}}$ is twisted by $\Ol(degree(\mathcal{B}))$.} In particular, in the Serre spectral sequence the obstruction to lift the classes of $S^1$ to that of the total space of the bundle is $d_2^{p,1}: E^{p,1}_2\rightarrow E^{p+2,0}_2$. It acts as the cup product by the Chern class of the $U(1)$ bundle.\\ 
For $D_\infty$ we have that $H^{2j}(\mathbb P^{n-1},\mathbb C)=\mathbb C$ for $j=0,1,\ldots,n-1$, while are zero otherwise. Note that if we call ${\rm H}$ the hyperplane at infinity of $\mathbb P^{n-1}$, then, the generator of $H^{2j}$ is ${\rm H}^j$. On the other hand, ${\rm H}$ is exactly the first Chern class of the $U(1)$ bundle. Thus, $d_2^{p,1}=\cdot \cup H$ is the cup product by $H$. It maps $H^{2j}(\mathbb P^{n-1},\mathbb C)\rightarrow H^{2j+2}(\mathbb P^{n-1},\mathbb C)$ injectively for $j=0,1,\ldots,n-2$ and $H^{2n-2}(\mathbb P^{n-1},\mathbb C)$ to  $0$. Therefore, the only surviving terms are $E^{0,0}_2\equiv \mathbb C\equiv H^0(D_\infty^{\mathbb R},\mathbb C)$ and $E^{2n-2,1}_2\equiv \mathbb C\equiv H^{2n-1}(D_\infty^{\mathbb R},\mathbb C)$. So
\begin{equation}
    H^k(D^\R_\infty,\C)\cong \begin{cases} \C \quad \mbox{for}\, k=0,2n-1\\ 0 \quad \mbox{otherwise}, \end{cases}
\end{equation}
\nn
and we finally obtain:

\begin{equation}
        H^\bullet (\widetilde{X},D^\R_\infty,\C)\cong \C^{m_2}\oplus..\oplus\C^{m_n},
\end{equation}
concentrated in degrees $2\leq k\leq n$.\\
If $p=D_\infty\cap D_{\overline{\mathcal{B}}}$ is a critical point for $\overline{\mathcal{B}}$, $\La$ contains the monodromy $M_\infty$ around $p$, and the twisted cohomology could be affected by its action. However, even in that case, because $\pi_1(S^{2n-1})=0$, any action is impossible. Therefore: 

\begin{equation}
    H^k(D^\R_\infty,\La)\cong  H^k(S^{2n-1},\C)\otimes\La.
    \label{InfinityCohology}
\end{equation}

\nn
In the following subsection we will use these results to explicitly compute \eqref{GlobalBettiLog} in a concrete example.\\
More examples will be provided in \cite{Angius:2025acm}.

\subsection{Elliptic fibers} \label{SI}
In this paragraph we apply the generalized framework for multivalued functions described above in the following family of integrals:
\begin{equation}
    \mathcal{I}= \int_{\Gamma} \frac{dx \wedge dy}{\left[ y^2+x (x-1)(x- \lambda) \right]^{ \gamma}} = \int_{\Gamma} e^{- \gamma \log \left[ y^2+x (x-1)(x- \lambda) \right]} dx \wedge dy =\int_{\Gamma} e^{- \gamma \log \mathcal{B}(x,y;\lambda)} dx \wedge dy,
\end{equation}
\nn
associated to the Legendre family of elliptic curves $\mathcal{E}_\lambda\equiv\{(x,y)\in \C^2| \mathcal{B}(x,y,\lambda)=0\}$, where $\lambda \in \mathcal{M}_{cs}=\mathbb{C} \setminus \left\lbrace 0,1,\infty \right\rbrace$ is a complex structure parameter. In the physical interpretation, $\lambda$ can be thought of as parameterizing the masses of internal particles or the external momenta.\\
In this example we have $X=\C^2\backslash \mathcal{E}_\lambda$, $\gamma \in \mathbb{C}^{\ast}_{\gamma}$ and the integration contour $\Gamma \in H_2 \left( \C^2, D_0 \right)$ is a singular Borel-Moore $2-$chain with boundaries on the divisor $D_0$ defined as
\begin{equation}
    D_0 = \lim_{N \mapsto \infty} X_N = \lim_{N \mapsto \infty} \left\lbrace (x,y) \in X\ \vert\ Re \left( \gamma \log \mathcal{B}(x,y;\lambda) \right) \geq N \right\rbrace.
\end{equation}
\nn
The natural choice for the compactification is
\begin{equation}
  \overline{X} \, = \, \Pro^2 \,  = \, \mathbb{C}^2 \setminus \left\lbrace \mathcal{E}_{\lambda}=0 \right\rbrace \, \cup \, \left\lbrace \overline{\mathcal{E}}_{\lambda}=0 \right\rbrace  \, \cup \, \mathbb{P}^1,
\end{equation}
 where $\mathcal{B}(x,y;\lambda)$ extends to
\begin{equation}
    \overline{\mathcal{B}}(x,y,\eta;\lambda)=y^2\eta-x(x-\eta)(x-\eta\lambda), 
\end{equation}
\nn
with 
\begin{equation}
    d\,\log \overline{\mathcal{B}} =\frac{2\eta y dy+[y^2+x^2+x\lambda(x-2\eta)]d\eta+[-3x^2-\eta^2\lambda+2x\eta(1+\lambda)]dx}{y^2\eta-x(x-\eta)(x-\eta\lambda)}.
\end{equation}
\nn
By analyzing the behavior of $\overline{\mathcal{B}}$ on $\Pro^2$ we can identify the types of divisors introduced in our compactification. In particular, we find

\begin{equation}
    D_h=D_v=\emptyset \quad\mbox{and}\quad D_{log}=D_{\overline{\mathcal{B}}}\cup D_\infty,
\end{equation}
\nn
with 
\begin{equation}
\begin{split}
    &D_{\overline{\mathcal{B}}}=  \overline{\mathcal{E}}_{\lambda}  = \{[x:y:\eta]\in \Pro^2| \overline{\mathcal{B}}=0\},\\
    &D_\infty = \mathbb{P}^1 = \{[x:y:0]\in \Pro^2\}, 
    \end{split}
    \label{intersectionpoint}
\end{equation}
\nn
intersecting at $D_{\overline{\mathcal{B}}}\cap D_\infty =[0:1:0]$.\\
Since the divisors $D_v$ and $D_h$ are empty, the $1$-form $\alpha$ receives a singular contributions only from the divisor $D_{\log}$. Therefore, we can write
\begin{equation}
    \alpha = \alpha_{\log} + \alpha_{reg}.
\end{equation}
Around $\mathcal{E}_\lambda\subset D_{log}$, the closed $1$-form $\alpha$ is the holomorphic form
\begin{equation}
    \alpha_{\log} = \frac{d \mathcal{B}}{\mathcal{B}},
\end{equation}
whose zeros are
\begin{equation}
\begin{split}
    & \mathcal{Z}(\alpha)= \left\lbrace z_i=(x_i,y_i) \in X \vert d \mathcal{B} (x,y;\lambda)/ \mathcal{B}(x,y,\lambda)=0 \right\rbrace =\\
    &  = \left\lbrace \left( \frac{1}{3} \left[ 1+ \lambda - \sqrt{1 + \lambda (\lambda-1)}\right] ,0 \right), \left(  \frac{1}{3} \left[ 1+ \lambda + \sqrt{1 + \lambda (\lambda-1)}\right], 0 \right) \right\rbrace, \\
\end{split}
\end{equation}
with corresponding critical values
\begin{equation}
 S= \left\lbrace \log t_i \in \mathbb{C} \vert \log \mathcal{B} \left( z_i; \lambda\right) \right\rbrace,
\end{equation}
with
\begin{equation}
     \begin{cases} & t_1= \frac{1}{27} \left(\sqrt{\lambda^2-\lambda+1}-\lambda-1\right) \left(\lambda^2-4 \lambda-(\lambda+1)\sqrt{\lambda^2-\lambda+1}+1\right)\\
    & t_2=  -\frac{1}{27} \left(\sqrt{\lambda^2-\lambda+1}+\lambda+1\right) \left(\lambda^2-4\lambda+(\lambda+1)\sqrt{\lambda^2-\lambda+1}+1\right).\\ \end{cases}
\end{equation}
The map 
\begin{equation}
    \mathcal{B} \, : \quad X \quad \longmapsto \quad \mathbb{C}_t^{\ast},
\end{equation}
defines a non-trivial Lefschetz fibration over $\mathbb{C}_t^{\ast}= \mathbb{C} \setminus \left\lbrace t_1, t_2 \right\rbrace$ for each $\lambda \in \mathcal{M}_{cs}=\C_\lambda\setminus \{0,1,\infty\}$. \\
At fixed $\lambda$, the generic fiber $\mathcal{B}^{-1}(t)$ is the elliptic curve:
\begin{equation}
   \mathcal{F}_t \, : \quad  y^2+x (x-1)(x- \lambda)=t.
\end{equation}
The badness of the fibration at the critical values in $\left\lbrace t_1, t_2 \right\rbrace$ is measured in terms of the local monodromies acting on the homology group $H_{1}(\mathcal{F}_t)$ through the matrices:
\begin{equation}
    M_{i} \, \, : \, \, H_1 (\mathcal{F}_t) \quad \longmapsto \quad H_1 (\mathcal{F}_t) \, \, \quad , \,\,\, i=1,2.
\end{equation}
To determine a basis for this homology and the therein representation of the monodromies we follow the description revisited in Appendix \ref{Appendix}. We fix a non-critical point $t_0$ in $\mathbb{C}_t$ and construct two paths
\begin{equation}
    u_i \, \, : \, \, \left[0;1 \right] \quad \longmapsto \quad \mathbb{C}_t,
\end{equation}
each connecting the non-critical value $t_0=u_i(0)$  to a critical value $t_i =u_i(1)$ without crossing any other critical point. For each path $u_i(t)$ we can define a family of $1$-dimensional spheres in the level manifolds $\mathcal{F}_{u_i}$:
\begin{equation}
    S_i(s)= \sqrt{u_i(s)-t_i} S^1,
\end{equation}
that shrink to zero radius as we approach the critical point $t_i$. The homology classes $\Delta_i \in H_1 (\mathcal{F}_{t_0})$ represented by these spheres are the Picard-Lefschetz vanishing cycles along the paths $u_i$ and they form a basis for the homology $H_1 (\mathcal{F}_{t_0})$. \\
To provide a clear visualization of the construction, we fix the parameter $\lambda=3$ and carry out the explicit computations for this case. As long as $\pi_0 \left( \mathcal{Z} (\alpha) \right)$ remains invariant under variations of  $\lambda \in \mathcal{M}_{cs}$ the wall crossing structures associated with the pairs $(X_{\lambda}, \alpha_{\lambda})$ remain continuously connected to that of $\left( X_3, \alpha_3 \right)$.  The sets $\mathcal{Z}(\alpha)$ and $S$ are:
\begin{equation}
\begin{split}
    &\mathcal{Z}(\alpha) = \left\lbrace z_1=\left( \frac{1}{3} \left(\sqrt{7}+4\right), 0 \right) , z_2 = \left(\frac{1}{3} \left(-\sqrt{7}+4\right), 0 \right)\right\rbrace \, , \\
    & S=\left\lbrace t_1= - \frac{2}{27} \left(7 \sqrt{7}+10\right), t_2= \frac{2}{27} \left(7 \sqrt{7}-10\right) \right\rbrace. \\
\end{split}
\end{equation}
The level manifold $\mathcal{F}_{t_0}$ at the regular point $t_0$ is the graph of the two-valued function
\begin{equation}
    y= \pm \sqrt{t_0-(x-3) (x-1) x},
\end{equation}
namely, the double-covering of the $x$ plane, branched at the points:
{\footnotesize{
\begin{equation*}
\begin{split}
   & x_1= \frac{1}{6} \left(-2^{2/3} \sqrt[3]{3 \sqrt{3} \sqrt{27 t^2+40 t-36}-27 t-20}-\frac{14 \sqrt[3]{2}}{\sqrt[3]{3 \sqrt{3} \sqrt{27 t^2+40 t-36}-27 t-20}}+8\right), \\
   & x_2 = \frac{1}{12} \left(2^{2/3} \left(1-i \sqrt{3}\right) \sqrt[3]{3 \sqrt{3} \sqrt{27 t^2+40 t-36}-27 t-20}+\frac{14 \sqrt[3]{2} \left(1+i \sqrt{3}\right)}{\sqrt[3]{3 \sqrt{3} \sqrt{27 t^2+40 t-36}-27 t-20}}+16\right), \\
   & x_3= \frac{1}{12} \left(2^{2/3} \left(1+i \sqrt{3}\right) \sqrt[3]{3 \sqrt{3} \sqrt{27 t^2+40 t-36}-27 t-20}+\frac{14 \sqrt[3]{2} \left(1-i \sqrt{3}\right)}{\sqrt[3]{3 \sqrt{3} \sqrt{27 t^2+40 t-36}-27 t-20}}+16\right).\\
\end{split}
\end{equation*}}}
Let us choose the first cut from $x_1$ to $x_3$ and the second cut from $x_2$ to infinity.\\
As we move the value of $t$ from $t_0$ to one of the critical values, the level manifold $\mathcal{F}_t$ is deformed and it becomes singular. In particular, when we approach $t_1$ we have that the branch point $x_2$ moves until overlaps with $x_3$, while when we approach $t_2$ the point $x_1$ moves towards the point $x_3$. From this construction we can draw the vanishing cycles $\Delta_1$ and $\Delta_2$ in $\mathcal{F}_{t_0}$ associated to the paths $u_1$ and $u_2$, respectively. The cycle $\Delta_1$ encircles the points $x_2$ and $x_3$, while $\Delta_2$ encircles $x_1$ and $x_3$.

Tracing the change in the positions of the three points $x_j$ as we follow the counterclockwise-oriented closed loop $\tau_i \in \pi_1 \left( \mathbb{C}_t \setminus \left\lbrace t_1, t_2\right\rbrace, t_0 \right)$ encircling the critical point $t_i$ we can deduce the corresponding monodromy action on the ordered basis $\left\lbrace \Delta_1, \Delta_2 \right\rbrace$ of vanishing cycles. In particular, we obtain
\begin{equation}
    M_1 = \left( \begin{matrix}  1 & 0 \\ 1 & 1  \end{matrix} \right) \quad \quad , \quad \quad M_2 = \left( \begin{matrix}  1 & -1 \\ 0 & 1  \end{matrix} \right).
    \label{monodromies:cubic_points}
\end{equation}
 Using the Picar-Lefschetz theorem \eqref{PicardLefschetzFormula}, we can derive the intersection form on $H_1 \left( \mathcal{F}_t, \mathbb{Z}  \right)$ from these monodromies, expressed with respect to the chosen basis of vanishing cycles:
\begin{equation}
     \Delta_i \circ \Delta_j = \left( \begin{matrix} 0 & 1 \\ -1 & 0 \end{matrix}\right).
    \label{intersection:cubic_vanishing_cycles}
\end{equation}
These local monodromies characterize the type of singularity occurring at the critical points $t_1$ and $t_2$. In this particular case, where the fiber is an elliptic curve, we can refer to Kodaira classification of singular fibers \cite{KodairaI,KodairaII} from which we deduce that the singularity at  $t_1$ is of type $II$ (a MUM-point)  while the singularity at $t_2$ is of type $I$ (a conifold point). \\

\paragraph{Betti cohomology.}
The divisors \eqref{intersectionpoint} intersect with normal crossing, thus the real oriented blow-up of $\Pro^2$ along $D_{log}$ is given by union of blow-ups along the two divisors. The normal bundles of $D_\infty\cong S^2$ and $D_{\overline{\mathcal{B}}}\cong T^2$ in $\Pro^2$ are respectively the complex line bundles $\Ol(1)$ and $\Ol(3)$. The corresponding circle bundles respectively are the Hopf fibration $S^3\rightarrow S^2$ and the 3-dimensional Heisenberg nilmanifold $Nil^3\cong H_3(\R)/H_3(\Z) $, so $\widetilde{X}$ has boundary

\begin{equation}
    \de \widetilde{X}=S^3 \cup Nil^3,
\end{equation}
\nn
where $\de D^\R_\infty$ and $ \de D^\R_{\overline{\mathcal{B}}}$ are glued along the corner $S^1\times S^1$, preimage of the intersection point \footnote{Notice that it has multiplicity 3.}.
We want to compute 
\begin{equation}
\begin{split}
    &H^\bullet_{Betti,glob,\gamma}(X,\alpha)(\widetilde{X},S^3,\Pi_\ast(\La_{\alpha,\gamma})), \quad\quad Re(\gamma)>0,\\  &H^\bullet_{Betti,glob,\gamma}(X,\alpha)(\widetilde{X},D^R_{\overline{\mathcal{B}}},\Pi_\ast(\La_{\alpha,\gamma})), \quad \,\,\,Re(\gamma)<0.
    \end{split}
\end{equation}
\nn
using the results discussed in section \ref{SubsectionBetti} and the monodromy matrices obtained in \eqref{monodromies:cubic_points}. 
The contributions coming from $D^{\mathbb R}_\infty \simeq S^3$, can be easily computed by the straight application of \eqref{InfinityCohology}, yielding: 
\begin{align}
    H^{\bullet}(D^\R_{\infty},\mathcal L)\simeq H^{\bullet}(S^3,\mathbb C)\otimes \mathcal L\simeq \mathbb C^2\oplus 0\oplus 0\oplus C^2.
    \label{HS}
\end{align}
\nn
The situation is much more involved in the case of $D^\R_{\overline{\mathcal{B}}}$, defined by the fibration 

\begin{align}
    S^1\hookrightarrow D^\R_{\overline{\mathcal{B}}} \rightarrow D_{\overline{\mathcal{B}}},
\end{align}
which is nontrivial (since is generated by a nontrivial normal bundle of degree 9). The monodromies are nontrivial around the elliptic curve, so we can choose to assign them to a basis of generators of its homotopy $\pi_1$, say $\rho(a)=M_1$, $\rho(b)=M_2$. The Heisenberg structure gives a central extension such that the commutator (in the group theoretical sense) $[a,b]=\iota$ generates the homotopy of the fibre. This means that the representation must respect this relation and we must have
\begin{align}
    \rho(\iota)=M_1 M_2 M_1^{-1}M_2^{-1} =\begin{pmatrix} 2 & 1 \\ 1 & 1.
    \end{pmatrix}
\end{align}
To compute the twisted cohomology we now use the isomorphism
\begin{align}
    H^{\bullet}(D^\R_{\overline{\mathcal{B}}},\mathcal L)=H^{\bullet}(\pi_1(D^\R_{\overline{\mathcal{B}}}),V_\rho),
\end{align}
where $V_\rho$ is the representation space of $\rho$ seen as left $\rho$-module. The Nilmanifold can be represented by a CW-complex obtained by gluing a 3-cell to three 2-cells, next to three 1-cells and finally to a 0-cell. So we have that the j-chains $C^j$ satisfy
\begin{align}
    C^0(D^\R_{\overline{\mathcal{B}}},\mathcal L)\simeq V_\rho, \qquad C^1(D^\R_{\overline{\mathcal{B}}},\mathcal L)\simeq V_\rho^{3}, \qquad C^2(D^\R_{\overline{\mathcal{B}}},\mathcal L)\simeq V_\rho^{3}, \qquad C^3(D^\R_{\overline{\mathcal{B}}},\mathcal L)\simeq V_\rho,
\end{align}
with $V_\rho\simeq \mathbb C^2$. Also, we can use that $H^j(\pi_1(D^\R_{\overline{\mathcal{B}}}),V_\rho)\simeq H^{3-j}(\pi_1(D^\R_{\overline{\mathcal{B}}}),V_{\rho^*})$, where $\rho^*$ is is the dual representation. These representations are irreducible so one finds that 
\begin{align}
   H^j(\pi_1(D^\R_{\overline{\mathcal{B}}}),V_{\rho^*})\simeq H^j(\pi_1(D^\R_{\overline{\mathcal{B}}}),V_\rho), \qquad j=0,1, 
\end{align}
so we can reduce to the computations for $j=0,1$. The differential in the group cohomology is the standard one. We have to consider explicitly the differentials. If 
\begin{align}
    f=\begin{pmatrix}
        f_1\\ f_2
    \end{pmatrix} \in C^0\equiv \mathbb C^2
\end{align}
is a 0-chain, one has that $d^0f$ has to be a 1-chain so, for $g=a,b,\iota$, is defined by
\begin{align}
    d^0f(g)=\rho(g)v-v.
\end{align}
This means that the elements of $H^0$ are the invariant vectors. So
\begin{align}
    0=d^0f(a)=\begin{pmatrix}
        0\\ f_1
    \end{pmatrix}, \qquad 0=d^0f(b)=\begin{pmatrix}
        -f_2\\ 0
    \end{pmatrix}, \qquad 0=d^0f(\iota)=\begin{pmatrix}
        f_1+f_2\\ f_1
    \end{pmatrix},
\end{align}
which gives $H^0(\pi_1(D^\R_{\overline{\mathcal{B}}}),V_\rho)=0$.\\
Similarly, for $f\in C^1\simeq \mathbb C^2\oplus \mathbb C^2\oplus \mathbb C^2$, one has (2-chains acts on pairs $(g,g')$ of elements of $\pi_1$)
\begin{align}
    d^1f((g,g'))=\rho(g)f(g')-f(gg')+f(g).
\end{align}
The explicit calculation of the kernel of $d^1$ is tedious but direct, and we leave the details to the readers. After quotienting by the image of $d^0$ one gets:

\begin{align}
    H^{\bullet}(D^\R_{\overline{\mathcal{B}}},\La)\cong 0\oplus \C^2\oplus \C^2\oplus 0.
    \label{HDB}
\end{align}
\nn
The last piece we have to compute is $H^{\bullet}(X,\mathcal L)$, for which we can apply \eqref{Primi}.
Therefore, we have to compute $H^k(\mathcal{E}_\lambda,\mathcal L)$. $\mathcal{E}_\lambda$ is an affine elliptic curve so $H^2(\mathcal{E}_\lambda,\mathcal L)=0$. $H^0(\mathcal{E}_\lambda,\mathcal L)$ is determined by the vectors of $V_\rho$ invariant under the action of $M_1-I$ and $M_2-I$ which, like before, give $H^0(\mathcal{E}_\lambda,\mathcal L)=0$. The cellular decomposition of $\mathcal{E}_\lambda$ consists in a 2-cell, an two 1-cells (the 0-cell is missing. In cohomology this corresponds to $H^2=0$). Thus the chains are $C^0\equiv V_\rho$, $C^1=V_\rho^2$, $C^2=0$. It follows that $\ker d^1=V_\rho^2\simeq \mathbb C^4$, while ${\rm Im}\, d^0\simeq \mathbb C^2$. We conclude that
\begin{align}
   H^{\bullet}(X,\mathcal L)=0\oplus 0\oplus \mathbb C^2\oplus 0\oplus 0. 
   \label{Hx}
\end{align}
Finally, we can you use \eqref{HS},\eqref{HDB} and \eqref{Hx} in the long exact sequence \eqref{LES}, getting:
\begin{equation}
\begin{split}
     &H^\bullet_{Betti,glob,\gamma}(X,\alpha)(\widetilde{X},S^3,\Pi_\ast(\La_{\alpha,\gamma}))\cong 0\oplus \C^2\oplus \C^2\oplus 0\oplus \C^2,  \\
     &H^\bullet_{Betti,glob,\gamma}(X,\alpha)(\widetilde{X},D^\R_{\overline{\mathcal{B}}},\Pi_\ast(\La_{\alpha,\gamma}))\cong 0 \oplus 0 \oplus \C^2\oplus \C^4\oplus 0.
     \label{LegendreCohmology}
     \end{split}
\end{equation}
\nn
The final result provided by \eqref{LegendreCohmology} shows the middle cohomology, in this case independently on the sign of $\gamma$, is two dimensional, as we expected it to be. 

\begin{figure}[h!]
    \centering
    \includegraphics[width=0.9\linewidth]{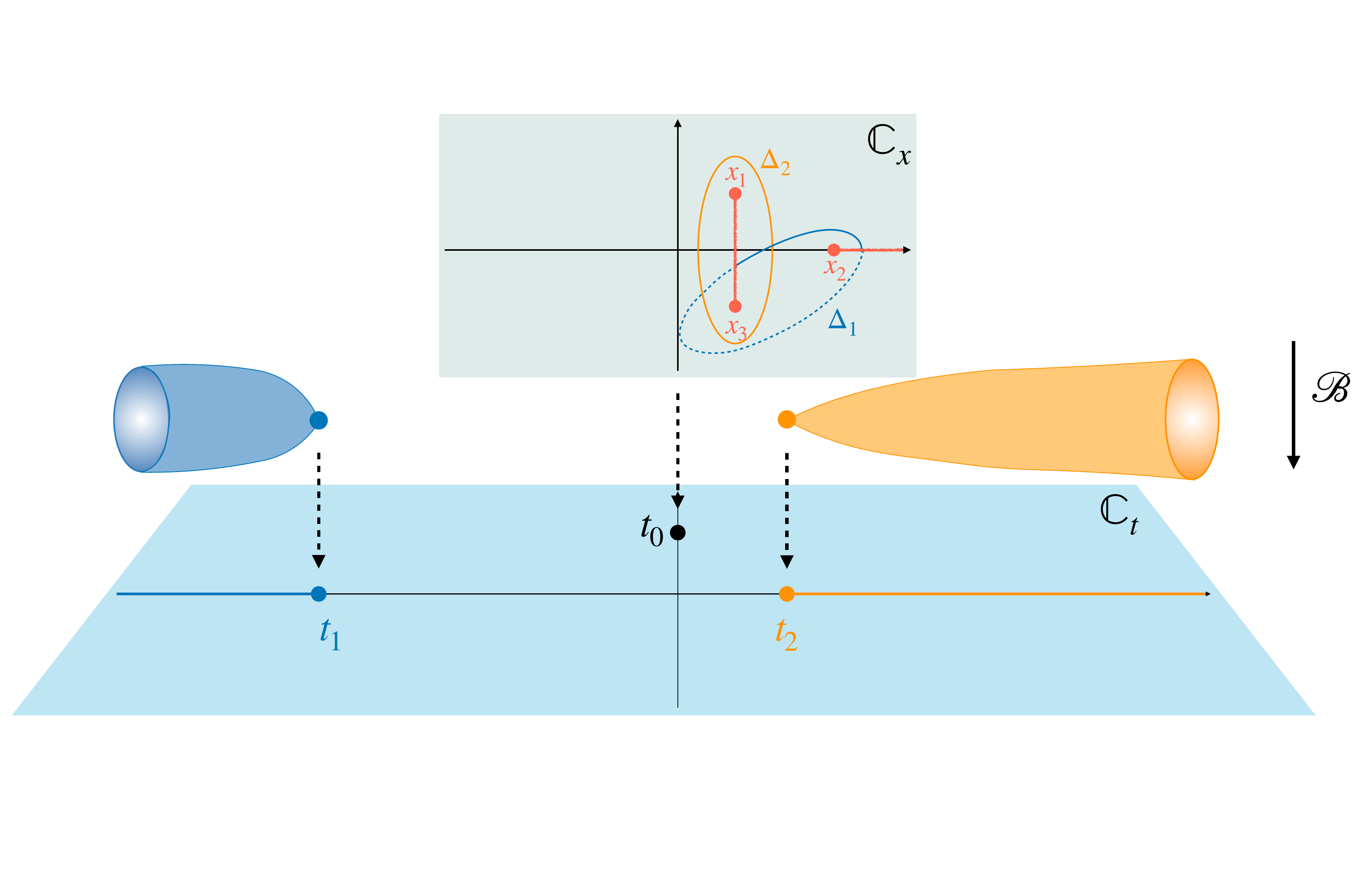}
    \caption{Representation of the construction of the thimbles in the double fibration $log: (P:X \mapsto \mathbb{C}_t) \mapsto \mathbb{C}_T$.}
    \label{fig:thimbles_elliptics}
\end{figure}

\paragraph{Thimbles construction.}
At this stage, we have all the necessary ingredients to construct the thimbles associated with the vanishing cycles $\Delta_1$ and $\Delta_2$, which form a basis for the local Betti homology groups $H_2^{Betti, local,z_i, \gamma} \left( X, \alpha \right)$. As in the case of holomorphic functions, we begin by studying the homology for a fixed $\gamma \in \mathbb{C}^{\ast}_{\gamma}$, and then we analyze its analytic continuation, equipped with a wall-crossing structure.\\
Let us fix $\gamma =1$. The step-ascend thimble associated with the vanishing cycle $\Delta_i$ is defined as the trace over a path in $\mathbb{C}_t^{\ast}$, starting from $t_i$, along which the imaginary part $\rm{Im}(T)=\rm{Im}(\gamma \log t)$ remains constant to the value $\rm{Im}(\gamma \log t_i)=\rm{Im}(\log t_i)=arg(t_i)$, while the real part $Re(\gamma \log t)=Re(\log t)$ increases monotonically from $\log \vert t_i \vert$ to $+ \infty$. A graphical illustration of the results is provided in Figure \ref{fig:thimbles_elliptics}.
\nn
The Stokes rays in the plane $\mathbb{C}_{\gamma}$ are
\begin{equation}
\begin{split}
    & s_{\theta_1} = \left\lbrace \gamma \, \vert \, \arg{\gamma}= \arctan{\frac{\pi}{\log \vert t_1 \vert - \log \vert t_2 \vert}} \right\rbrace, \\
    & s_{\theta_2} = \left\lbrace \gamma \, \vert \, \arg{\gamma}= \arctan{\frac{\pi}{\log \vert t_1 \vert - \log \vert t_2 \vert}} + \pi \right\rbrace, \\
\end{split}
\end{equation}
that never stay along the real axis.\\
Now, if $\gamma$ does not belong to a Stokes ray we can consider the collection of integrals evaluated along the thimbles:
\begin{equation}
    I_i (\gamma)= \int_{th_{i, \theta_{\gamma}}} e^{-\gamma \log \mathcal{B}(x,y)} dx \wedge dy,
\end{equation}
where $\log \mathcal{B}$ is exactly the function $f$ such that $df= \alpha_{\log}$.\\
Using the parameterization $th_{i, \theta_{\gamma}} = \Delta_i (s) \times \mathbb{R}_{s \geq 0} $, we have 
\begin{equation}
    I_i (\gamma) = e^{- \gamma \log t_i} \int_0^{+ \infty} e^{- \gamma s} vol_{\Delta_i}(s) ds,
\end{equation}
where $vol_{\Delta_i} (s)$ is the volume of the vanishing cycle in the fiber $\mathcal{B}^{-1}(s)$ defined with respect to the Gelfand-Leray form $\frac{dx \wedge dy}{d\mathcal{B}/\mathcal{B}}$, namely:
\begin{equation}
    vol_{\Delta_i} (s)= \int_{\Delta_i (s)} \frac{dx \wedge dy}{d \mathcal{B}/ \mathcal{B}} = \int_{\Delta_i(s)} (t_i+s) \frac{dx}{2y(s)}.
\end{equation}
In the integrand we can recognize the holomorphic form $\omega^{1,0}= dx/y$,  so the resulting integrals are precisely the periods of this form with respect to the basis of vanishing cycles for the family of varieties $\mathcal{F}_t$:
\begin{equation}
    \mathbf{\Pi}(t) = \left( \begin{matrix} \int_{\Delta_1} \frac{dx}{y} \\ \int_{\Delta_2} \frac{dx}{y} \end{matrix} \right).
\end{equation}
Since we know the monodromies \eqref{monodromies:cubic_points} around the critical values $t_i$ we can compute the expansions for $0 \leq s \leq \epsilon$ using the Nilpotent Orbit Theorem:
\begin{equation}
    \mathbf{\Pi} (\tilde{s}) = e^{\tilde{s} N_i} \left( \mathbf{a}_0 + \mathbf{a}_1 e^{2 \pi i \tilde{s}}+\dots \right),
\end{equation}
where $N_i$ is the Nilpotent matrix encoding the unipotent part of the monodromy $M_i$:
\begin{equation}
    N_i = \log \left( M_i^{(u)}\right),
\end{equation}
and
\begin{equation}
    \tilde{s}= \frac{1}{2 \pi i} \log s.
\end{equation}

\section{Conclusions}\label{C}
The analysis of physical systems across various domains, from quantum mechanics to statistical physics and from quantum field theory to string theory, often necessitates the computation of increasingly complex integrals. Developing systematic methods to address their computation remains a central challenge in theoretical physics. One of the most powerful techniques for simplifying certain classes of integrals, those expressible as periods of de Rham cocycles over closed cycles on smooth manifolds, is provided by Stokes' theorem. By fixing appropriate bases in the relevant cohomology and homology spaces, the integration of basis cocycles over basis cycles yields a set of simpler integrals encoded in the period matrix. Stokes' theorem then allows any integral within the family to be reduced to a linear combination of these fundamental integrals, whose coefficients are interpretable as intersection numbers in either cohomology or homology.

A natural and compelling extension of this framework would be to encompass broader classes of integrals, particularly those encountered in physics. However, a significant obstacle arises in cases involving multivalued or otherwise intricate integrals, where geometric intuition is lost, and the appropriate cohomology/homology needed to define the pairing required to interpret the integral as a period is no longer evident.

In this work, we propose a systematic approach to identifying the appropriate (co)-homological structures to apply in a large classes of physical integrals. Leveraging recent mathematical developments, we employ twisted de Rham cohomology and Betti homology over complex manifolds to rigorously treat exponential-type integrals as periods. This framework accommodates a wide range of physically relevant integrals, including quantum mechanical partition functions, conformal correlators, and, importantly, Feynman integrals. In the latter case this interpretation becomes viable through a generalization of established techniques for exponential integrals involving holomorphic functions in the exponent, extended to accommodate multivalued functions. Indeed, Feynman integrals expressed in the Baikov representation, as in any parametric representation as well, naturally admit such a reformulation, where the role of the multivalued function is played by the logarithm of a polynomial, the Baikov polynomial.

A key ingredient in this framework is the study of the complex analytic continuation of a real parameter $\gamma$, which appears as a prefactor in the exponent. This continuation induces a wall crossing structure, known in physics as the Cecotti-Vafa wall crossing structure \cite{Cecotti:1992ccv}, on the complex $\gamma$-plane, $\mathbb{C}_{\gamma}^{\ast}$, where four distinct local systems can be defined: local and global versions of twisted de Rham and Betti (co)homologies. The complex plane $\mathbb{C}_{\gamma}$ is partitioned into sectors by Stokes rays, and within each sector, one can define a canonical basis for each of the four (co)homologies. As $\gamma$ crosses a Stokes ray in this fan, these bases undergo discontinuous transformations encoded by Stokes automorphisms.

After a concise review of the mathematical tools required for this analysis, along with a reformulation of the formalism suited to the physical contexts of interest, we present our main ideas for applying these techniques in physics and outline the objectives we aim to achieve.
 The main results of the present work are the following:
\begin{itemize}
    \item We perform an explicit analysis of the wall-crossing structure and associated Stokes phenomena for the Lefschetz thimble decomposition of a class of exponential Pearcey integrals arising in the grand-canonical partition function of gauged Skyrme models, which describe nuclear matter in various pasta phases.
    \item We introduce, using a language accessible to physicists, the recent mathematical framework developed by Kontsevich and Soibelman \cite{Kontsevich:2024mks} for studying wall-crossing structures in exponential integrals involving multivalued functions. We argue that this is the appropriate framework for interpreting Feynman integrals in the Baikov representation as periods.
    \item We present a concrete example of this generalization applied to the Legendre family of elliptic curves. The associated thimble decomposition yields a basis of simplified integrals, expressible in terms of standard elliptic integrals of the first and second kind.
    \item We propose that the decomposition of Feynman integrals into simpler components via this formalism matches the standard notion of Master Integral decomposition. This correspondence offers a more geometric and algebraic perspective on the structure of these integrals. Moreover, we argue that the analytic continuation in the parameter $\gamma$ and the study of the associated wall crossing structure provide a principled method for enumerating Master Integrals, thereby avoiding ambiguities linked to Stokes phenomena.
    \item We analyze the large-parameter asymptotic expansion of exponential integrals expressed over a basis of Lefschetz thimbles, where the expansion coefficients correspond to periods of standard (co)homology classes associated with families of algebraic varieties. The existence of a well-defined pairing between the global de Rham and Betti (co)homology imposes a constraint on these periods: the volume growth of these standard cycles must not exceed an exponential rate.  
\end{itemize}

We believe that this initial analysis already provides a solid background to motivate a more systematic investigation into the applications of these methods in physics. In particular, we propose the following directions for future work:
\begin{itemize}
    \item A detailed study of the implementation of the method to perform Master Integral decompositions in families of Feynman integrals with free external kinematic parameters (such as masses of the internal particles and external momenta). A thorough analysis, including several concrete examples, will be presented in the companion paper \cite{Angius:2025acm}.
    \item An extension of the framework to the analysis of string amplitudes, where the contribution of external states, implemented by the insertion of vertex operators, is controlled by the exponential terms of Koba-Nielsen.   
\end{itemize}

\acknowledgments{We are pleased to thank Vsevolod Chestnov, Wojciech Flieger, Thomas Grimm, Pierpaolo Mastrolia, Simone Noja, Riccardo Re, \'Angel Uranga and Roberto Volpato for helpful discussions and/or comments on the manuscript. 
The work of R.A. has been supported by the ERC Starting Grant QGuide-101042568- StG 2021.}
\newpage
\appendix
\section{Picard-Lefschetz Theory}\label{Appendix}

\begin{figure}[h!]
    \centering    
    \includegraphics[scale=0.15]{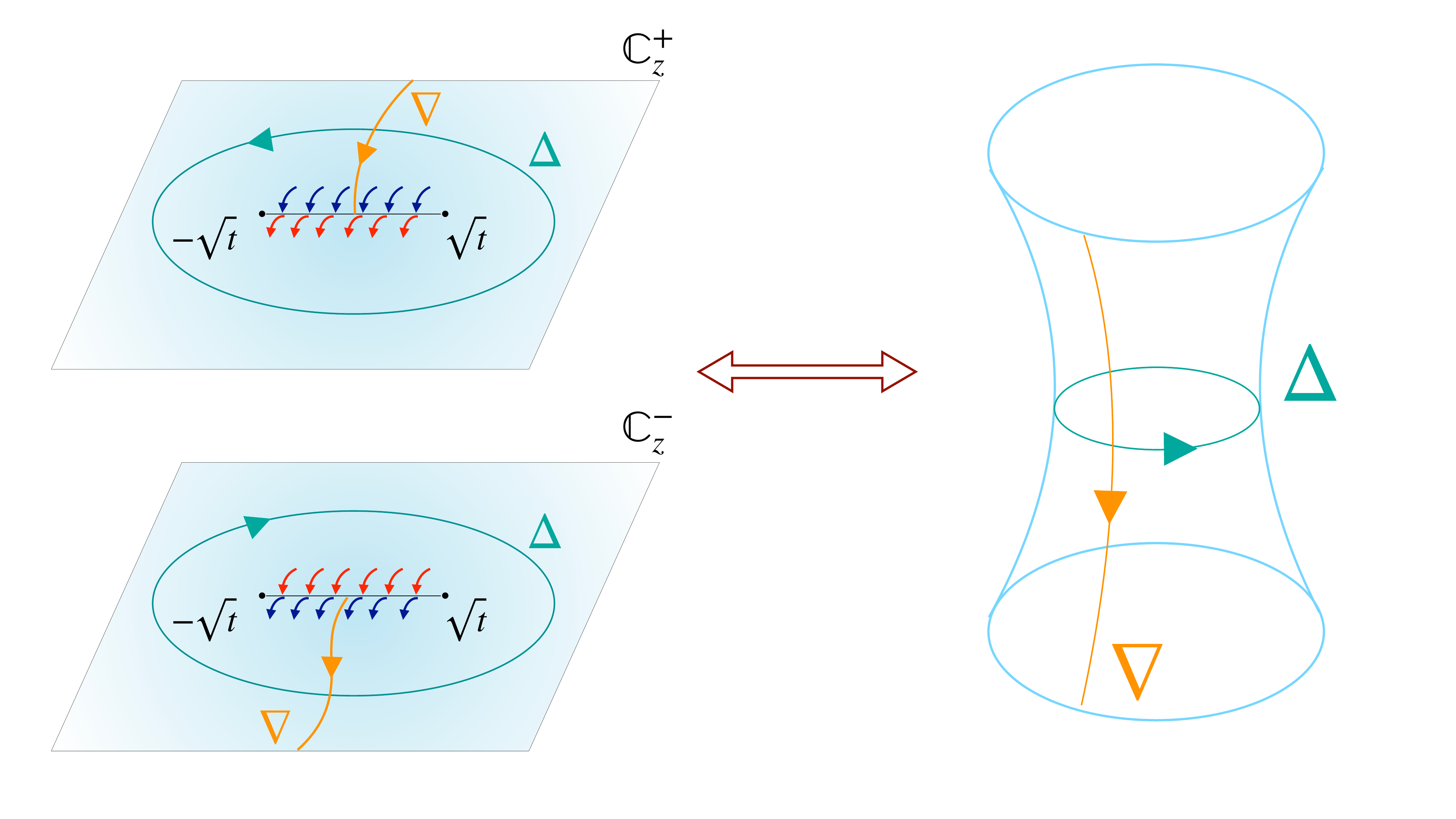}
    \caption{Gluing of the two Riemann sheets along the two edges of the cut from $- \sqrt{t}$ to $\sqrt{t}$. The resulting surface is topologically equivalent to a cylinder.}
    \label{fig1}
\end{figure}

The \textit{Picard-Lefschetz theory} \cite{HuseinSade} is the complex analogous of the Morse theory that studies the topology of level sets of complex analytic functions.\\
Let us begin by considering the following simple example of a two variable function:
\begin{equation}
    f(z,w)=z^2+w^2.
    \label{function_ex1}
\end{equation}
The function $f(z,w)$ has a unique critical point:
\begin{equation}
    \begin{cases} \partial_z f(z,w)=0 \\ \partial_w f(z,w)=0 \end{cases} \quad \quad \longmapsto \quad \quad (z,w) =(0,0).
\end{equation}
We refer to the value of the function $f(z,w)$ at a critical point as a \textit{critical value}, in the present case $f(0,0)=0$. The \textit{critical set} is the set of points in $\mathbb{C}^2$ where the function $f$ takes the critical value:
\begin{equation}
    V_0= \left\lbrace (z,w) \vert z^2+w^2=0 \right\rbrace. 
\end{equation}
For any other value $f(z,w)=t$, different from the critical one, we call \textit{level sets} the loci
\begin{equation}
    V_{t} = \left\lbrace (z,w) \vert z^2 + w^2 = t \right\rbrace.
\end{equation}
In order to figure out the topology of these level sets we can consider the Riemann surfaces associated with the function \eqref{function_ex1}
\begin{equation}
    w = \sqrt{(t - z^2)}.
\end{equation}
\nn
These surfaces can be obtained gluing together two copies of the complex plane $z$ with a cut along the segment $(- \sqrt{t}, \sqrt{t})$, as showed in Figure \ref{fig1}, resulting in a surface topologically equivalent to a cylinder. When $t=0$, the corresponding critical level set consists of two lines intersecting at the point $0$.

\nn
Consider now the fibration $f: \mathbb{C}^2 \mapsto \mathbb{C}_{t} \setminus \left\lbrace 0 \right\rbrace$ over the space $\mathbb{C}_{t}^{\ast} = \mathbb{C} \setminus \left\lbrace 0 \right\rbrace $, which fibers are Riemann surfaces representing the non-critical level sets $V_{t}$. Note that we removed from the base the point $t=0$, which correspond to the singular fiber $V_0$. Let us now consider the circular path around $t=0$
\begin{equation}
    t (\tau)= e^{2 \pi i \tau} \alpha \quad \quad 0 \leq \tau \leq 1 , \quad \alpha >0.
    \label{closed_curve}
\end{equation}

\begin{figure}[h!]
    \centering    
    \includegraphics[scale=0.2]{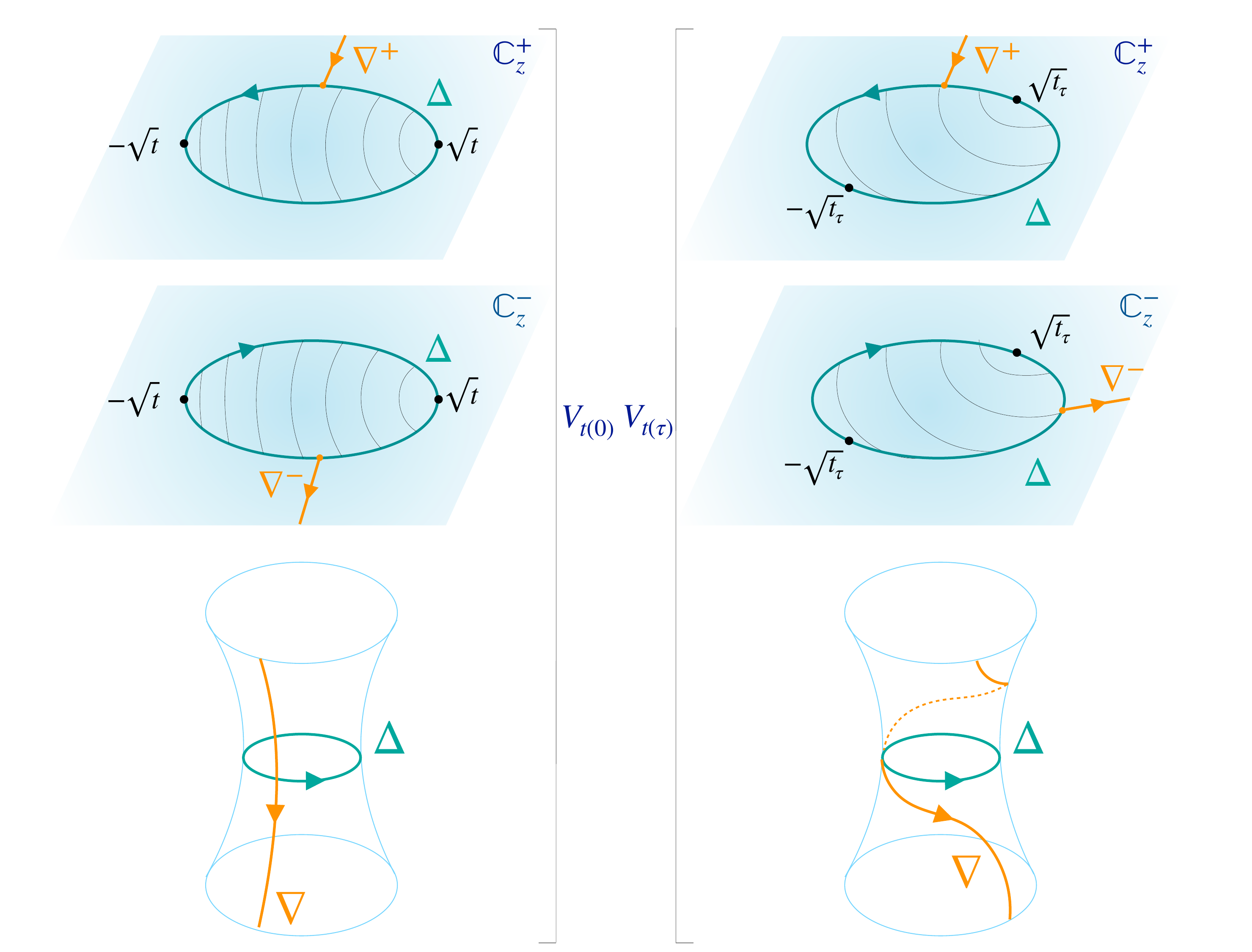}
    \caption{Action of $\Gamma_\tau$ on the vanishing and covanishing cycles. }
    \label{fig2}
\end{figure}
\nn
We aim to study how the fiber $V_{t}$ varies along this path  by tracking the motion of the branch points $z= \pm \sqrt{t(\tau)}$ in the complex planes $\mathbb{C}_z^\pm$ as the parameter $\tau$ evolves. We observe that these branch points rotate counterclockwise around $z=0$ undergoing a half-turn (rotation by $\pi$) at $\tau=1$ (see Figure \ref{fig2}). 
Considering the fibration $V\longrightarrow I$, with $I\equiv [0,1]$, defined by $V_{t(\tau)}\mapsto \tau$, we can associate to the closed curve $t(\tau)$ in \eqref{closed_curve} a continuous map
\begin{align}
    \Gamma: V\longrightarrow V_{t(0)},
\end{align}
such that for any $\tau$, the map $\Gamma_\tau: V_{t(\tau)}\longrightarrow V_{t(0)}$ defined by $\Gamma_\tau(\cdot)=\Gamma(\tau,\cdot)$ is a diffeomorphism, and $\Gamma_0=id$. Since $t(1)=t(0)$, the corresponding map $h=\Gamma_1: V_{t(0)}\longrightarrow V_{t(1)}$ is called \textit{monodromy map}.

\begin{figure}[h!]
    \centering    
    \includegraphics[scale=0.2]{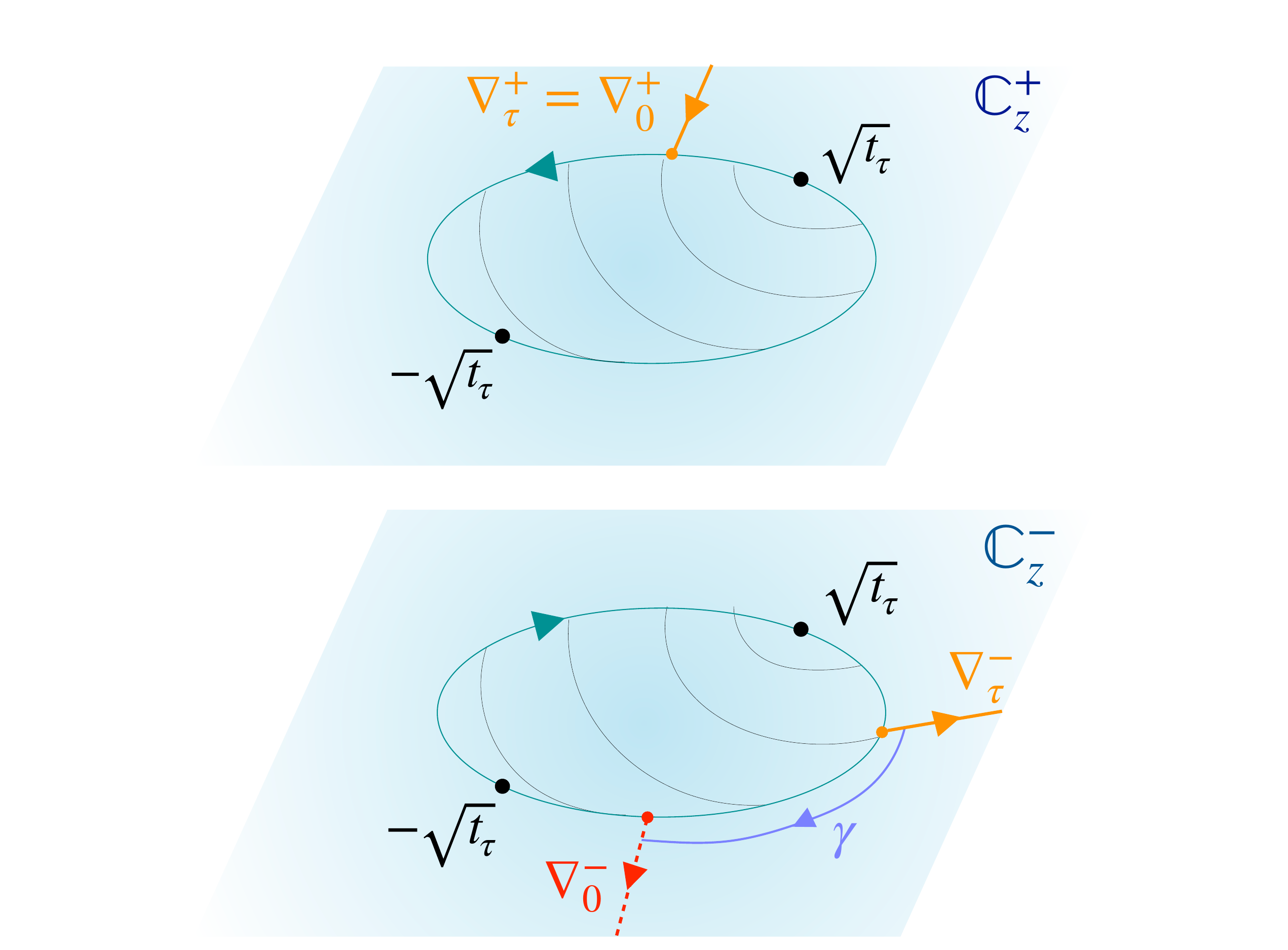}
    \caption{Homotopically deformation of $\nabla^-_\tau$ over $\nabla^-_0$. }
    \label{fig3}
\end{figure}

We are interested to know how this map acts on the first homology group of the fiber $V_{t}$, which is generated by the 1-cycle $\Delta$ represented in Figure \ref{fig1}. This $1$-cycle $\Delta$ is called  \textit{Picard-Lefschetz vanishing cycle}, due to the fact it shrinks to a point when $t \rightarrow 0$. Its transversely intersecting cycle $\nabla$ is called \textit{covanishing cycle} and it generates the first homology group $H^{BM}_1 \left( V_{t}, \mathbb{Z} \right) \simeq \mathbb{Z}$. Here, $H^{BM}_1 (V_{t})$ is the first Borel-Moore homology of the non-compact space $V_{t}$. This homology admits chains that may be infinite in extent but are restricted to be finite in any compact region. A vertical line in the cilinder $V_{t}$ is locally finite because, in any compact sub-interval of the vertical direction the line is finite. Even though the line can extend indefinitely along the cylinder, within any small, bounded region it is just a finite segment. Since $H^{BM}_1 \left( V_{t}, \mathbb{Z} \right) \simeq H_1 \left( C, \partial C ; \mathbb{Z} \right)$, where $C$ is the compact cylinder, it is easy to prove that $H^{BM}_1 \left( V_{t}, \mathbb{Z} \right) \simeq \mathbb{Z}$ with generator $\nabla$.  \\
In Figure \ref{fig2} we show how the diffeomorphism $\Gamma_\tau$ acts on these two cycles. In the $\mathbb{C_z^-}$ foil, after the action of $\Gamma_\tau$, $\nabla^-$ is rigidly transported along the counterclockwise direction of the rotation. As depicted in Figure \ref{fig3} we can deform homotopically the support of $\nabla^-_{\tau}$ to the support of $\nabla^-_0$ through a connected path $\gamma$ in $\mathbb{C}_z^-$ such that:
\begin{equation}
    \nabla_\tau = \nabla^+_\tau + \nabla^-_\tau  \thickapprox \nabla^+_0 + \tilde{\nabla}^- = \nabla^+_0 + \gamma + \nabla_0^-.
\end{equation}
In terms of the homology we have:
\begin{equation}
    \left[ \nabla_\tau \right] \sim \left[ \nabla_0 \right] + \left[ \gamma \right] \sim \left[ \nabla_0 \right],
\end{equation}
if $\gamma$ in contractible. If $\tau=1$ the path $\gamma$ closes around the hole and 
\begin{equation}
    \left[ \nabla_1 \right] = \nabla_0 + \Delta.
\end{equation}

\nn In particular, one gets that, up to homotopies, the monodromy map $h$ acts as the identity outside a compact set around $z=0$ and non-trivially inside this set. More precisely, it maps the vanishing cycle $\Delta$ into itself and it acts as the (homotopically equivalent) identity  in the part of $\nabla$ extending outside the compact set and with the following transformation inside the set:
\begin{equation}
    h \nabla = \nabla - \Delta.
\end{equation}
It is worth to mention that while the map $h$ defined on the fibration $V\to I$ depends on the choice of the diffeomorphisms $\Gamma$, when induced to the (co)homology it becomes independent on such a choice.
This action allows us to define a function, called \textit{variation map}, mapping a cycle with closed support to a cycle with compact support:
\begin{equation}
    \begin{split}
         Var \quad : \quad H_1^{BM} \left( V_{t}, \mathbb{Z} \right) \quad & \longmapsto \quad H_1 \left( V_{t}, \mathbb{Z} \right), \\
         \nabla  \quad & \longmapsto \quad (\Delta \circ \nabla) \Delta, \\
    \end{split}
\end{equation}
where $(\Delta \circ \nabla)$ denotes the intersection pairing between the cycles $\Delta$ and $\nabla$, defined as the number of topological intersections counted with a sign depending by the relative orientation of the two cycles.

The main objects we defined so far are the \textit{vanishing cycles}, the \textit{monodromy} and the \textit{variation map} for a two-variables function. Defining these objects for arbitrary functions of several variables is a challenging problem that remains unsolved in general. Picard-Lefschetz theory offers a powerful method to address this problem by using \textit{deformation theory} techniques. 

\subsection{Monodromy, variation operators and vanishing cycles}
In this paragraph, we extend the above discussion to the case of holomorphic functions in several complex variables and provide formal definitions of the concepts introduced in the previous example.\\
Let
\begin{equation}
    f : \mathcal{M}^n \quad \longmapsto \quad \mathbb{C}
\end{equation}
be a holomorphic function on a $n-$dimensional complex manifold $\mathcal{M}^n$.  Let $U$ be a contractible compact region in the target space with smooth boundary $\partial U$, and let us assume $f$ has a finite number of critical points $\Sigma= \left\lbrace \sigma_i \right\rbrace_{i=1}^{\mu}$ with critical values $t_i = f \left( \sigma_i \right)$ on $\overline{U}$. Let us indicate with $F_t$ the level set of the function $f$ at $t \in U$:
\begin{equation}
    F_t = \left\lbrace \mathbf{z}\in \C^n \vert f(\mathbf{z})=t \right\rbrace.
\end{equation}
If $t \in U$ is not a critical value, then $F_t$ is a $(n-1)-$dimensional complex manifold with smooth boundary. Let $t_0$ be a non-critical value in the boundary $\partial U$ and let us construct for each class of loops $\left[  \gamma \right] \in \pi_1 \left( U \setminus \left\lbrace t_i \right\rbrace , t_0 \right)$ a continuous family of mappings $\Gamma_\tau : F_{t(\tau)} \mapsto F_{t_0}$,  for which $\Gamma_0= id$. Then, $h_{\gamma} = \Gamma_1$, transforming the non-singular level $F_{t_0}$ into itself, defines the \textit{monodromy map}  along the loop $\gamma$. Note that the map $h_{\gamma}$ depends on the specific path $\gamma$ we are considering.\\\\
\textbf{Definition} \textsc{[Monodromy operator]:} \\
\textit{We call monodromy operator of the loop $\gamma$ the action $h_{\gamma \ast} = h_{\left[ \gamma \right]}$ of the transformation $h_{\gamma}$ on the homology of the non-singular level set $H_{\ast} \left( F_{t_0} \right)$.} \\\\
The transformation $h_{\gamma}$ also induces an automorphism $h_{\left[ \gamma \right]}^{(r)}$ in the relative homology group $H_{\bullet} \left( F_{t_0}, \partial F_{t_0} \right)$ of the non-singular level set $F_{t_0}$ modulo its boundary. This homology is isomorphic to the homology of cycles with closed support:
\begin{equation}
    H_{\bullet} \left( F_{t_0}, \partial F_{t_0} \right) \simeq H^{BM}_{\bullet} \left( F_{t_0} \setminus \partial F_{t_0} \right).
\end{equation}
Since the action $h_{\gamma}$ is trivial on the boundary $\partial F_t$, then the difference between $h_{ \left[ \gamma \right]}^{(r)} \delta$ and $\delta \in H_{\bullet} \left( F_{t_0}, \partial F_{t_0} \right)$ is a cycle in $H_{\bullet} \left( F_{t_0} \right)$.\\\\
\textbf{Definition }\textsc{[Variation]}: \\
{\it The homomorphism 
\begin{equation}
    Var_{\gamma} \quad : \quad H_{\bullet} \left( F_{t_0}, \partial F_{t_0} \right) \quad \longmapsto \quad H_{\bullet} \left( F_{t_0}\right)
\end{equation}
is called the variation operator over the loop $\gamma$.} \\\\
Using the natural homomorphism
\begin{equation}
    i_{\ast} \quad : \quad H_{\bullet} \left( F_{t_0}\right) \quad \longmapsto \quad H_{\bullet} \left( F_{t_0}, \partial F_{t_0} \right)
    \label{map1}
\end{equation}
induced by the inclusion $F_{t_0} \subset \left( F_{t_0}, \partial F_{t_0} \right)$, we can write the following relations connecting the automorphisms $h_{\left[ \gamma \right]}$ and $h_{\left[ \gamma \right]}^{(r)}$:
\begin{equation}
  \begin{split}
      & h_{\left[ \gamma \right]} = id + Var_{\gamma} \cdot i_{\ast} \\
      & h_{\left[ \gamma \right] }^{(r)} = id + i_{\ast} \cdot Var_{\gamma}.
  \end{split}  
\end{equation}
If the class $\left[ \gamma \right] \in \pi_1 \left( U \setminus \left\lbrace z_i \right\rbrace , z_0 \right)$ is given by $\left[ \gamma \right] = \left[ \gamma_1 \right] \cdot \left[ \gamma_2 \right]$,\footnote{Note that in the composition of homology classes, we follow the convention of right multiplication.} then 
\begin{equation}
\begin{split}
    & Var_{\gamma} = Var_{\gamma_1} + Var_{\gamma_2} + Var_{\gamma_2} \cdot i_{\ast} \cdot Var_{\gamma_1} \\
    & h_{\left[ \gamma \right]} = h_{\left[ \gamma_2 \right]} \cdot h_{\left[ \gamma_1 \right]}\\
    & h_{\left[ \gamma \right]}^{(r)} = h_{\left[ \gamma_2 \right] }^{(r)} \cdot h_{\left[ \gamma_1 \right] }^{(r)}.\\
\end{split}
\end{equation}
\nn
Let us suppose all the critical points are non-degenerate and the corresponding critical values are different:\footnote{This second requirement is not strictly necessary to define a Morse function.} such a function is said to be \textit{Morse}.\\[0.5cm]
\textbf{Definition} \textsc{ [Monodromy group]}:\\
\textit{The map }
\begin{align}
    \pi_1 \left( U \setminus \left\lbrace z_i \right\rbrace , z_0 \right) \quad &\longrightarrow \quad Aut \left( H_{\bullet} (F_{t_0})\right),\\
    [\gamma] \quad & \longmapsto h_{\gamma \ast}
\end{align}
\textit{is called monodromy representation of $\pi_1 \left( U \setminus \left\lbrace t_i \right\rbrace , t_0 \right)$. The imagine of this map defines what we call the Monodromy group of the Morse function $f$.}\\[0.5cm]
Now, we construct a path $u: \left[0,1 \right] \mapsto U$ joining the non-critical value $t_0=u(0) \in \partial U$  to some critical value $t_i = u(1) \in U$ without crossing any other critical value. The Morse lemma tells us that, given a holomorphic Morse function $f$, it always exists a local set of coordinates in a neighbourhood of the non-degenerate critical point $p_i$ such that the function takes the form
\begin{equation}
    f(z_1 ,\dots, z_n) = t_i + \sum_{j=1}^n  z_j^2.
\end{equation}
Then, for each path $u$, we can define a family of $(n-1)-$dimensional spheres in the level manifolds $F_{u(\tau)}$. For each point of the path $u(\tau)$ the level set $F_{u(\tau)}$ is a hyperboloid equivalent to a trivial fibration with base a $(n-1)-$dimensional sphere of radius$\sqrt{|u(\tau)-t_i|}$:
\begin{equation}
    S(\tau)= \sqrt{u(\tau)-t_i} S^{n-1}.
\end{equation}
In particular, we have that the sphere $S(1)$ reduces to the critical point $p_i$. \\[0.5cm]
\textbf{Definition} \textsc{[Vanishing cycle of Picard-Lefschetz]}:\\ \textit{The homology class $\Delta \in H_{n-1} \left( F_{t_0} \right)$ represented by the $(n-1)-$dimensional sphere $S(0)$ in $F_{t_0}$ is called vanishing cycle of Picard-Lefschetz along the path $u$.} \\[0.5cm]
Note that the homotopy class of $u \in U$ uniquely defines the homology class of the vanishing cycle $\Delta$ modulo orientation. \\[0.5cm]
\textbf{Definition} \textsc{[Distinguished Basis]}:\\
\textit{The set of cycles $\Delta_1, \dots, \Delta_{\mu} \in H_{n-1} \left(F_{t_0} \right)$, with $t_0$ non-singular, is called distinguished if:
\begin{itemize}
    \item[(i)] The cycles $\Delta_i$ are vanishing along non-self-intersecting paths $u_i$ reaching the critical values $t_i$;
    \item[(ii)] The unique common point of $u_i$ and $u_j$ for $i \neq j$ is $u_i(0)=u_j(0)=t_0$; 
    \item[(iii)] The paths $u_i, \dots, u_{\mu}$ are numbered in the order in which they enter to the point $t_0$ counting clockwise starting from the boundary $\partial U$ of $U$. \\
\end{itemize}}

\begin{figure}[h!]
    \centering    
    \includegraphics[scale=0.15]{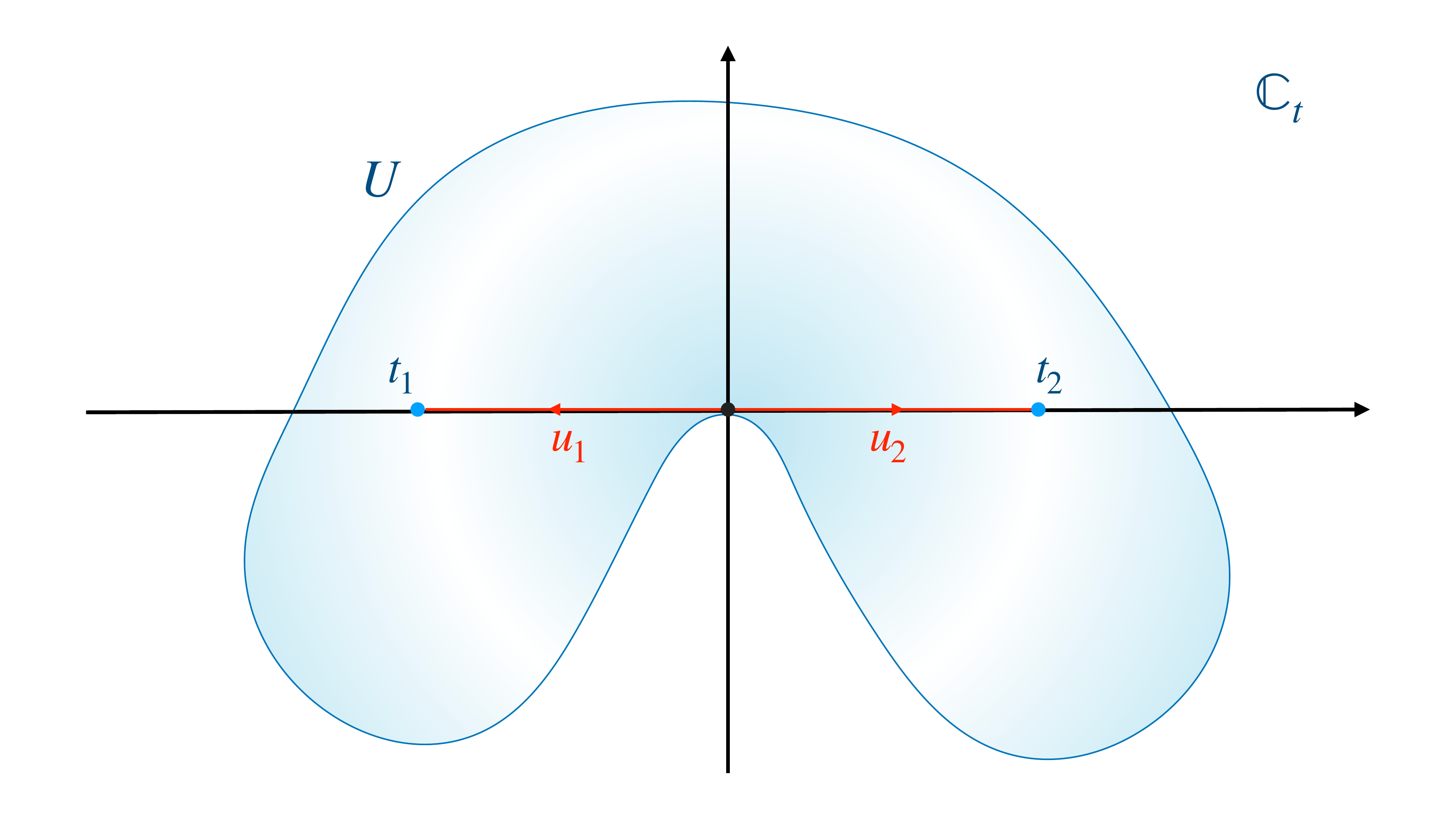}
    \caption{Choice of the set U and the paths $u_1$ and $u_2$ in the codomain of the function $f(z)=z^3-3 \lambda z$.}
    \label{fig4}
\end{figure}

\noindent
\textbf{Example 1:} $f(z)=z^3-3 \lambda z$.\\
Let us consider the Morse function $f(x)=z^3-3 \lambda z$, with $\lambda \in \mathbb{R}_+$, and let us construct a distinguished basis of vanishing cycles. This function is a deformation of the function $f(z)=z^3$ and it has two critical points in the real line 
\begin{equation}
    \mathcal{C}: \quad \bar z_1= \sqrt{\lambda}, \quad \quad \bar z_2 = -\sqrt{\lambda},
\end{equation}
with corresponding critical values are
\begin{equation}
    t_1= - 2 \lambda \sqrt{\lambda}, \quad \quad t_2 = 2 \lambda \sqrt{\lambda}.
\end{equation}
Let us choose as non-critical reference point $t_0=0$ and let us construct the paths $u_1$ and $u_2$ connecting the critical values with $t_0$. The level manifold at $t_0=0$ consists of three points:
\begin{equation}
    F_{t_0}: \quad z^3-3 \lambda z =0 \quad \rightarrow z_1=- \sqrt{3 \lambda}, \quad z_2=0, \quad z_3= \sqrt{3 \lambda}.
\end{equation}
In this example the level manifold for a generic regular point $t$ is given by the condition $f(z)=t$ which admits three point solutions. The vanishing cycles, when we approach the critical values $t_1$ and $t_2$, are the differences
\begin{equation}
    \Delta_1 = \left\lbrace z_3 \right\rbrace - \left\lbrace z_2 \right\rbrace, \quad \quad \quad \Delta_2 = \left\lbrace z_2 \right\rbrace - \left\lbrace z_1 \right\rbrace 
\end{equation}
between the zeroth homology classes represented by the points.\\
Choosing the set $U$ as depicted in Figure \ref{fig4} the cycles $\Delta_1$ and $\Delta_2$ form a distingushed basis for $H_{1} \left(F_{t_0} \right)$.\\[0.5cm]

\noindent
\textbf{Example 2:} $f(x,y)=x^3-3 \lambda x + y^2$.\\
In this second example we consider the function of two variables $f(x,y)=x^3-3 \lambda x + y^2$, which is a deformation through the small real parameter $\lambda$ of the function $f(x,y)=x^3+y^2$. The set of critical points in $\mathbb{C}^2$ with their corresponding critical values is
\begin{equation}
    \mathcal{C}: \quad \begin{cases}  \, P_1 \, : \, \, \left( x, y\right) = \left( \sqrt{\lambda} , 0 \right)  \quad &\rightarrow \quad t_1= -2 \lambda \sqrt{\lambda}\\ 
     \, P_2 \, : \, \, \left( x, y\right) = \left( - \sqrt{\lambda} , 0 \right)  \quad & \rightarrow \quad t_2= 2 \lambda \sqrt{\lambda}.\\
    \end{cases}
\end{equation}

\begin{figure}[h!]
    \centering    
    \includegraphics[scale=0.15]{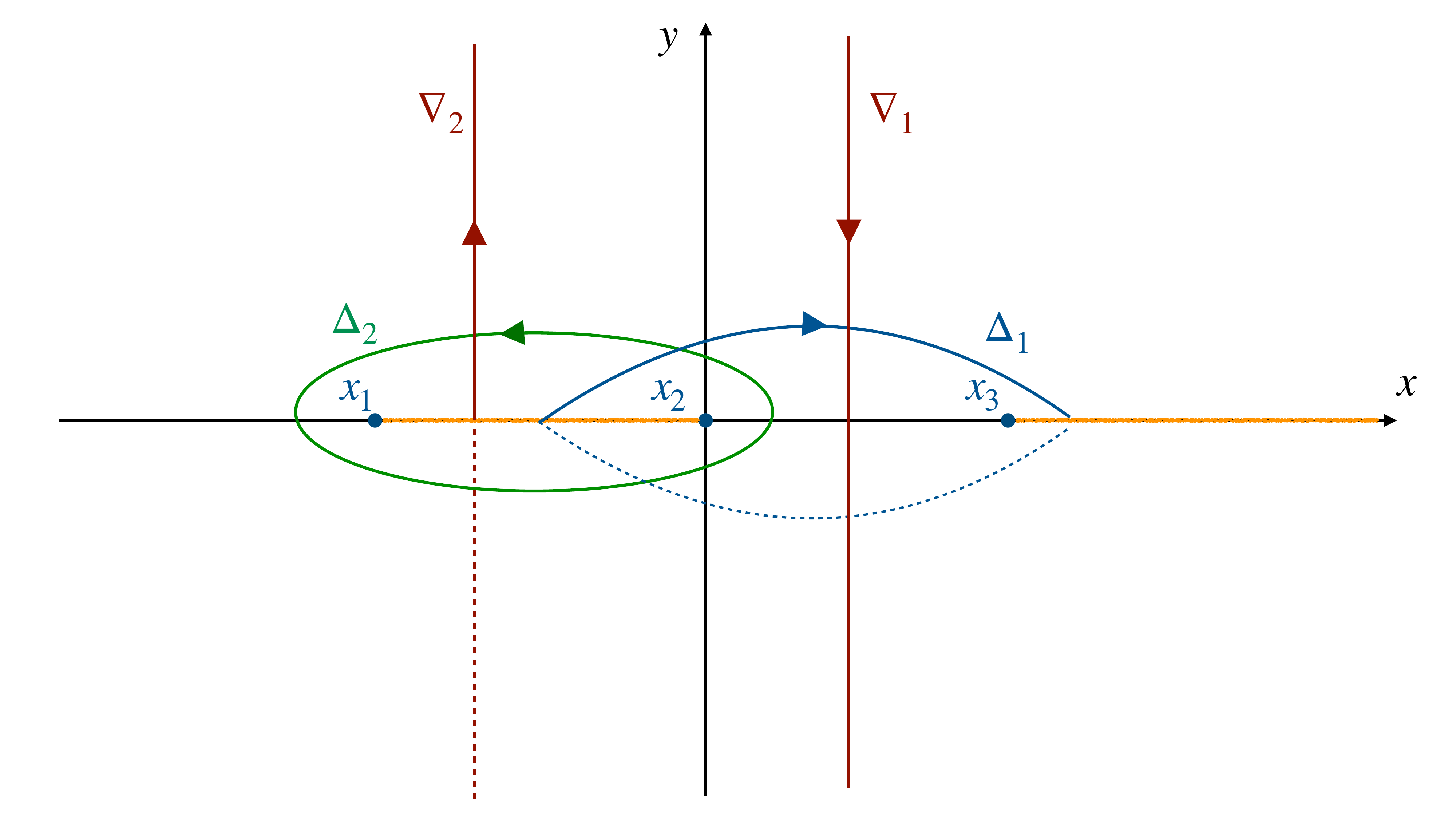}
    \caption{Vanishing and Covanishing cycles.}
    \label{fig5}
\end{figure}

\nn As in the previous example we can consider the paths $u_1$ and $u_2$ joining the two critical values with the non-critical value $t_0=0$. 
The level manifold in this regular point is the graph of the two-valued function $y= \pm \sqrt{x^3-3 \lambda x}$, namely the double-covering of the $x$ complex plane branched between the points $x_1= - \sqrt{3 \lambda}$ and $x_2=0$ and $x_3= \sqrt{3 \lambda}$ and infinity. \\
As we move the value of $t$ from $0$ to one of the two critical values, the level manifold $f(x,y)=t$ is deformed and becomes singular at $t=t_1$ and $t=t_2$. In particular, when we approach $t_1$ we have that the branch point $x_2$ moves until it overlaps $x_3$, while, when we approach $t_2$ the point $x_2$ moves towards the point $x_1$. From this construction, we can draw the vanishing cycles corresponding to the paths $u_1$ and $u_2$: we obtain $\Delta_1$ encircling the points $x_2$ and $x_3$, and $\Delta_2$ encircling the points $x_1$ and $x_2$ (see Figure \ref{fig5}).

\nn\textbf{Definition} \textsc{[Simple loops]}: \\ \textit{A simple loop is an element $\tau_i$ of $\pi_1 \left( U \setminus \left\lbrace t_i \right\rbrace, t_0 \right)$ represented by the loop going along the path $u_i$ from $t_0$ to $t_i$, then encircling $t_i$ with a anticlockwise path and returning along $u_i$ to $t_0$.}\\[0.5cm]
The region $\left( U \setminus \left\lbrace t_i \right\rbrace_{i=1}^{\mu}, t_0 \right)$ is homotopically equivalent to a bouquet of $\mu$ circles. Then, the fundamental group $\pi_1 \left( U \setminus \left\lbrace t_i \right\rbrace, t_0 \right)$ is a free group with $\mu$ generators $\tau_1, \tau_2,\dots,\tau_{\mu}$. \\[0.5cm]
\textbf{Definition} \textsc{[Weakly distinguished]}:\\ \textit{The set of vanishing cycles $\Delta_1,\dots,\Delta_{\mu}$ defined by the paths $u_1,\dots,u_{\mu}$ is called weakly distinguished if $\pi_1 \left( U \setminus \left\lbrace t_i \right\rbrace, t_0 \right)$ is the free group generated by the simple loops $\tau_1,\dots,\tau_{\mu}$ associated to the paths $u_1,\dots,u_{\mu}$.}\\[0.5cm]
We have that if the paths $\left\lbrace u_i \vert i=1,\dots, \mu \right\rbrace$ define a weakly distinguished set of vanishing cycles $\Delta_i$ in the $(n-1)-$homology group of the non-singular level manifold, then, the monodromy group of the function $f$ is generated by the monodromy operators $h_{\tau_i \ast} = h_{\left[ \tau_i \right]}$. Hence, the monodromy group of $f$ is always a group generated by $\mu$ generators. \\[0.5cm]
\textbf{Definition} \textsc{[Picard-Lefschetz operator]}: \\ \textit{The monodromy operator}
\begin{equation}
    h_i =h_{\tau_i \ast} \, : \, \, H_{\bullet} \left( F_{t_0} \right) \, \longmapsto \, H_{\bullet} \left( F_{t_0} \right) 
    \label{monodromy_ops_simple_loops}
\end{equation}
\textit{of the simple loop $\tau_i$ is called the $i^{\rm th}$ Picard-Lefschetz operator.}\\[0.5cm]
In the Example 1, we can take trace of the change of the position of the three points $z_i$ when we move $t$ along the paths $\tau_1$ and $\tau_2$. We observe that along the path $\tau_1$ the point $z_2$ approaches the point $z_3$, then they make a half-turn around a common centre and move again away one from the other. The point $z_1$ stays fixed. Then, we deduce the following monodromy action on the vanishing cycles:
\begin{equation}
    h_1 \Delta_1 = - \Delta_1, \quad \quad \quad \quad \quad \quad h_1 \Delta_2 = \Delta_1 + \Delta_2.
\end{equation}
In the same way we can deduce
\begin{equation}
    h_2 \Delta_1 = \Delta_1 + \Delta_2, \quad \quad \quad \quad \quad \quad h_2 \Delta_2 = - \Delta_2.    
\end{equation}
The cycles $\Delta_i$ are in the homology group $H_{n-1} \left( F_t \right)$ of the non singular level manifold $F_t$. Moreover, we are interested also in the homology group $H_{n-1} \left( F_t , \partial F_t \right)$, which is dual to the group $H_{n-1} \left( F_t \right)$. In the present case it is generated by two cycles $\nabla_i$ such that
\begin{equation}
    \left( \nabla_i \circ \Delta_j \right) = \delta_{ij}.
\end{equation}
We can choose
\begin{equation}
    \nabla_1 =  \left\lbrace z_3 \right\rbrace \quad , \quad \nabla_2 = - \left\lbrace z_1 \right\rbrace,
\end{equation}
for which we have the following variations:
\begin{equation}
    \begin{split}
         Var_{\tau_1} \nabla_1 = \left\lbrace z_2 \right\rbrace - \left\lbrace z_3 \right\rbrace = - \Delta_1&, \quad \quad Var_{\tau_1} \nabla_2 = 0, \\
         Var_{\tau_2} \nabla_1 = 0&, \quad \quad Var_{\tau_2} \nabla_2 = - \left\lbrace z_2 \right\rbrace + \left\lbrace z_1 \right\rbrace = - \Delta_2. \\        
    \end{split}
\end{equation}
We can now consider the loop $\tau = \tau_2 \tau_1$ that turns around the point $t_0=0$ encircling the two critical values $t_1$ and $t_2$ in a positive counterclockwise direction. The monodromy transformation associated to this loop permutes the points $z_3 \mapsto z_2 \mapsto z_1 \mapsto z_3$, then,
\begin{equation}
    \begin{split}
        & h_{\tau} \Delta_1 = \left\lbrace z_2 \right\rbrace -\left\lbrace z_1 \right\rbrace = \Delta_2, \\
        & h_{\tau} \Delta_2 = \left\lbrace z_1 \right\rbrace -\left\lbrace z_3 \right\rbrace =- \Delta_1 - \Delta_2, \\
    \end{split}
\end{equation}
and
\begin{equation}
    \begin{split}
        & var_{\tau} \nabla_1 = \left\lbrace z_2 \right\rbrace -\left\lbrace z_3 \right\rbrace = - \Delta_1,\\
        & var_{\tau} \nabla_2 = - \left\lbrace z_3 \right\rbrace +\left\lbrace z_1 \right\rbrace = - \Delta_1 - \Delta_2.
    \end{split}
\end{equation}
The monodromy group of the Morse function $f$ is generated by the Picard-Lefschetz operators $h_{\tau_1}$ and $h_{\tau_2}$. All the elements of this group preserve the intersection product of the group $H_0 \left( F_t \right)$, for $t$ non-critical, generated by the vanishing cycles $\Delta_i$. The monodromy group is the group $S_3$ of permutations of three elements.\\[0.5cm]
Now, let us construct the monodromy group for the Example 2. Drawing the analogous of the Figure \ref{fig3}, we can deduce the action of the Picard-Lefschetz operator on the vanishing cycles to be
\begin{equation}
    \begin{split}
       h_1 \Delta_1 = \Delta_1& , \quad  h_1 \Delta_2 = \Delta_1 + \Delta_2, \\
       h_2 \Delta_1 = \Delta_1 - \Delta_2& , \quad h_2 \Delta_2 = \Delta_2,
    \end{split}
\end{equation}
and the following variation on the dual cycles:
\begin{equation}
    \begin{split}
        Var_{\tau_1} \nabla_1 = - \Delta_1 \quad & , \quad Var_{\tau_1} \nabla_2 = 0, \\
        Var_{\tau_2} \nabla_1 =0 \quad & , \quad Var_{\tau_2} \nabla_2 =- \Delta_2.
    \end{split}
\end{equation}
The monodromy group of the Morse function $f(x, y)$ is isomorphic to the group of non-singular $2 \times 2$ integer matrices with determinant $1$. The group is generated by the action of the Picard-Lefschetz operators on the vanishing cycles, given by
\begin{equation}
    M_1 = \left( \begin{matrix} 1 & 0 \\ 1 & 1 \end{matrix} \right), \quad \quad \quad \quad M_2 = \left( \begin{matrix} 1 & -1 \\ 0 & 1 \end{matrix} \right).
\end{equation}

These methods, explicitly shown in one or two complex dimensions, can in principle be extended to higher dimensions to determine vanishing cycles, their duals, and the action of the monodromy group on them. 

\subsection{Picard-Lefschetz Theorem}
The Picard-Lefschetz theorem establishes a relation between the variation of (co)-vanishing cycles due to the action of the monodromy operator with their intersection product in $H_{n-1} \times H_{n-1}^{\vee} \mapsto \mathbb{Z}$.\\
Let us start considering the simple loop $\tau_i$ associated to the path $u_i$ connecting  the non-critical reference point $t_0 \in \mathbb{C}_t$ with the critical value $t_i \in \mathbb{C}_t$. Let us assume that the critical value is $t_i=0$, so that in some local coordinates around the critical point $P_i \in \mathbb{C}^{n}$, we can write the function $f$ in the form
\begin{equation}
    f(z_1, \dots, z_n) = \sum_{j=1}^n z_j^2.
\end{equation}
If we intersect $f^{-1} (t_0)$ with the ball $\sum_j \vert z_j \vert^2 \leq 4 \epsilon^2$, the non-critical value $t_0$ is sufficiently close to the critical value $0$, say $\vert t_0 \vert = \epsilon^2$. We can suppose that all other critical values of $f$ are outside the disk of radius $4 \epsilon^2$ in $\mathbb{C}_t$, so that our simple loop encircles just one singularity.\\
Let us define the ball $\overline{B}_{2\epsilon}$ of radius $2\epsilon$ in the space $\mathbb{C}^n$,
\begin{equation}
    \overline{B}_{2\epsilon} = \left\lbrace \left( z_1 , \dots , z_n \right) \vert r \leq 2\epsilon \right\rbrace
\end{equation}
and let us call $\tilde{F}_t$ the intersection of the level set $F_t$ with this ball. \\[0.5cm]
\textbf{Lemma 1:} \\ \textit{For $\vert t \vert < 4\epsilon^2$, the level set $F_t$ is transverse to the $(2n-1)$-dimensional sphere $\partial \overline{B}_{2\epsilon}$.}\\[0.5cm]
From this lemma it follows that for $0 < \vert t \vert <4\epsilon^2$ the sets $\tilde{F}_t = F_t \cap \overline{B}_{2\epsilon}$ are diffeomorphic manifolds with boundary, while $\tilde{F}_0$ is a cone with vertex in zero. \\[0.5cm]
\textbf{Lemma 2:} \\ \textit{For $0 < \vert t \vert < 4 \epsilon^2$, the manifold $\tilde{F}_t$ is diffeomorphic to the disk sub-bundle of the tangent bundle of the standard $(n-1)$ dimensional sphere $S^{n-1}$.}\\[0.5cm]
From this second lemma follows the following result: \\[0.5cm]
\textbf{Lemma 3:} \\
\textit{The self-intersection number of vanishing cycle $\Delta$ in the complex manifold $\tilde{F}_{\epsilon^2}$ is equal to}
\begin{equation}
\left( \Delta \circ \Delta \right) = (-1)^{(n-1)(n-2)/2} \left( 1 + (-1)^{n-1}\right) \, = \, \begin{cases}
    0 \quad & \text{for} \quad n=0 \, \text{mod} \, 2, \\
    +2 \quad & \text{for} \quad n=1 \, \text{mod} \, 4, \\
    -2 \quad & \text{for} \quad n=3 \, \text{mod} \, 4. \\
\end{cases}
\label{SelfIntersection1}
\end{equation}

\nn Poincar\'e duality for a compact manifold $X$ of dimension $n$ states that $H^k(X)\sim H_{n-k}(X)$. If $X$ is noncompact, while for cohomology it is not a problem, for homology one has to introduce Borel-Moore homology for which one has $H^k(X)\sim H^{BM}_{n-k}(X)$, see \cite{Bredon}. Hence, in our case, we get $H_{k} \left(  \tilde{F}_{\epsilon^2}, \partial \tilde{F}_{\epsilon^2} \right)\sim H^{BM}_{n-k} \left(  \tilde{F}_{\epsilon^2} \right)$, and $H_{k} \left(  \tilde{F}_{\epsilon^2}, \partial \tilde{F}_{\epsilon^2} \right)\sim H^{k} \left(  \tilde{F}_{\epsilon^2} \right)\sim H^k(S^{n-1})$.
Therefore, the relative homology group $H_k  (\tilde{F}_{\epsilon^2} , \partial \tilde{F}_{\epsilon^2})$ is zero for $k \neq n-1$, while $H_{n-1}(\tilde{F}_{\epsilon^2} , \partial \tilde{F}_{\epsilon^2})$ is isomorphic to the $\mathbb{Z}$. Moreover the latter is generated by the relative cycle $\nabla$ dual to $\Delta$ such that $\Delta \circ \nabla =1$. \\
In general, a relative cycle $\delta \in H_k\left( F_{\epsilon^2}, \partial F_{\epsilon^2}\right)$ can be represented in the form
\begin{equation}
    \delta = \delta_1 + \delta_2
\end{equation}
\nn
where $\delta_1 \in H_k \left( \tilde{F}_{\epsilon^2}, \partial \tilde{F}_{\epsilon^2} \right)$ and $\delta_2$ is a chain in $F_{\epsilon^2} \setminus B_{2\epsilon}$. The transformation $h_{\tau}=\Gamma_1$ is the identity in $F_{\epsilon^2} \setminus B_{2\epsilon}$, hence, it acts non-trivially only on the cycle $\delta_1$. Therfore, $Var_{\tau} (\delta) = Var_{\tau} (\delta_1)$. \\
Since $H_k \left( \tilde{F}_{\epsilon^2} , \partial \tilde{F}_{\epsilon^2} \right)= \langle \nabla \rangle \simeq \mathbb{Z} $, then $\delta_1 = m \cdot \nabla$, with $m \in \mathbb{Z}$ and $m= \delta \circ \Delta $, and, in order to compute the action of the variation operator on $H_k\left( \tilde{F}_{\epsilon^2} , \partial \tilde{F}_{\epsilon^2} \right) $, it is sufficient to calculate its action on $\nabla$.\\[0.5cm]
\textbf{Theorem} \textsc{[Picard-Lefschetz]}:\\ {\it Under the above hypotheses}
\begin{equation}
    Var_{\tau} \left( \nabla \right) = (-1)^{n(n+1)/2} \Delta.
\end{equation}
\nn
It follows from this:\\[0.5cm]
\textbf{Corollary:} \\
\textit{For $a \in H \left( F_{t_0}, \partial F_{t_0}\right)$:}
\begin{align}
    Var_{\tau} (a) &= (-1)^{n(n+1)/2} \left( a \circ \Delta \right) \Delta,\\
    h^{(r)}_{\tau} (a) &= a+(-1)^{n(n+1)/2} \left( a \circ \Delta \right) i_{\ast}\Delta;
\end{align}
\textit{for $a \in H_{n-1} \left( F_{t_0}\right)$}
\begin{equation}
    h_{\tau} (a) = a+(-1)^{n(n+1)/2} \left( a \circ \Delta \right) \Delta,
    \label{PicardLefschetzFormula}
\end{equation}
\textit{where $i_{\ast}$ is the homomorphism \eqref{map1}.}

\bibliographystyle{unsrt}  
\bibliography{ZZ_bibliography}

\end{document}